\documentclass[twocolumn,samsmath,amssymb,floatfix,unsortedaddress,aps]{revtex4-1}
\usepackage{amsmath}
\usepackage{fixmath}
\usepackage{latexsym,amsfonts,mathrsfs,color,graphicx}
\begin{document}

\title{Orientational order and morphology of clusters of self-assembled Janus swimmers}%
\author{Francisco Alarcon}
\affiliation{Departamento de Estructura de la Materia, F\'isica T\'ermica y Electr\'onica and GISC,
Universidad Complutense de Madrid 28040 Madrid, Spain}
\author{Eloy Navarro-Argem\'i}
\affiliation{Departament de F\'isica de la Mat\`eria Condensada, Universitat de Barcelona, C. Mart\'i Franqu\'es 1, 08028 Barcelona, Spain}
\affiliation{University of Barcelona Institute of Complex Systems (UBICS), Universitat de Barcelona, Barcelona, Spain}%
\author{Chantal Valeriani}
\email{cvaleriani@ucm.es}
\affiliation{Departamento de Estructura de la Materia, F\'isica T\'ermica y Electr\'onica and GISC,
Universidad Complutense de Madrid 28040 Madrid, Spain}
\author{Ignacio Pagonabarraga}
\affiliation{Departament de F\'isica de la Mat\`eria Condensada, Universitat de Barcelona, C. Mart\'i Franqu\'es 1, 08028 Barcelona, Spain}
\affiliation{University of Barcelona Institute of Complex Systems (UBICS), Universitat de Barcelona, Barcelona, Spain}
\affiliation{CECAM, Centre Europ\'een de Calcul Atomique et Mol\'eculaire, \'Ecole Polytechnique F\'ed\'erale de Lasuanne, Batochime, Avenue Forel 2, 1015 Lausanne, Switzerland}%

\date{\today}

\begin{abstract}

Due to the combined effect of anisotropic interactions and activity, Janus swimmers are capable to self-assemble in a wide variety of structures, many more than their equilibrium counterpart. 
This might lead to the  development of novel active materials
capable of performing tasks without any central control. 
Their potential application in designing  such materials endows 
trying to understand the fundamental mechanism in which these swimmers self-assemble. 
 In the present work, we study a quasi-two dimensional semi-dilute suspensions of two classes of amphiphilic spherical swimmers whose direction of motion can be tuned: 
 either swimmers propelling in the direction of the  hydrophobic patch (WP), or swimmers propelling in the opposite direction (towards the hydrophilic side) (AP). 
 In both systems we have systematically  tuned   swimmers' hydrophobic strength and  signature, and 
  observed that
  the anisotropic interactions, characterized by the angular attractive potential and its interaction range, 
in  competition with the active stress,  pointing towards or against the attractive patch  
    gives rise to a  rich aggregation phenomenology. 
\end{abstract}

\maketitle

\section{\label{Intro}Introduction}
In the last decade,  the pioneer experiments by Dombrowski and Cisneros \cite{Dombrowski, Cisneros} 
inspired the numerical work  aimed at   understanding  the influence of  hydrodynamics  in the formation of coherent structures of micro-organisms, 
simulated as spherical squirmers \cite{Ishikawa2006}  in 3D  \cite{Ishikawa2007a, Ishikawa2008JFM} and  quasi-2D \cite{Ishikawa2008PRL}. 
The features of a 3D squirmer suspension have been  numerically characterized 
 using the smoothed profile method \cite{molinaSoftMatt13,molina_PRE16,JPS_yamamoto} 
and  Lattice Boltzmann (LB), either in bulk  \cite{alarcon2013} or in response to  a steady Couette flow  \cite{LlopisSoftMatter13}.
While Evans et al.  \cite{evansPoF} used Stokesian Dynamics to detect the instability of the isotropic state in a 3D suspension of weak pullers, 
Delmotte and coworkers \cite{Delmotte} applied a force-coupling method to investigate, by large-scale simulations, 
the emergence of a polar order in a 3D squirmer suspension.  


More recently, experiments by Ref.~\cite{SoftMatter_35, SoftMatter_33, PRL_bacteriaFastest,Cecille, Thutupalli} have inspired 
numerical simulations of  2D and quasi-2D  suspensions of squirmers interacting via a repulsive potential. 
When a suspension (simulated by means of multi-particle collision dynamics) was confined  between parallel walls (quasi-2D geometry) \cite{Zottl_prl}, 
active particles were forming clusters whose size depended on  the system's concentration with the largest being made of 
pullers. 
The suspension prepared in such confined geometry would undergo 
 phase separation into a  dilute  and a dense phase \cite{Zottl_softmatter16}, differently to 
 the 2D suspension of disk-like squirmers  \cite{PRE_Ricard}, 
where hydrodynamics suppressed 
  phase separation. 
  
  Inspired by the self-assembly observed in 
 experiments with  either bacteria \cite{PRL_bacteriaFastest} or active colloids \cite{Cecille} 
 and by numerical works on 2D dilute suspensions of 
active brownian particles (ABP) interacting via isotropic short-range attractions \cite{SoftMatter_26,SoftMatter_27,SoftMatter_28,SoftMatter_43}, 
 we have  studied a semi-diluted quasi 2D  suspensions of attractive squirmers  
 \cite{SoftMatter17}, to unravel the role played by hydrodynamics  in suspensions of  attractive active particles. 

{Most of the nowadays synthetic active colloids are colloids partially coated with Pt. These colloids self-propell by a catalytic reaction of H$_2$O$_2$ and O$_2$ taking place at the Pt surface. This mechanism of colloid's propulsion is due to phoresis.
On the one side, properly modeling the phoretic interactions can be quite complicated \cite{phoretic39,phoretic40,phoretic41,phoretic42,phoretic43,phoretic44}, eventhough there are experiments, simulations and theories that consider them relevant for particles' propulsion. 
On the other side, which interactions dominate in these systems are still a matter of debate \cite{JCP_Benno2019}.}

{From the experimental point of view, several studies have focused on understanding the effect of hydrophobicity in catalytic Janus colloids. Modifying  the surface of the particles, the authors of Ref. \cite{Manjare_HydrophobicJanus} found that particles with an hydrophobic patch would move faster than the ones with an  hydrophilic patch.}
{Similarly, Gao et al \cite{Gao_OTSJanusExp}, studied self-assembly of amphiphilic particles made with hydrophobic
octadecyltrichlorosilane (OTS)-modified silica microspheres capped with a catalytic Pt hemisphere patch.}
{Additionally, Yan et al \cite{Dipolar_Granick} have recently reported different collective states of active dipolar Janus colloids, whose motion is originated by an induced-charge electrophoresis. These colloids self-assemble in different morphologies, ranging from: gas, swarms, chains and clusters, by modifying both the charge imbalance of the colloids and the electric field frequency. Moreover, in \cite{Dipolar_Granick}, the authors were able to numerically reproduce all observed states by means of an overdamped molecular dynamics simulation whose particles were interacting via directed imbalanced interactions.}

{Motivated by these two experimental works \cite{Gao_OTSJanusExp,Dipolar_Granick}, simulations of a dilute quasi-2D suspension of ABP  
  interacting via  anisotropic amphiphilic Janus interactions have been reported in  
Ref.~\cite{Stewart_17}. To the best of our knowledge, 
this is the first numerical attempt to unravel the collective behaviour of amphiphilic active Janus particles, taking into account their anisotropic nature. In Ref.~\cite{Stewart_17} both the effect of interaction directionality and the propulsion speed have been
used to control the physical properties of the assembled active aggregates. One might suggest that these two parameters could be easily tuned also in experiments, given that the concentration of a cationic surfactant could be used to reverse the propulsion's direction and adding pH neutral salts could be used to control the propulsion speed \cite{Brown_Poon_SaltJanus}.} 

In order to establish the relevance of hydrodynamics in this suspension, we study the collective behaviour   of quasi 2D suspensions of spherical squirmers 
interacting via an anisotropic Janus  potential, 
considering different interaction ranges,  interaction strengths and hydrodynamic signatures.
When clusters appear, we 
 characterise  their morphology  as a function of the    above mentioned features. 
{The objective of this manuscript is the analysis of  minimal models that contain the essential ingredients of activity and steric interactions for heterogeneous  particles. The scenarios we identify can be clearly correlated with the competition between self-propulsion, stress generation, and  attraction among particles. It is true that  phoresis is more complex and can also involve long range interactions among the active elements. Our results serve as a benchmarking scenario if phoresis is controlled by near field effects.  In this sense, the systematic analysis we have carried out  provides a valuable contribution in this general perspective. In a more specific perspective, this study also helps to understand the effect of the hydrodynamic interactions in the collective dynamics of active Janus colloids, inspired by experiments showing flow fields around a single catalytic Janus sphere \cite{Campbell_FlowFieldJanus}.}

 The manuscript is organised as follows. In section \ref{model} we present  the anisotropic interaction model, 
the squirmer model, the lattice Boltzmann methodology and simulations details. 
In section \ref{sec:Tools} we describe the tools  used to characterise the Janus squirmers. 
In section \ref{sec:Results} we present all  results and  discuss our conclusions in section \ref{sec:Conclusions}.

\section{\label{model}Simulation details}


In this work, we  have studied  a quasi two dimensional dilute suspension of squirmers~\cite{Lighthill,Blake,Pedley:2016} 
interacting via a Janus anisotropic  potential 
similar to the one used in Ref. \cite{Stewart_17} for Active Brownian Particles (ABP). 
In order to model the Janus swimmers,  
 we have introduced a pair potential $V(r_{ij},\theta_i,\theta_j)$ which depends 
on the distance between two squirmers ($r_{ij}$) and their attractive patches' orientations
\begin{eqnarray}
\label{eq:JanusGralPotential}
V(r_{ij},\theta_i,\theta_j)= V_{rep}(r_{ij})+V_{att}(r_{ij})\phi(\theta_i,\theta_j)\,  
\end{eqnarray}
where 
$\theta_i$ is the angle  between the patch unit vector $\vec{p_i}$ and the inter-particle vector $\vec{r}_{ji}=\vec{r}_j - \vec{r}_i$; 
and $\theta_j$  the angle  between  $\vec{p_j}$ and  $\vec{r}_{ij}=-\vec{r}_{ji}$.

 $V_{rep}(r_{ij})$ represents  
a short range soft  repulsion  \cite{Allen_Tildesley} (to avoid  overlapping)  given by
\begin{flushleft}
\begin{eqnarray}\label{eq:SSPotential}
\begin{split}
&V_{rep}\left(r_s\right)= \\ 
&\begin{cases}
\epsilon_s \left[ \left(\frac{\sigma_s}{r_{s}}
\right)^{\nu}-\left( \frac{\sigma_s}{h_{0}}
\right)^{\nu}\left(1-\frac{(r_s-h_0)\nu}{h_0} \right) \right] & 0<r_{s}\leq h_0\\
0 & otherwise\,,\\
\end{cases}
\end{split}
\end{eqnarray}
\end{flushleft}

\noindent
where $\epsilon_s=1.5$  is the energy scale,  $\nu=2$, 
 $r_s \equiv r_{ij}-\sigma$  is the inter-particle distance
(with $r_s > 0$ to ensure that particles never overlap), 
 $\sigma_s=0.5$ is  the characteristic   separation length 
between particles, and $h_0$  the distance  at which the potential 
is shifted to ensure its continuity at $r_s=h_0$.
\footnote{Note that separation $h_0$ is relatively short, $h_0 < \sigma_s$, and its  value  depends 
on the location of the minimum of the attractive potential.}

The  attractive potential  consists of a radial $V_{att}(r_{ij})$ and an angular $\phi(\theta_i,\theta_j$)
contribution.  Concerning the radial part of the potential, we have considered two different truncated Lennard-Jones: 
one long  ($r_c =2.5 \sigma$) and another short  ($r_c =1.5 \sigma$) range.
While the following equation: 
\begin{eqnarray}\label{eq:VLJmidrange}
\begin{split}
&V_{att} (r_{ij}) = \\
&\begin{cases}
4 \epsilon \left[ \left( \frac{\sigma}{r_{ij}} \right)^{12} - \left( \frac{\sigma}{r_{ij}} \right)^6 \right] & (2^{1/6}\sigma)<r_{ij}<r_{c}\,,\\
-\epsilon & r_{ij}< (2^{1/6}\sigma)\,, \\
0 & otherwise\,, \
\end{cases}
\end{split}
\end{eqnarray}

\noindent
represents the former, it has been modified for the  latter 
by 
adding a smooth transition function, so that the potential  $V_{att}$ and its derivative 
 vanish continuously as $r_{ij}$ approaches  $r_c$ \cite{pra_smoothfunct} (see appendix \ref{Appendix:attra}). 
 Therefore, whenever  $(2.2^{1/6}\sigma)<r_{ij}<r_{c}$ 
\begin{widetext}
\begin{eqnarray}
\label{eq:VLJshortrange}
V_{att} (r_{ij}) = 4\epsilon \left[ \left( \frac{\sigma}{r_{ij}}
\right)^{12} - \left( \frac{\sigma}{r_{ij}} \right)^{6} \right] 
 + 4\epsilon \left[ 6 \left( \frac{\sigma}{r_{c}}
\right)^{12} - 3 \left( \frac{\sigma}{r_{c}}
\right)^{6} \right] \left( \frac{r_{ij}}{r_{c}}
\right)^{2}
 - 4\epsilon \left[ 7 \left(\frac{\sigma}{r_{c}}
\right)^{12} + 4 \left( \frac{\sigma}{r_{c}}
\right)^{6} \right]\,,   
\end{eqnarray}
\end{widetext}
\noindent
whereas for $r_{ij} < (2.2^{1/6}\sigma)$   $V_{att}=-0.305 \epsilon$   
and $V_{att}=0$ otherwise.

Inspired by the work of Hong {\sl et al.}~\cite{langmuir08}
for amphiphilic colloidal spheres,
 the anisotropic interaction is controlled by the angular  contribution of the attractive potential 
\begin{eqnarray}\label{eq:phi}
\begin{split}
&\phi(\theta_i,\theta_j)=&\\
&\begin{cases}
\cos\theta_i \cos \theta_j &\left( \hat{p}_i \cdot \hat{\mu}_{ji} \right) > 0 \cap \left( \hat{p}_j \cdot \hat{\mu}_{ij} \right) > 0\,,\\
0 & otherwise\,.\\
\end{cases}
\end{split}
\end{eqnarray}
where  $\hat{\mu}_{ij} = \vec{r}_{ij}/r_{ij}$ 
is the unit vector joining the squirmers' center of mass.
 This anisotropic potential  ensures a non-vanishing torque 
when  $\left( \hat{p}_i \cdot \hat{\mu}_{ji} \right) > 0 $~and~$ \left( \hat{p}_j \cdot \hat{\mu}_{ij} \right) > 0$
(a more detailed analysis  is discussed in appendix \ref{Appendix:JanusOrient}).

\begin{figure}[h!]
\begin{flushleft}
		\includegraphics[width=0.480\linewidth]{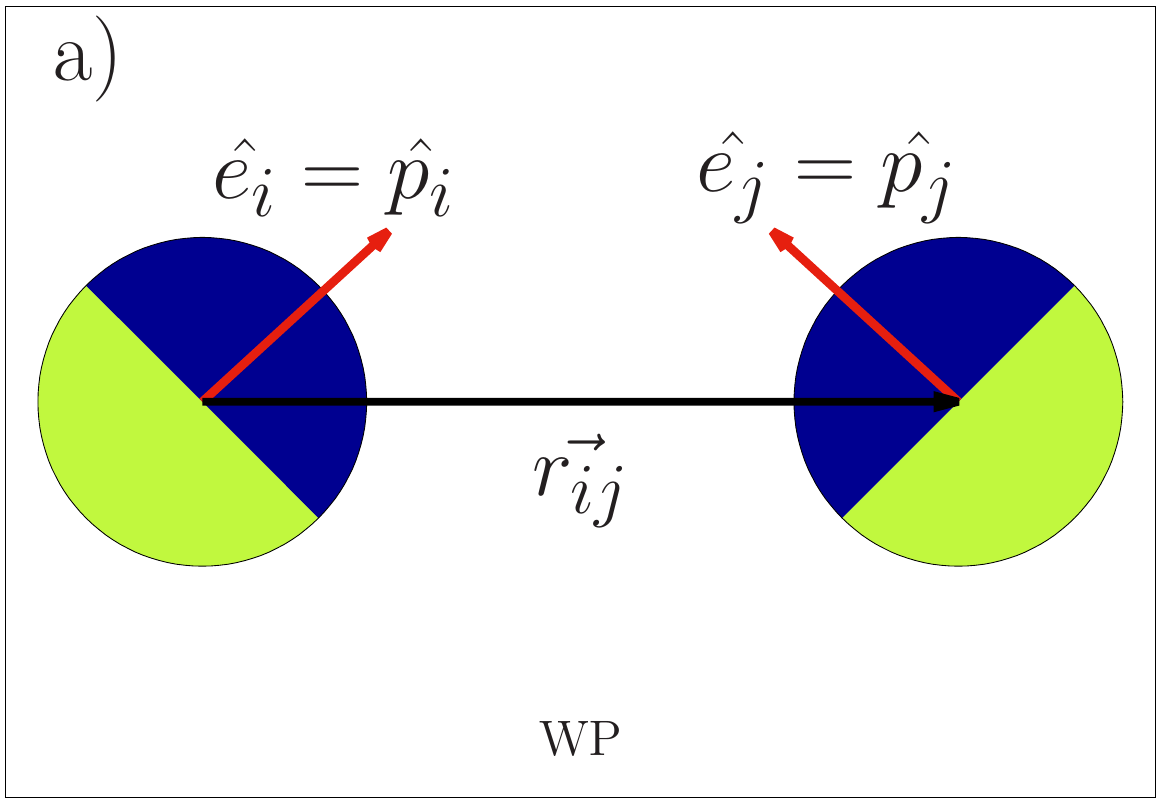}
	\includegraphics[width=0.480\linewidth]{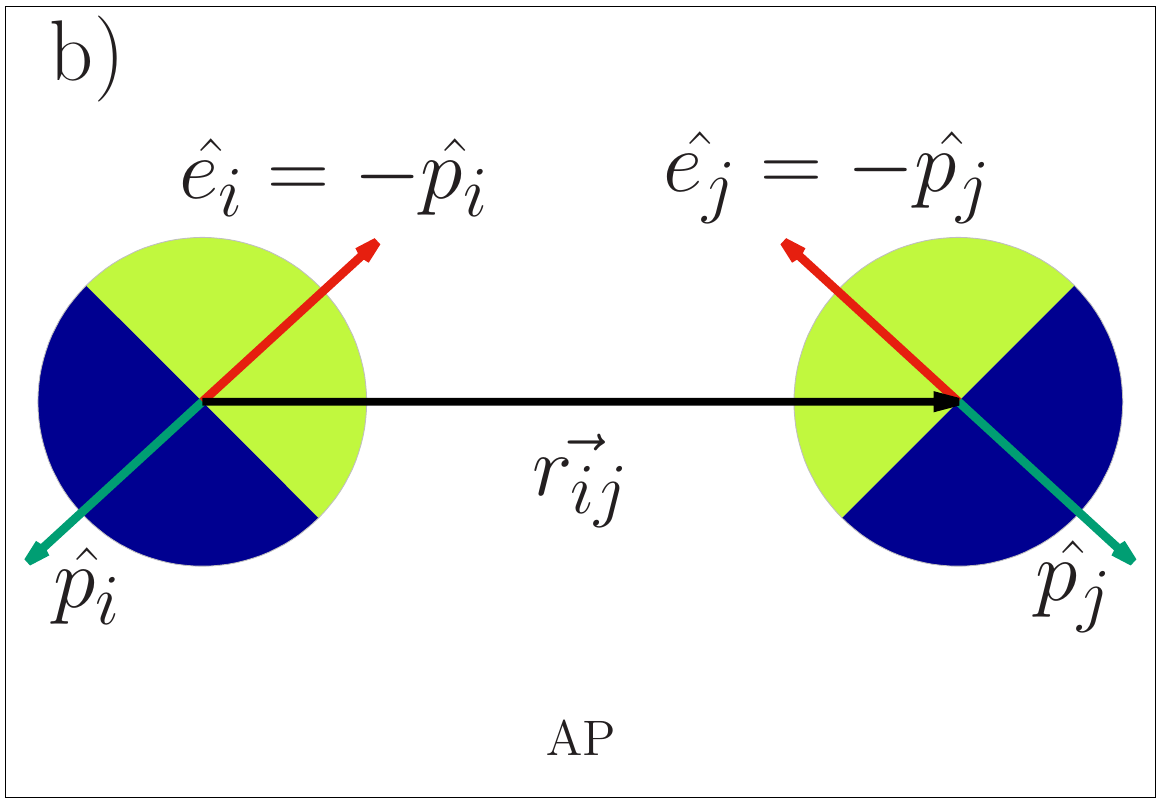}
	\end{flushleft}	\caption{Panel a) WP Janus swimmers. Panel b) AP Janus swimmers. 
	The attractive patch is represented in blue {(dark gray)}, and the non-attractive patch is in green {(light gray)}. 
	The  swimming direction $\hat{e}$ is the red arrow and direction of the attractive 
	patch $\hat{p}$ is the green one.  The center-to-center distance between two swimmers $\vec{r_{ij}}$ is the black vector.
	\label{fig:JanusSquirmers}}
\end{figure}

Within our model, each Janus particle self-propels along a direction  $\hat{e}_i$  rigidly bounded to the particle.
As in Ref.~\cite{Stewart_17}, we study two types of active Janus particles
depending of the orientation of $\hat{e}_i$   with respect to $\hat{p}_i$. 
When the  propulsion direction ($\hat{e}_i$, red arrow in figure~\ref{fig:JanusSquirmers}-left panel) is pointing towards 
the attractive patch $\hat{p}_i$ (red arrow in figure~\ref{fig:JanusSquirmers}-left panel),  swimmers are propelling "with the patch" (WP).  
When the  propulsion direction ($\hat{e}_i$, red arrow in figure~\ref{fig:JanusSquirmers}-right panel) is pointing  against   
the attractive patch  $-\hat{p}_i$ (green arrow in figure~\ref{fig:JanusSquirmers}-right panel), swimmers are propelling "against the patch" (AP).

\begin{figure*}
\centering
	\includegraphics[width=.32\textwidth]{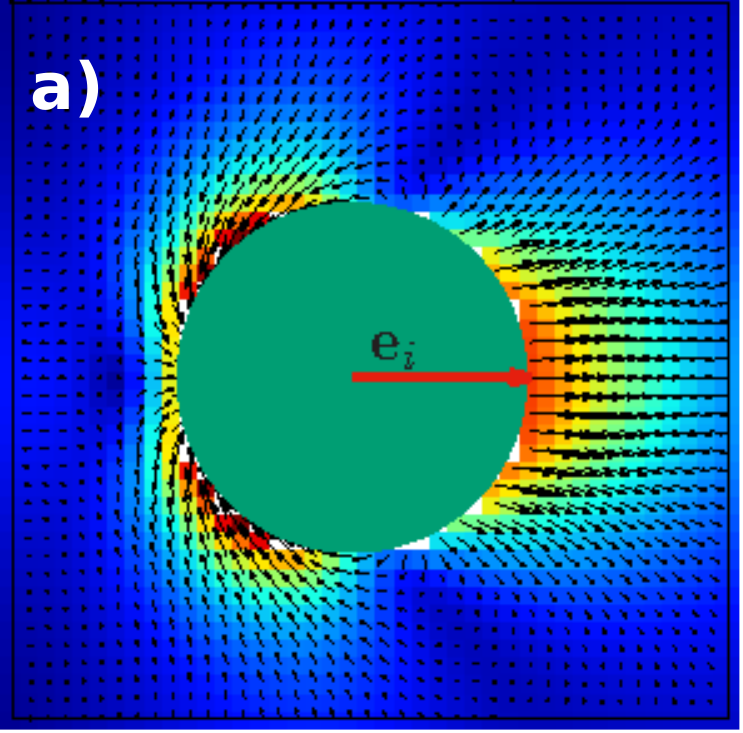}\includegraphics[width=.32\textwidth]{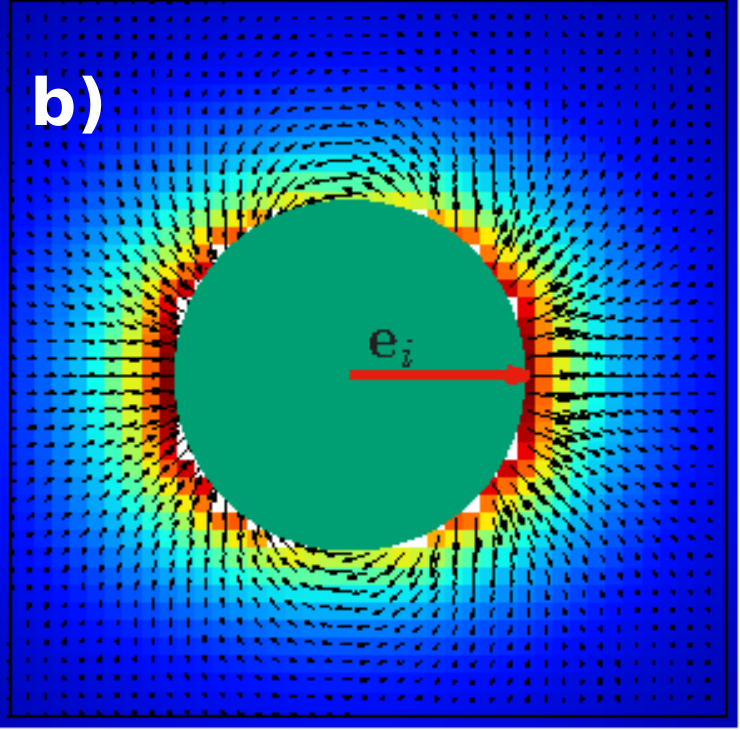}\includegraphics[width=.362\textwidth]{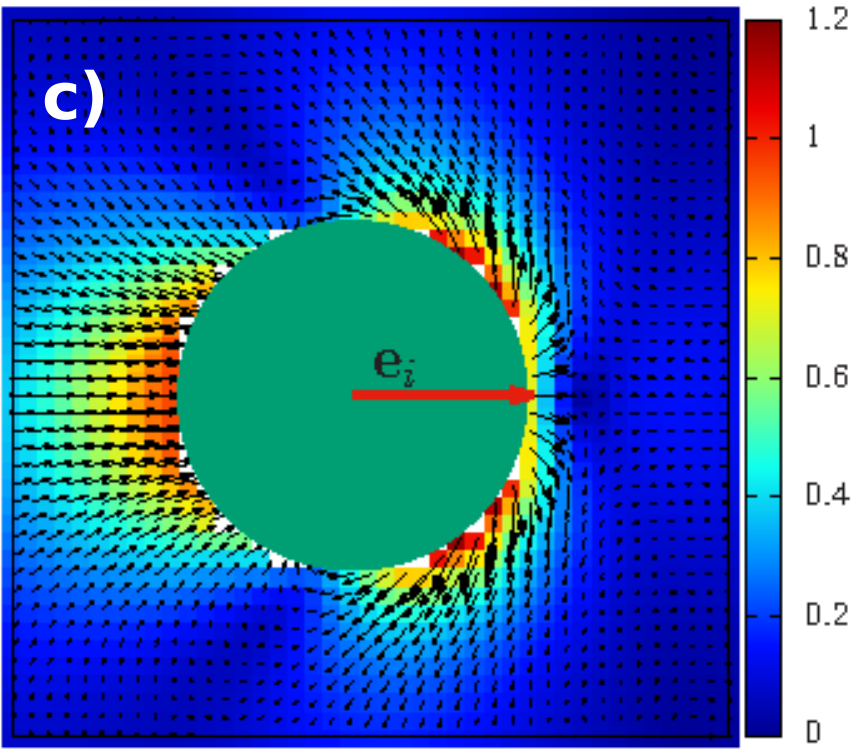}
	\caption{XY projection of a squirmer at $\beta=-1$ (pusher, left panel), $\beta=0$ (neutral, middle panel) and $\beta=1$ (puller, right panel).
	 The red arrow represents the swimming direction, the black arrows  the velocity directions of the fluid, the color code   the speed of the fluid normalized by the Stokes velocity ($v_s=2/3~B_1$). Velocities are computed  in the squirmer frame of reference.
	\label{fig:Squirmers}}
\end{figure*}

A spherical Janus squirmer    of 
radius $R_p$ is characterized by the tangential surface  slip velocity,  
\begin{eqnarray} \label{eq:surface_veloc}
\textbf{u}|_{R_p} = \left[ B_1 \sin \theta + B_2 \sin \theta \cos \theta \right] \boldsymbol{\tau}\, ,
\end{eqnarray}
\noindent where $\boldsymbol{\tau}$ is a unit vector tangential to the particle surface,  $B_1$   quantifies the asymptotic self-propelling speed of an isolated squirmer ($v_s= \frac{2}{3} B_1$)
   \cite{Prieve,Stone} and $B_2$ is proportional to the active stress generated on the surrounding fluid.
  The ratio between  active stress and self-propelling velocity, $\beta=B_2/B_1$~\cite{Ishikawa2006}, 
  quantifies the active state of the squirmer and its interaction with the fluid. A positive value of  
    $\beta$ corresponds to the case when thrust is generated in front of the squirmer's body (puller); 
    a negative value of   $\beta$  to the case when  thrust is generated at the back   (pusher)
  ~\cite{Ishikawa2007a} and $\beta=0$ corresponds to  a neutral swimmer (see Fig.~\ref{fig:Squirmers}). 
  In our simulations we   
  disregard thermal fluctuations; thus
velocity fluctuations are simply induced by the particles' activity, where $B_2$ 
 can be understood  as an effective temperature.


In order to characterize how hydrodynamics affects the assembly of a dilute suspension of 
Janus swimmers,  we 
  model  the embedding solvent using  a Lattice Boltzmann (LB) method 
with a D3Q19  three dimensional lattice~\cite{cates_lb,succi}, implemented in a highly parallelized code \cite{ludwig}  
    that exploit the excellent scalability of LB on supercomputing facilities. 
Spherical squirmers (with a diameter $\sigma$ of 4.6 lattice nodes~\cite{nguyen})   are individually resolved,  
 imposing a modified bounce-back rule  on the  one-particle velocity distribution functions  to  nodes  
 crossing the solid particle's boundary  
 (including the slip velocity
 to impose  the squirming motion~\cite{llopis_wall,ricard_epje}).
   The total force and torque the fluid exerts on the particle is obtained by
  imposing that the total momentum exchange between the  particle and the fluid nodes vanishes
    (as a result of   the modified bounce-back). While accounting for all forces acting on each squirmer allows to update 
  them dynamically, the torque exerted by the fluid determines the change of  the squirmer 's  
  self-propulsion direction.
  
The simulated system consists of  a   suspension of  $N=888$ spherical squirmers 
at   $\phi=0.10$ (where $\phi=\frac{\pi}{4} \rho\sigma^2$ and  $\rho=N/L^2$)
in a quasi two-dimension geometry with periodic boundary conditions  ($L\times L \times k \sigma$, with $k=5$ and  
$L \approx 83 \sigma$) needed to capture  three dimensional hydrodynamics effects, {whereas both colloids' position and orientation are confined to move in 2D}. 
Additionally, we have carried out bigger simulations with $N=3552$ squirmers at the same packing fraction and $L \approx 167 \sigma$, in order to proof that results are not due to finite size effects, specially cases with large cluster sizes.


As in Ref.~\cite{SoftMatter17}, we quantify the competition between attractive and self-propelling forces  via the dimensionless parameter
\begin{equation}
\xi = \frac{F_d}{F_{att}},
\end{equation}
\noindent where $F_d = 6 \pi \eta R_p v_s$ is  the  friction force associated to the squirmer self-propulsion and 
$F_{att}$ is the absolute value of the attractive force at its minimum, corresponding to  $r=(26/7)^{1/6}\sigma = 1.245\sigma$ for   the long-range
 potential in Eq.~\ref{eq:VLJmidrange} and to  $r=1.239\sigma$ for the 
  short-range potential in Eq.~\ref{eq:VLJshortrange}.
{On the one hand $\xi$ controls the competition between two mechanisms: the self-propulsion and the interaction strength. On the other hand the mechanisms that control the re-orientation of the particles  are the active stress given by the $B_2$ squirmer parameter and the orientational-dependence of the potential which is modulated by the range and the strength of the interaction.}  
  
    Given that a squirmer travels its own size in a time $\tau = \frac{\sigma}{v_s}$, we  perform simulations from $1450$ up to $3000$ $\tau$. 
 Once the system reaches steady state, in a time window between  
 $1000$ and  $2000$ $\tau$, we 
carry out a systematic analysis of the dynamics of the squirmer suspension,  considering  $\xi$  from 0.1 to 10 and $\beta$ from -3 up to 3.

\section{\label{sec:Tools}Analysis Tools}

In order to be able to establish the effect played by the anisotropy of the interaction between swimmers, 
we follow the procedure in Ref.~\cite{SoftMatter17} and study the  degree of aggregation as a function of $\beta$ and  $\xi$. 
Whenever the system forms a steady-state cluster distribution, 
we explore the clusters morphology by computing their mean  size, the cluster-size distribution, their radius of gyration, and both  the  
polar and  nematic order parameter.

When clusters are present, we identify them following  a distance criterion: 
at each time step, two particles belong to the same cluster of size $s$  whenever their distance is smaller 
than $r_{cl} = 1.75 \sigma$  (first minimum of the $g(r)$, see appendix \ref{Appendix:Clusters}).

 To calculate  the cluster-size distribution $f(s)$ we apply a 
criterion based on \cite{chantalbins2012}: (i) We arbitrarily subdivide the range of $s$-values into intervals $\Delta s_i = (s_{i,max} - s_{i,min})$, 
where $n_i^t$ is the total number of clusters within each interval $\Delta s$; (ii) we assign the value $n_i=n^t_i/\Delta s_i$ to every $s$ 
within $\Delta s_i$, and compute the fraction of clusters of size $s$ as $f(s)=n_i/N_c$, where $N_c=\sum_i n_i \Delta s_i$ is the total number of clusters.

To characterise the clusters' morphology, we  compute  the  radius of gyration
\begin{equation} \label{eq:Rg}
R_g(s) = \sqrt{\sum_{i,j=1}^s \frac{\left(\vec{r}_i-\vec{r}_j\right)^2}{2s}}\,,
\end{equation}
the average polar order parameter for each cluster size $s$ 
\begin{equation} \label{eq:PolarOrderCluster}
P(s)=\left\vert\frac{\sum_{i=1}^s \hat{e}_i}{s} \right\vert\,
\end{equation}
\noindent
 the  tensor order parameter    (for a $2D$ system \cite{Barci_2DNematic})
\begin{equation}\label{eq:lmbdaMtrx}
Q_{hk}(t)=\frac{1}{N}\sum_{i=1}^{N}\left(2 e_{ih}(t)e_{ik}(t)-\delta_{hk}\right), 
\end{equation}
(being $h$ and $k$ equal to ${x,y}$ and $N$ the total number of squirmers) 
and  the nematic order parameter $\lambda(t)$   being the 
largest eigenvalue of $Q_{hk}$  \cite{Eppenga1984}.

We have also computed   both global nematic and  global polar order parameters, 
substituting $s$ with $N$
in the global polar order parameter  (Eq.~\ref{eq:PolarOrderCluster}).
 Therefore,  the polar order in  steady-state is $P_{\infty} = P(t>>0)$ 
 while the nematic order in steady-state is  $\lambda_{\infty} = \lambda(t>>0)$. 

\section{\label{sec:Results} Results}

\subsection{Mean cluster size}

We start with  
computing the mean cluster size $\langle  s \rangle$ as a function of time
for different values of $\beta $ and $\xi$. This allows us 
 to distinguish coarsening from clustering,  establishing, in the latter case, when the system reaches a steady state. 
 \begin{figure}[h!]
\centering
\includegraphics[width=0.48\columnwidth]{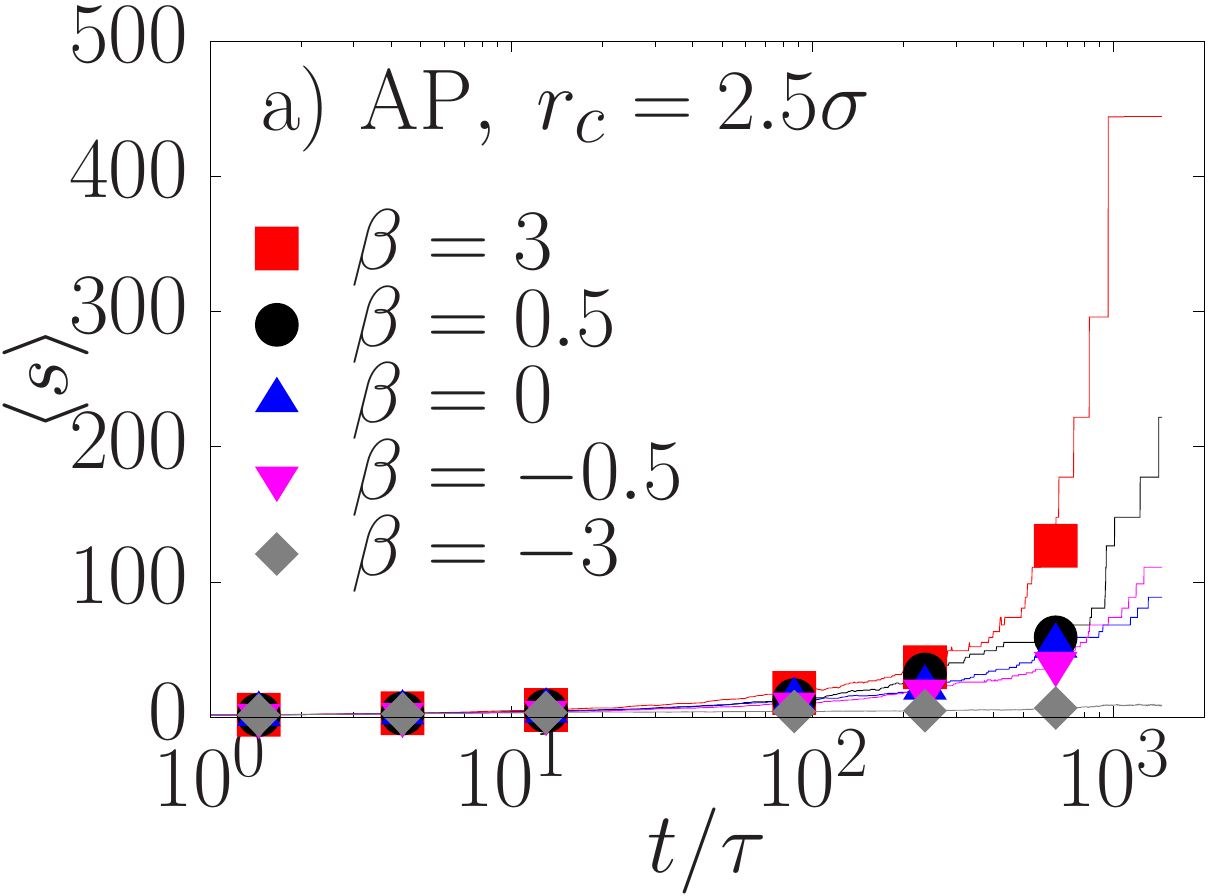} \includegraphics[width=0.48\columnwidth]{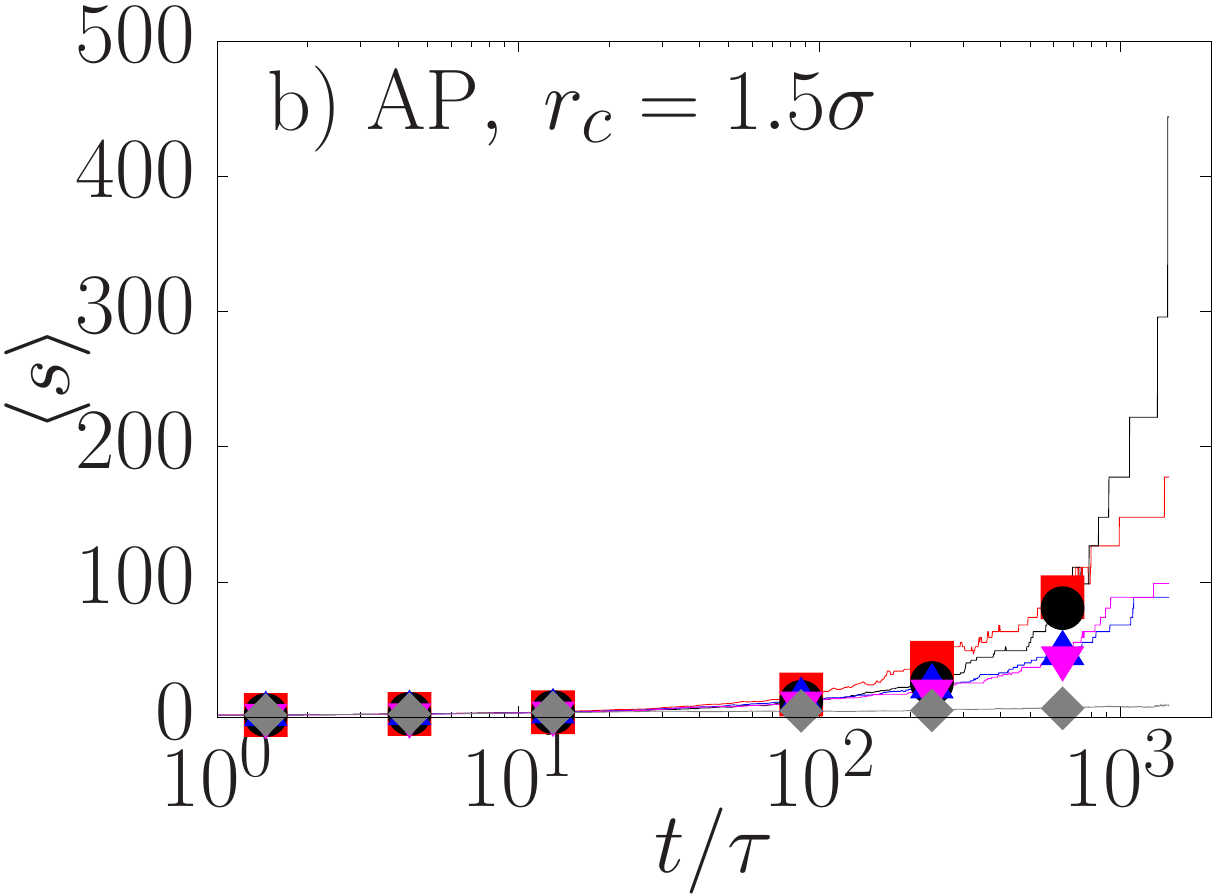}
\includegraphics[width=0.48\columnwidth]{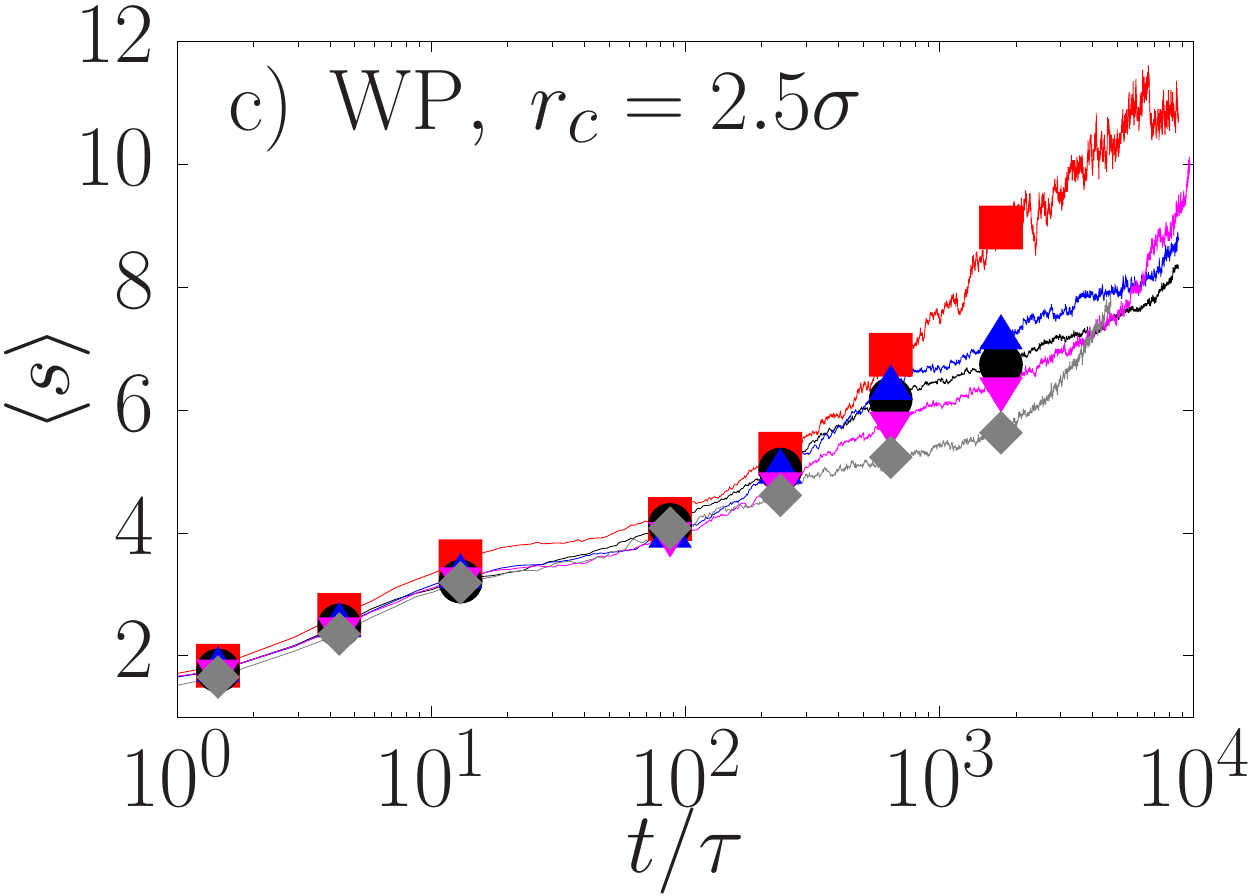} ~ \includegraphics[width=0.48\columnwidth]{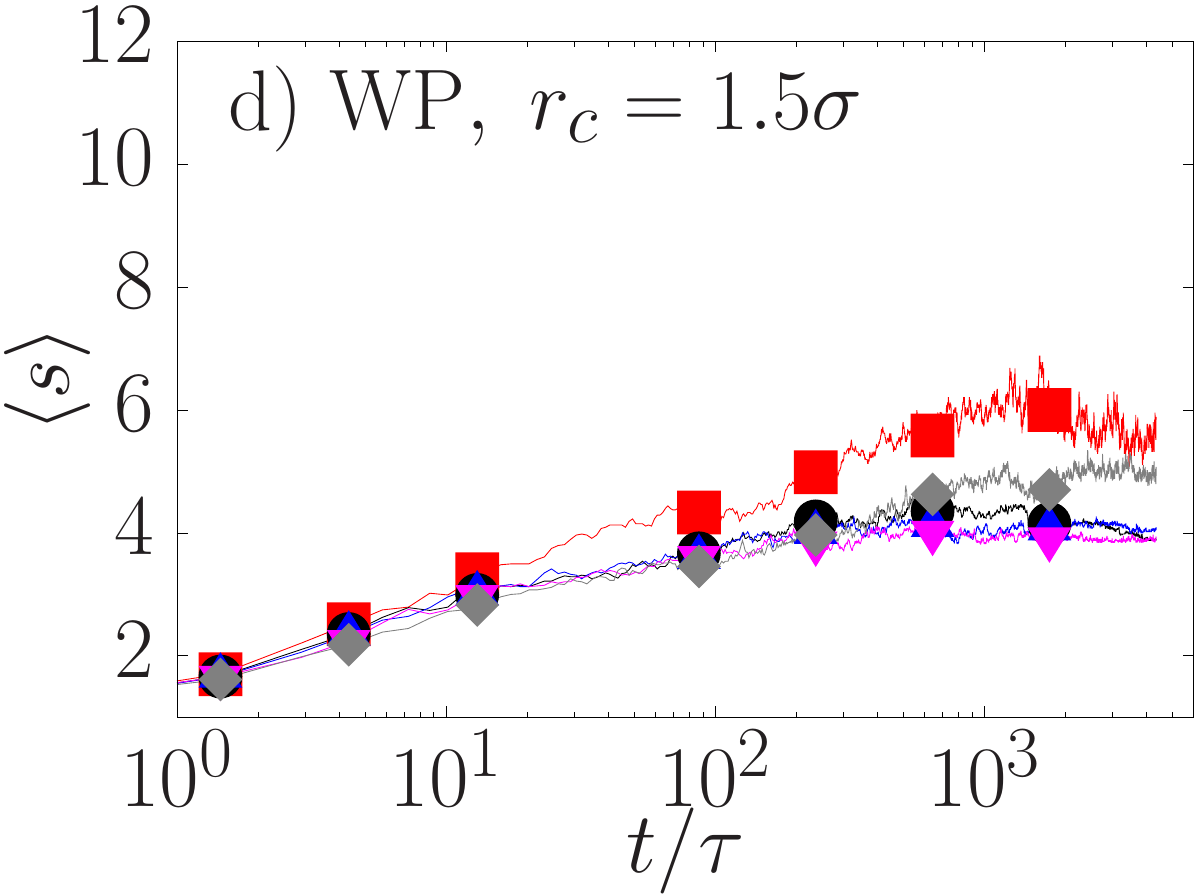}
\caption{\label{fig:nct_xi0_1}Time evolutions of the mean cluster size for squirmers when $\xi=0.1$ (pullers: red {squares} and black {circles}; neutral: blue {up triangles}; pushers: pink {down-triangles} and grey {diamonds}).
 (a) AP squirmers with  $r_c=2.5\sigma$ and (b) $r_c=1.5\sigma$, (c) WP squirmers with $r_c=2.5\sigma$ and 
(d) $r_c=1.5\sigma$. Note the different scale on the y-axis.}
\end{figure}
 
Fig.~\ref{fig:nct_xi0_1} represents the mean cluster size as a function of time 
  for  suspensions  where attraction dominates over propulsion ($\xi=0.1$) of AP squirmers 
   (interacting via  
 long-range (panel-a) or  short-range  (panel-b) attraction)  
 and of WP squirmers (interacting via  
 long-range (panel-c) or  short-range (panel-d) attraction).  
  In all AP suspensions the system coarsen, within the simulated time window as shown by the continuously growing 
   mean cluster size, with a speed dependent on $\beta$. Pullers coarsening  faster than pushers,  
      a part from the system with $\beta=-3$ where particles form clusters independently on the interaction range. 
WP suspensions only 
 coarsen 
when  interactions are long-range (panel-c) with a speed dependent on $\beta$ and  pullers coarsening  faster than pushers, 
even though coarsening is much slower than in the AP case. 
{In panel-c, the coarsening dynamics is so slow that $\langle s \rangle < 10 $ even when $t > 1000 \tau$ (it is worth noting that we have never observed a steady state despite having  run up to $6 \times 10^6$ LB-time steps, corresponding to  $\sim 8700 \tau$).}
Comparing panel-c and panel-d, we conclude that 
the attraction range affects  aggregation, given that 
 when   the interaction is short-range (panel-d),
coarsening is suppressed and 
$\langle s \rangle$ reaches a steady state at long $t$ for all values of $\beta$.
{A more detailed analysis is given in appendix \ref{Appendix:xi0_1WPrc1_5} for the case of the WP system  with $\xi=0.1$. }

\begin{figure}[h!]
\includegraphics[width=0.48\columnwidth]{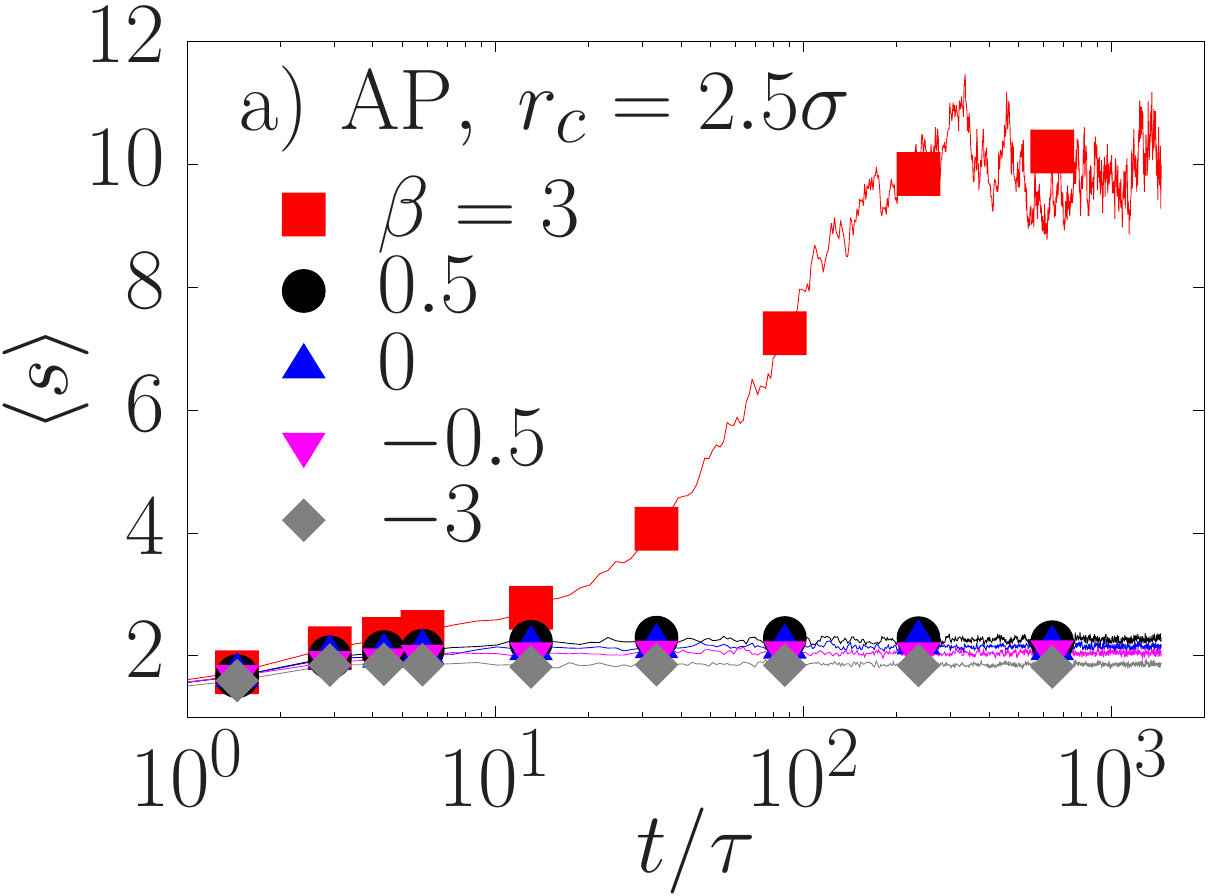}~\includegraphics[width=0.48\columnwidth]{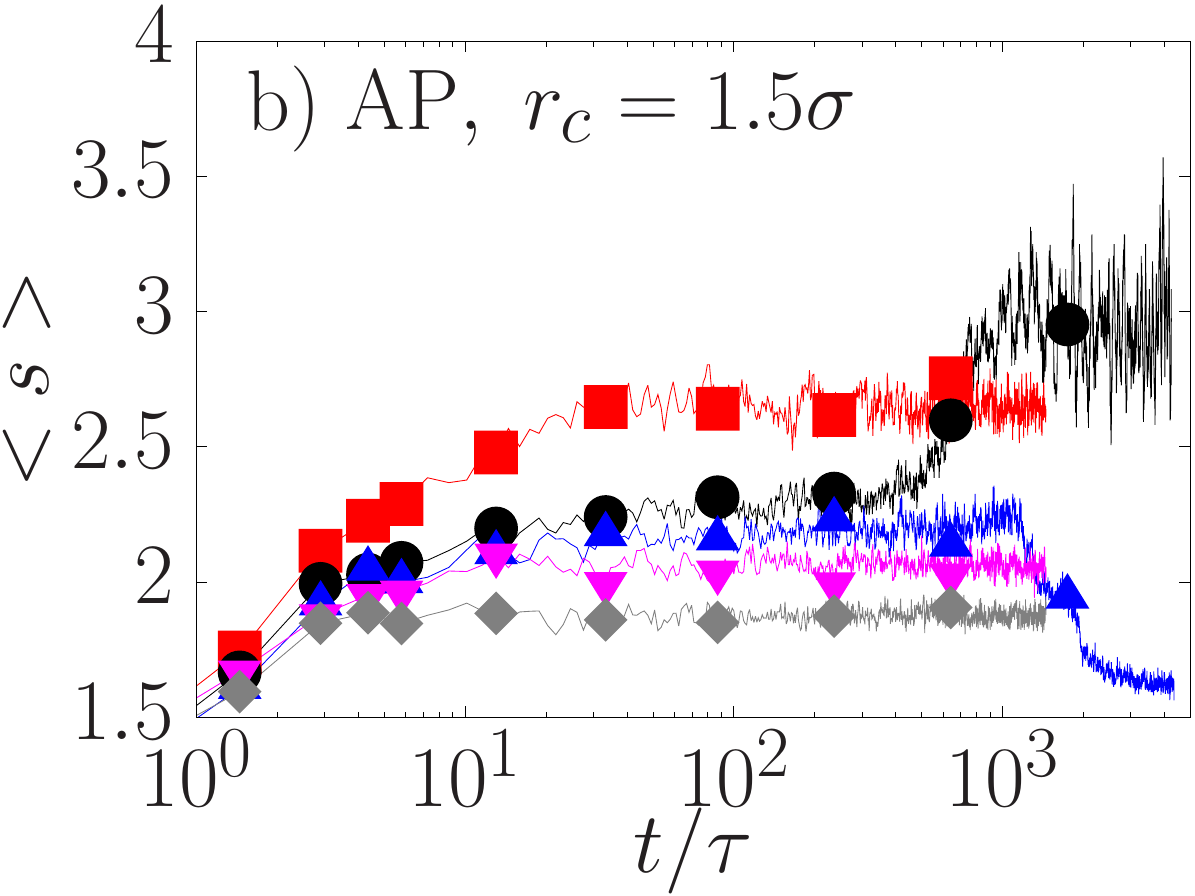}
\includegraphics[width=0.48\columnwidth]{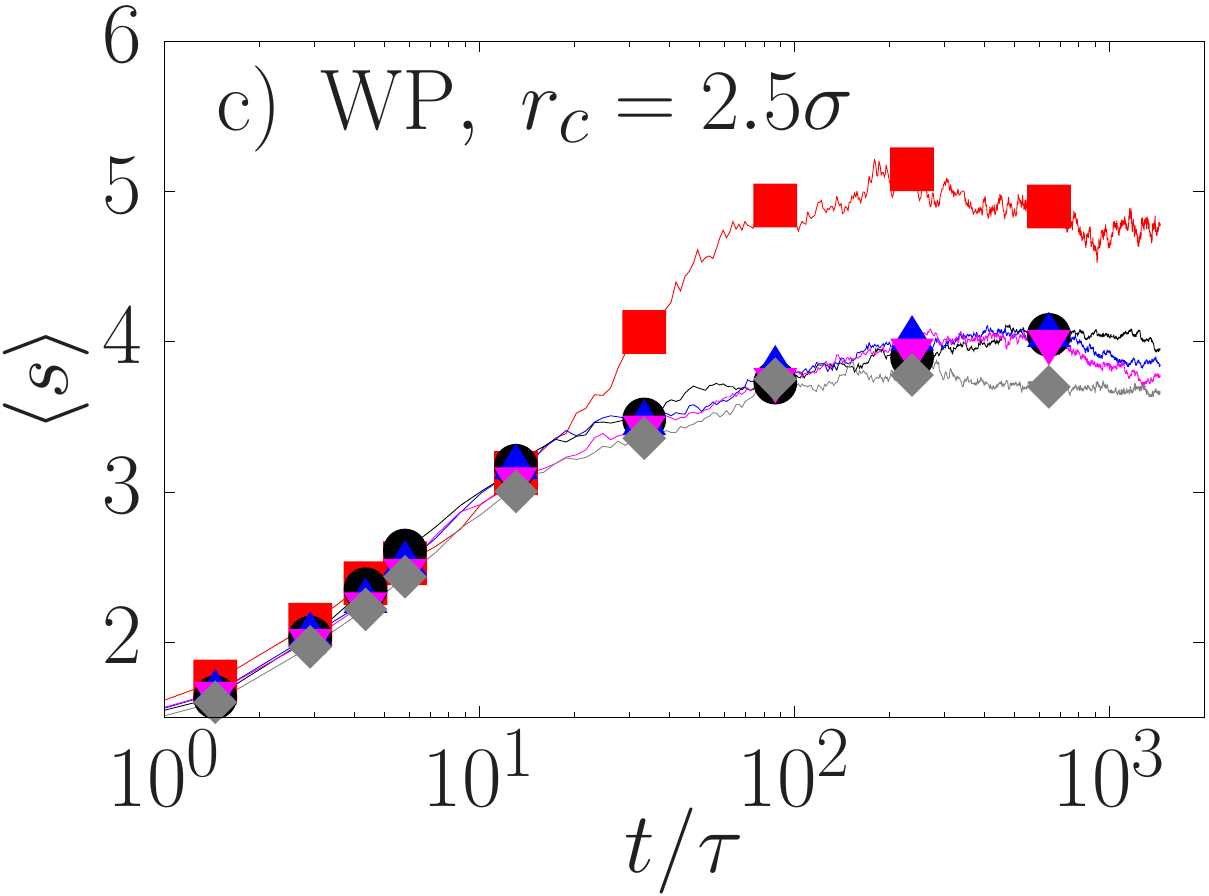}
\includegraphics[width=0.48\columnwidth]{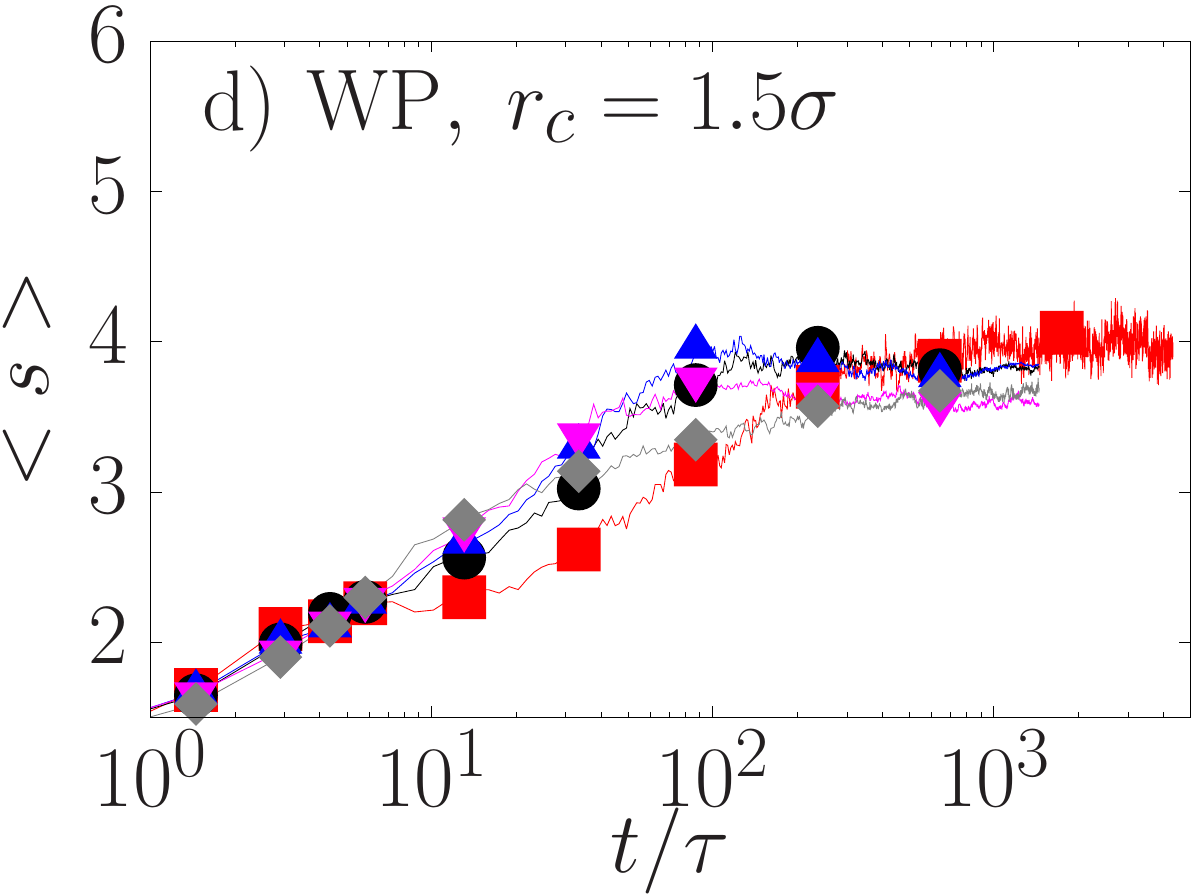}
\caption{\label{fig:nct_xi1_0} Time evolutions of the mean cluster size for squirmers when $\xi=1$ (pullers: red {squares} and black {circles}; neutral: blue {up triangles}; pushers: pink {down-triangles} and grey {diamonds}).
 (a) AP squirmers with  $r_c=2.5\sigma$ and (b) $r_c=1.5\sigma$, (c) WP squirmers with $r_c=2.5\sigma$ and                                                                           
(d) $r_c=1.5\sigma$. Note the different scale on the y-axis. }
\end{figure}

Fig.~\ref{fig:nct_xi1_0} represents the mean cluster size as a function of time 
  for  suspensions  where attraction competes with propulsion ($\xi=1$) of AP squirmers 
   (interacting via  
 long-range (panel-a) or  short-range  (panel-b) attraction)  
 and of WP squirmers (interacting via  
 long-range (panel-c) or  short-range (panel-d) attraction).  

 
{
AP squirmers interacting via long-range attractions (panel a) 
  quickly reach 
  a steady state   characterized by   
$\langle s \rangle \approx 2$ except for $\beta=3$ where $\langle s \rangle \approx 10$ (red curve in panel-a). Whereas when dealing with AP squirmers interacting via short-range attractions (panel b), pushers reach a steady state quite quickly with $\langle s \rangle \approx 2$ as well as pullers with $\beta=3$ (with a slightly larger mean cluster size of $\langle s \rangle \approx 2.5$). Interestingly, the mean cluster size for weak pullers (black curve in panel-b) first develops a metastable state at short times ($t \in (10,500) \tau $) and then another one for longer times ($t > 1000 \tau$): the latter  due to the formation of dynamic clusters with polar order. Similarly,}
 when $\beta=0$ the decay in $\langle s \rangle $ at long time corresponds to monomers  aligning and  
 moving in the same direction. 

WP squirmers with $\beta <3$ relatively quickly reach steady-state with  clusters with an average of 4 particles (panel-c).
{Whereas pullers with $\beta=3$ reach a slightly higher mean cluster size $\langle s \rangle \sim 5 $: such increase in the cluster size is observed only for this interaction range, since WP squirmers  with short-range attractions (panel d) develop a } mean cluster-size of $\sim 4$ particles, independently on their hydrodynamic signature.

\begin{figure}[h!]
\includegraphics[width=0.48\columnwidth]{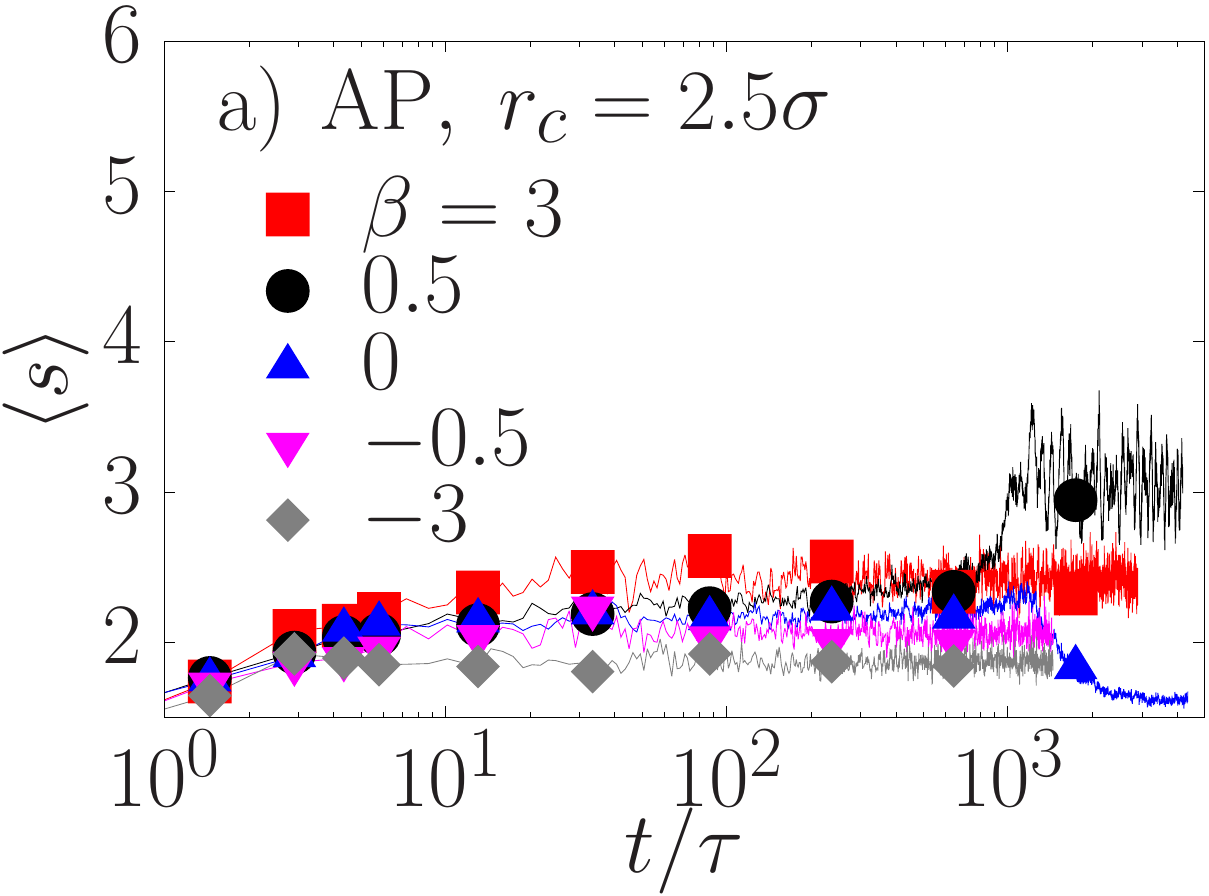} \includegraphics[width=0.48\columnwidth]{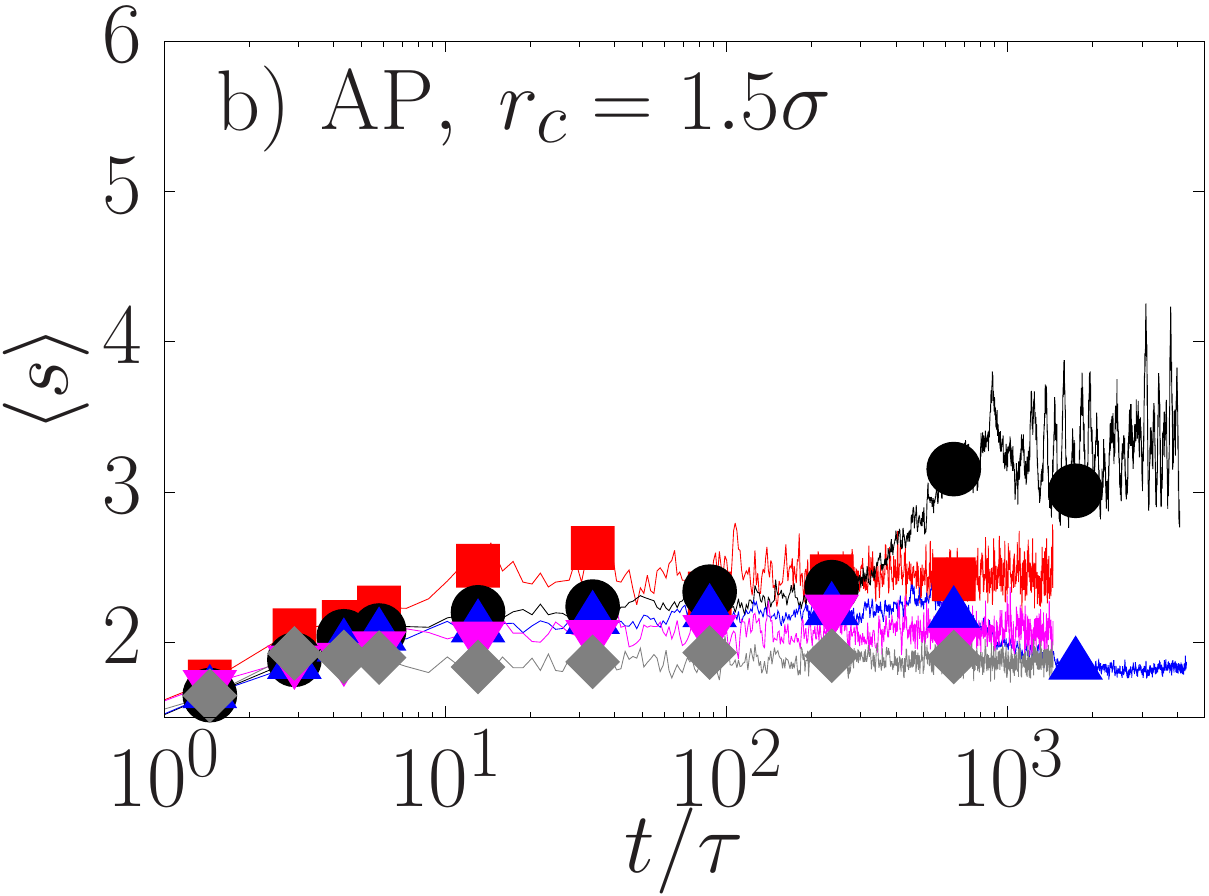}
\includegraphics[width=0.48\columnwidth]{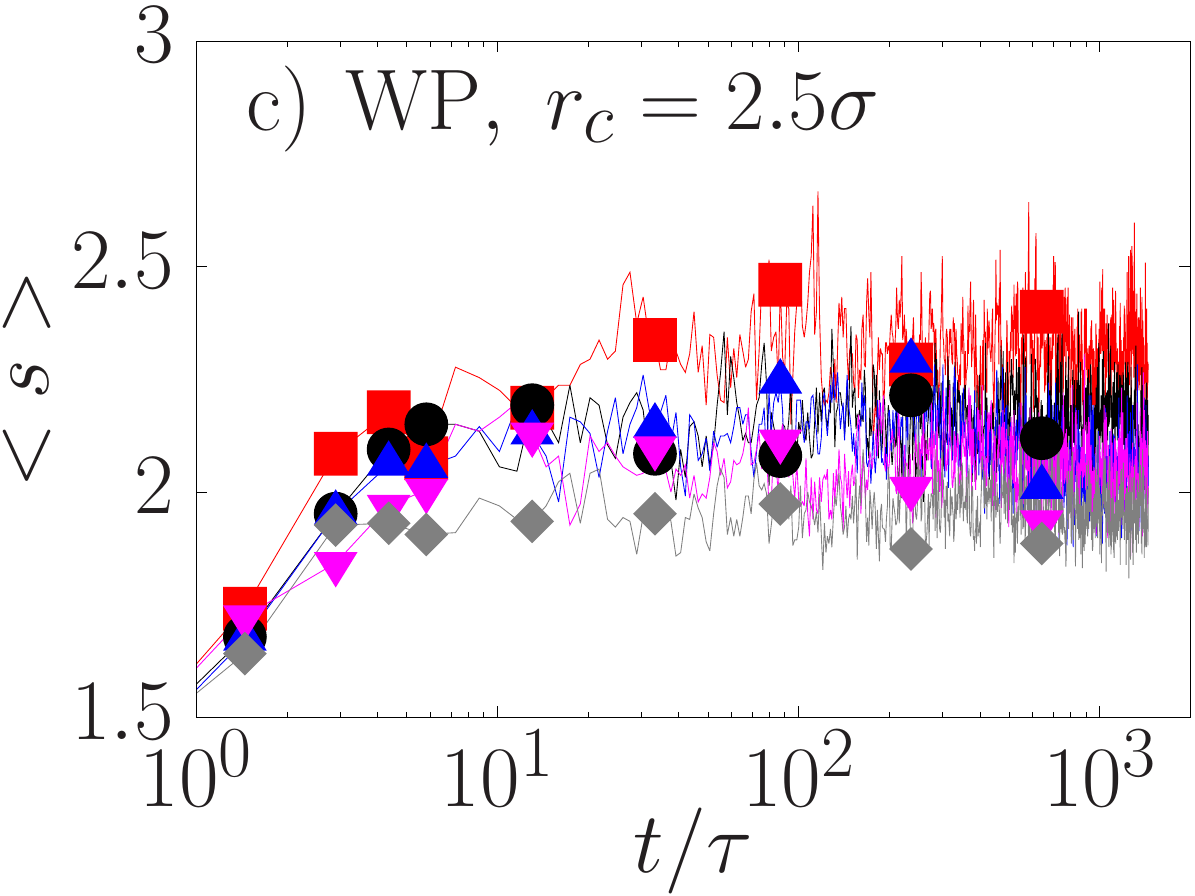}
\includegraphics[width=0.48\columnwidth]{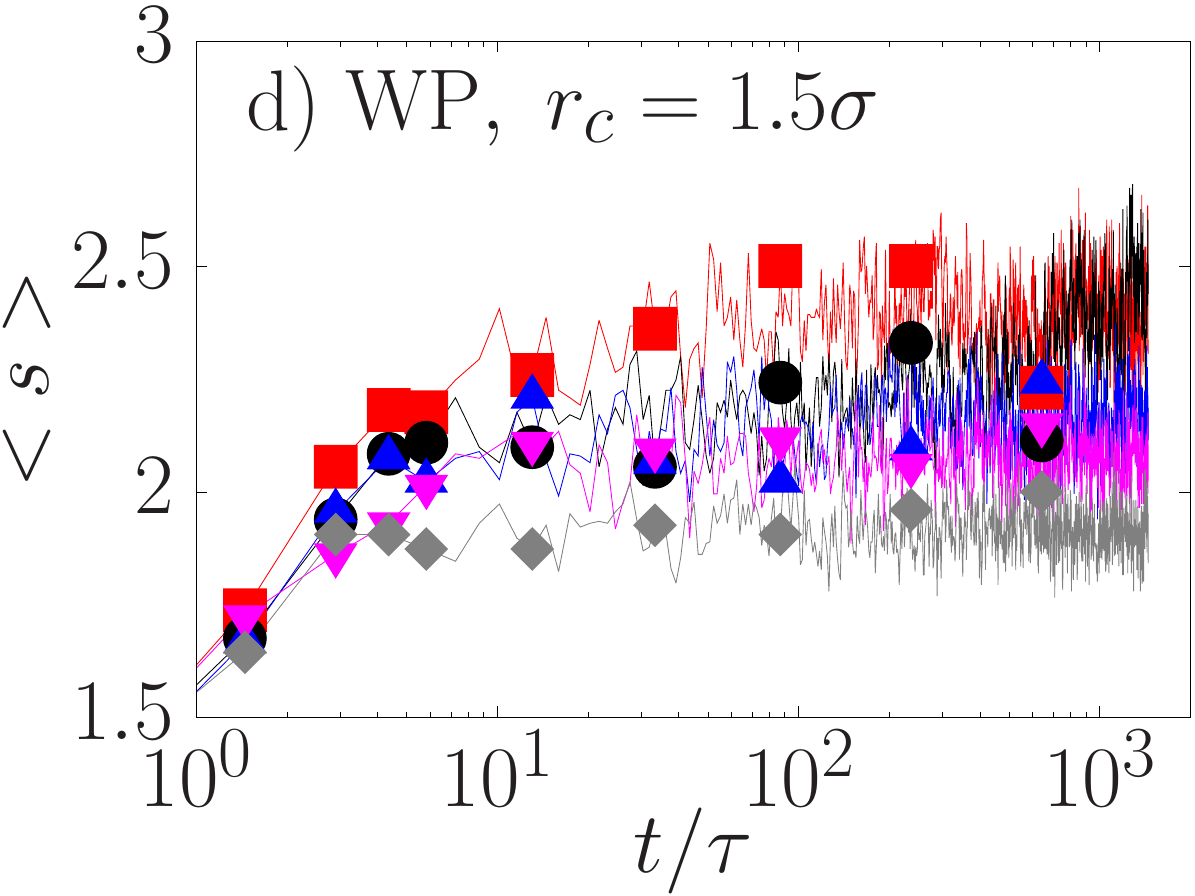}
\caption{\label{fig:nct_xi10}  Time evolutions of the mean cluster size for squirmers when $\xi=10$ (pullers: red {squares} and black {circles}; neutral: blue {up triangles}; pushers: pink {down-triangles} and grey {diamonds}).
 (a) AP squirmers with  $r_c=2.5\sigma$ and (b) $r_c=1.5\sigma$, (c) WP squirmers with $r_c=2.5\sigma$ and 
(d) $r_c=1.5\sigma$. Note the different scale on the y-axis.}
\end{figure}

   Fig.~\ref{fig:nct_xi10} represents the mean cluster size as a function of time 
  for  suspensions  where propulsion dominates over attraction ($\xi=10$) of AP squirmers 
   (interacting via  
 long-range (panel-a) or  short-range  (panel-b) attraction)  
 and of WP squirmers (interacting via  
 long-range (panel-c) or  short-range (panel-d) attraction).        

When self-propulsion dominates over attraction, squirmers in general form  smaller clusters {(never larger than 4 particles on average)}.
{
In either AP suspensions, the mean cluster size for weak pullers (black circles in panels-a and b) is larger than the one for
 any other $\beta$, fluctuating around a value of 3 for long times.  In the case of neutral squirmers and large $\xi$ and after a long time ($t > 1000 \tau$), particles in the suspension become strongly aligned and most of the clusters are monomers, for both interaction ranges.
 }


{When dealing with WP squirmers and high $\xi$, we have observed a small value of the  mean cluster size (around 2), which indicates that particles form dimers on average. A similar behavior is observed for WP squirmers with short-range interactions, with slightly larger fluctuations in the mean cluster size for weak pullers at long times.}

\begin{figure}[h!]
\begin{flushleft}
\includegraphics[width=0.48\columnwidth]{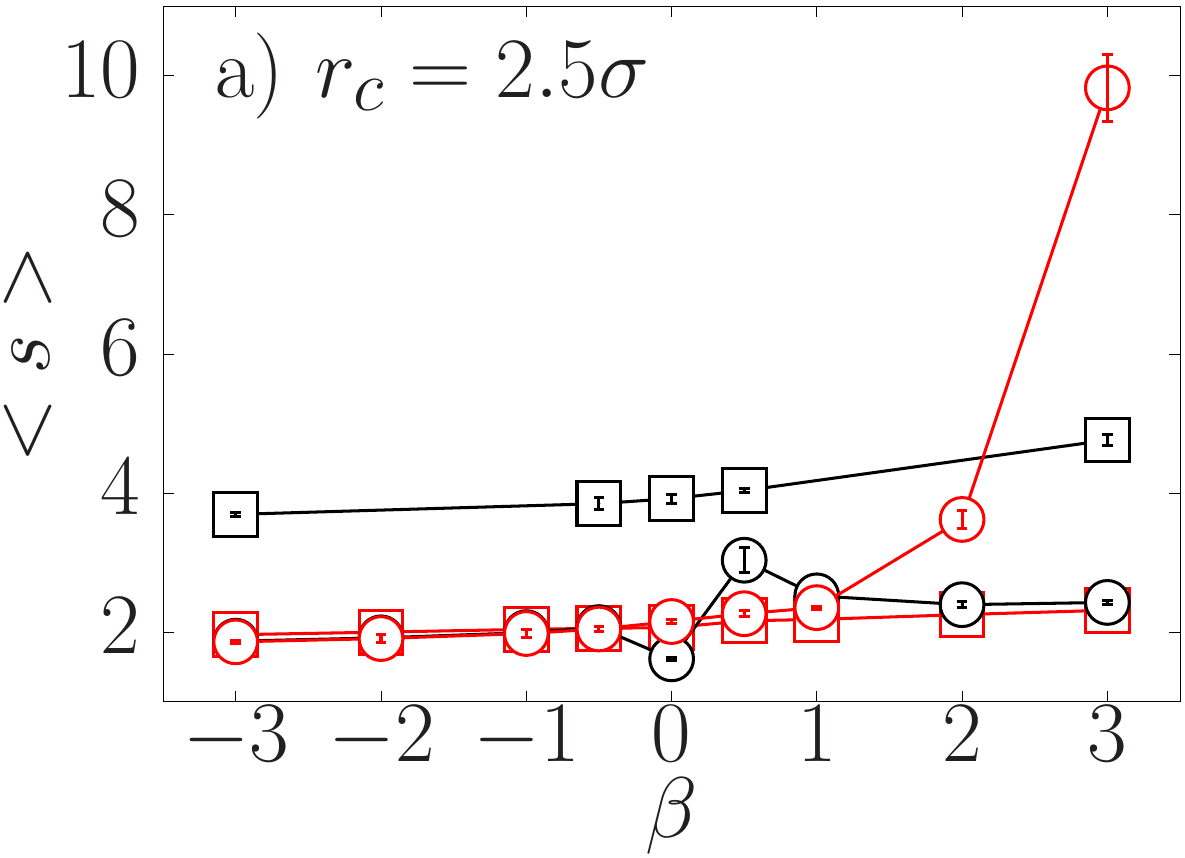}~\includegraphics[width=0.48\columnwidth]{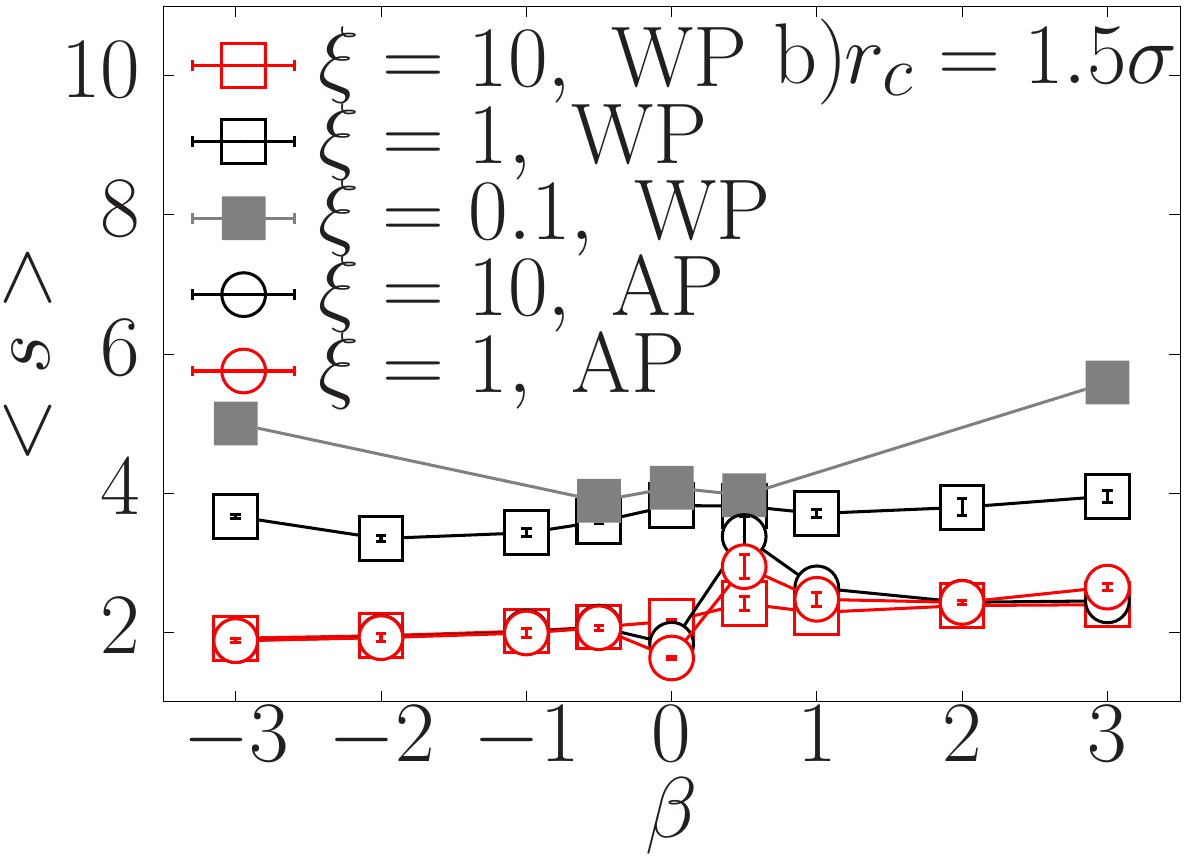}
\end{flushleft}
\caption{\label{fig:MeanClustersBetas}Mean cluster size for suspensions with (a)  $r_c=2.5\sigma$ and (b) $r_c=1.5\sigma$ in steady-state for WP ($\xi=10$, $\xi=1$ and $\xi=0.1$, square symbols) and AP ($\xi=10$ and $\xi=1$, circular symbols) }
\end{figure}

Having established the conditions needed for clustering to appear, we represent in  Fig.~\ref{fig:MeanClustersBetas} 
the mean cluster size as a function of $\beta$  for $r_c=2.5 \sigma$ (panel a) and   $r_c=1.5 \sigma$ (panel b). 
\footnote{The WP system  with $\xi=0.1$, where clustering is observed (panel d of Fig. \ref{fig:nct_xi0_1}), is studied in more detail  in appendix \ref{Appendix:xi0_1WPrc1_5}.}



{As shown in figure \ref{fig:MeanClustersBetas}, when $\xi=1$ WP swimmers mostly form tetramers and trimers, while WP swimmers with $\xi=10$ form dimers. The clusters observed for WP squirmers  have an average size rather insensitive to the squirmer hydrodynamic signature for these two interaction strength $\xi = \{1$,$10$\} and the two interaction ranges studied here.}

{ For lower values of $\xi$, the dependence on $\beta$ becomes more evident for
WP squirmers with  $\xi=0.1$ (gray squares on panel-b): while for extreme values of 
$\beta$ (either pushers or pullers) the clusters on average form pentamers, 
whereas neutral and weak pullers or weak pushers are mostly form tetramers.}
  
{AP squirmers with $r_c=2.5 \sigma$ (circular symbols panel-a) show that the mean cluster size increases as $\beta$ increases when $\xi=1$, whereas it does not change significantly when $\xi=10$ or when the interaction range is shorter (circular symbols panel b).} 

 As summary, on the one side, comparing the two panels in Fig.~\ref{fig:MeanClustersBetas}, when $\xi=1$ 
 and $10$, 
 the mean cluster size $\langle s \rangle$ for WP squirmers with interaction range of $r_c=1.5 \sigma$ 
 behaves  in  the same way as the one  with a larger cutoff ($r_c=2.5 \sigma$). 
 Therefore, when attraction does not dominate, the mean cluster size  for WP squirmers is independent on  the interaction range. 
On the other side, the mean cluster size $\langle s \rangle$ for AP squirmers with $\xi=1$ and  $r_c=1.5\sigma$ 
(red circles in figure \ref{fig:MeanClustersBetas}-(b))
 does not behave in the same way than when  $r_c=2.5\sigma$ (red cirlces in figure \ref{fig:MeanClustersBetas}-(a)), 
 the former showing a  systematic lower value of $\langle s \rangle$ for all $\beta$ while  
 the latter  a monotonic increase of the cluster size for pullers.  
 Finally, AP squirmers with $\xi=10$  (black circles  in figure \ref{fig:MeanClustersBetas})
{show the same behavior varying $\beta$, independently on the attraction range.} With a small peak for weak pullers, and a minimum at $\beta=0$.

\subsection{Coarsening, clustering and aligned states}

\begin{figure}[h!]
\includegraphics[width=0.49\columnwidth]{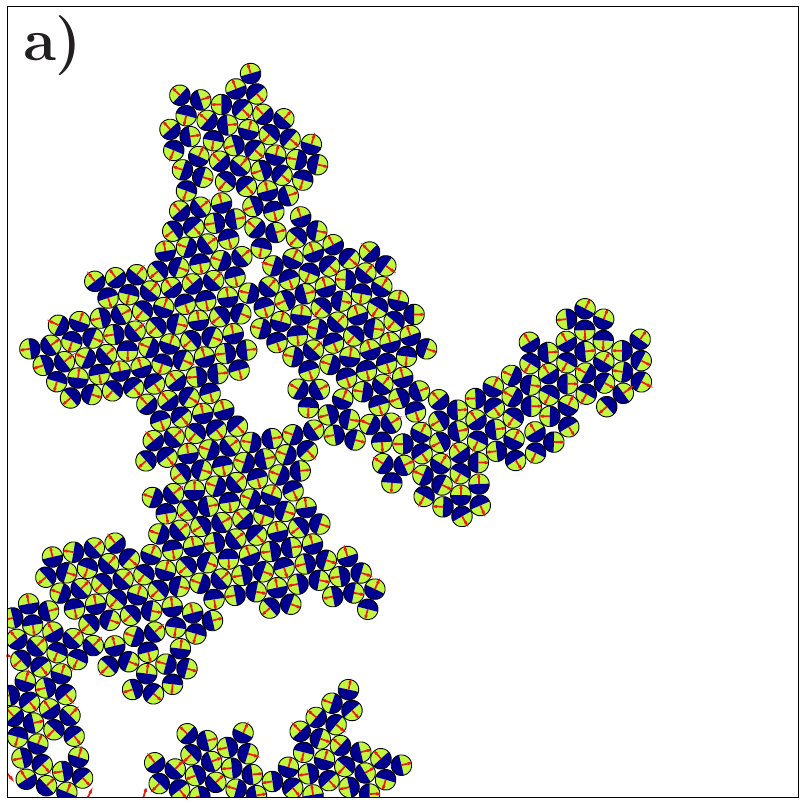}
\includegraphics[width=0.49\columnwidth]{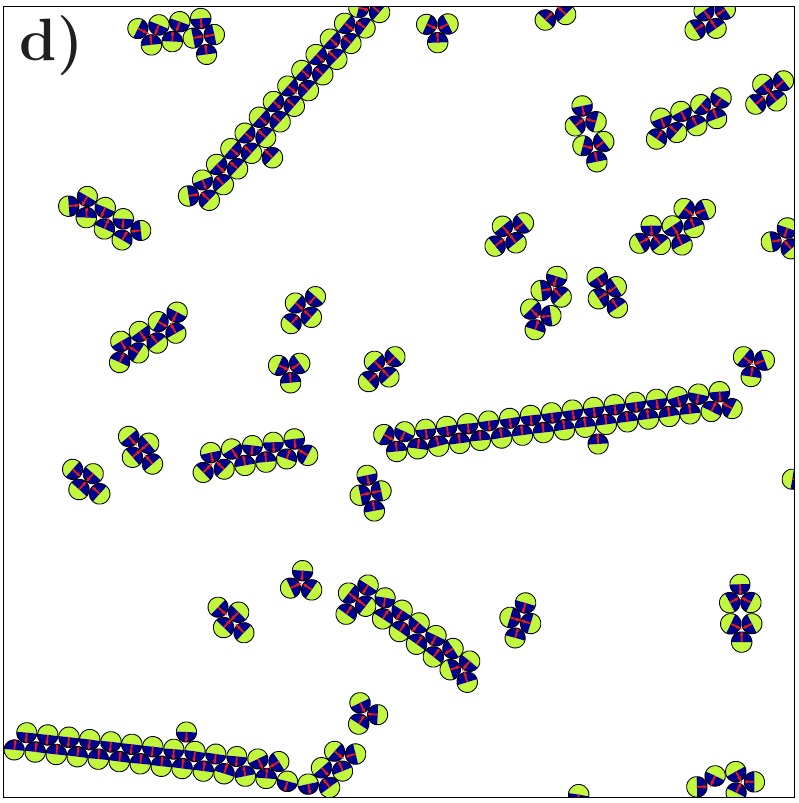}
\includegraphics[width=0.49\columnwidth]{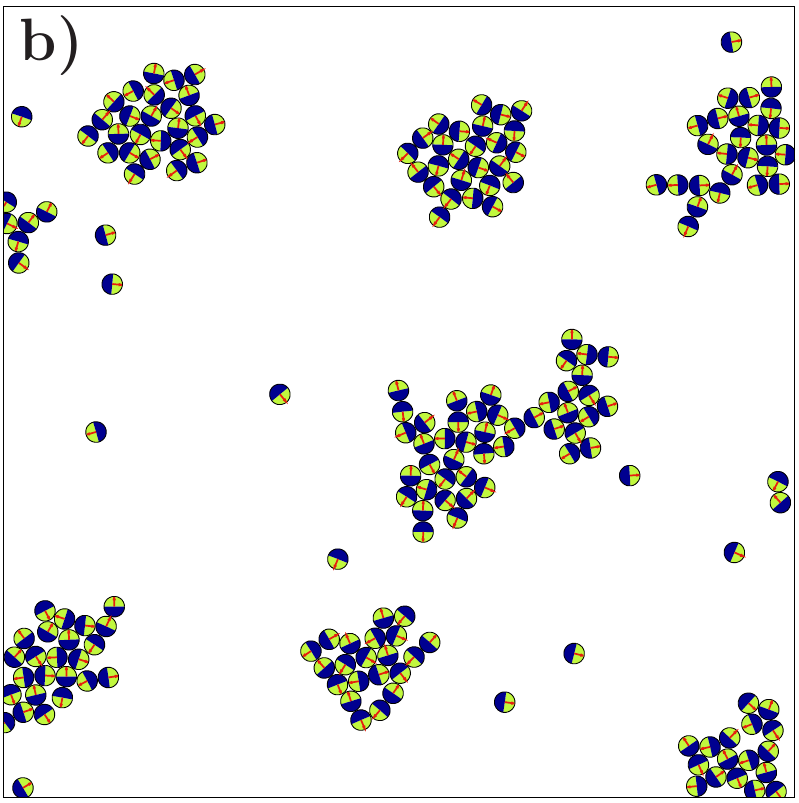}
\includegraphics[width=0.49\columnwidth]{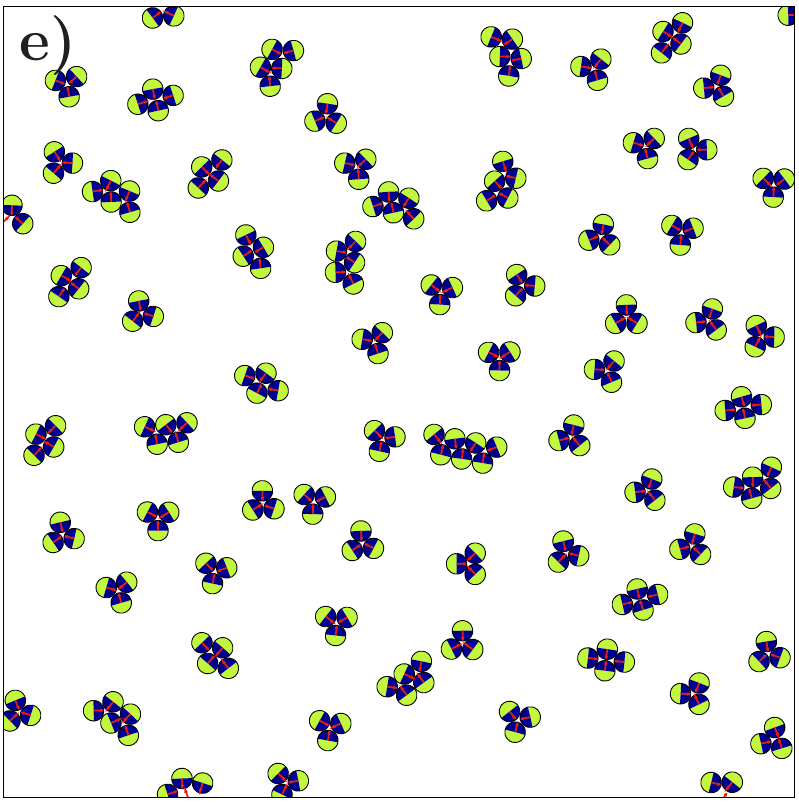}
\includegraphics[width=0.49\columnwidth]{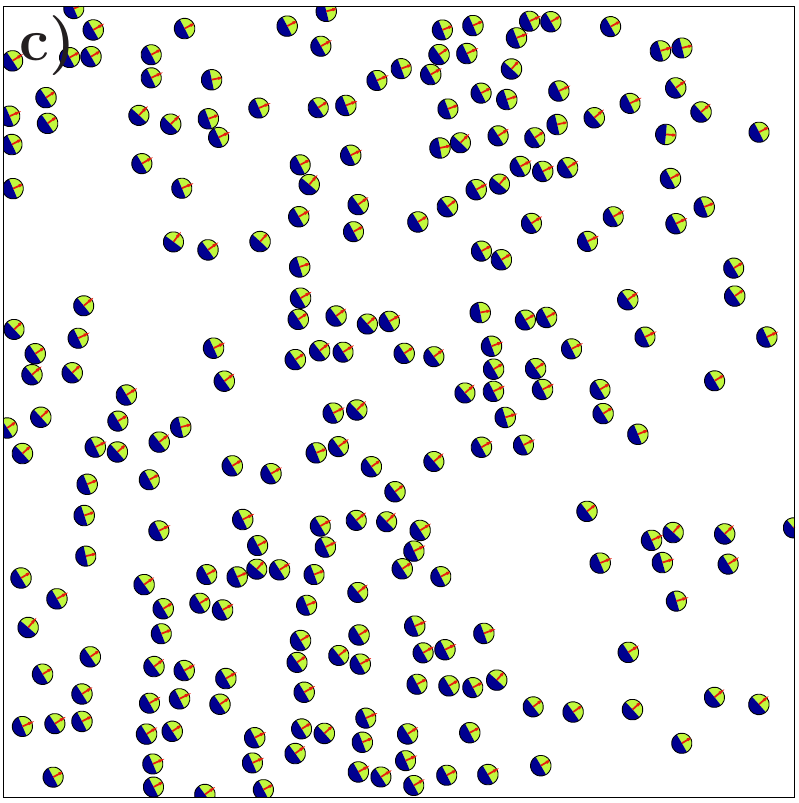}
\includegraphics[width=0.49\columnwidth]{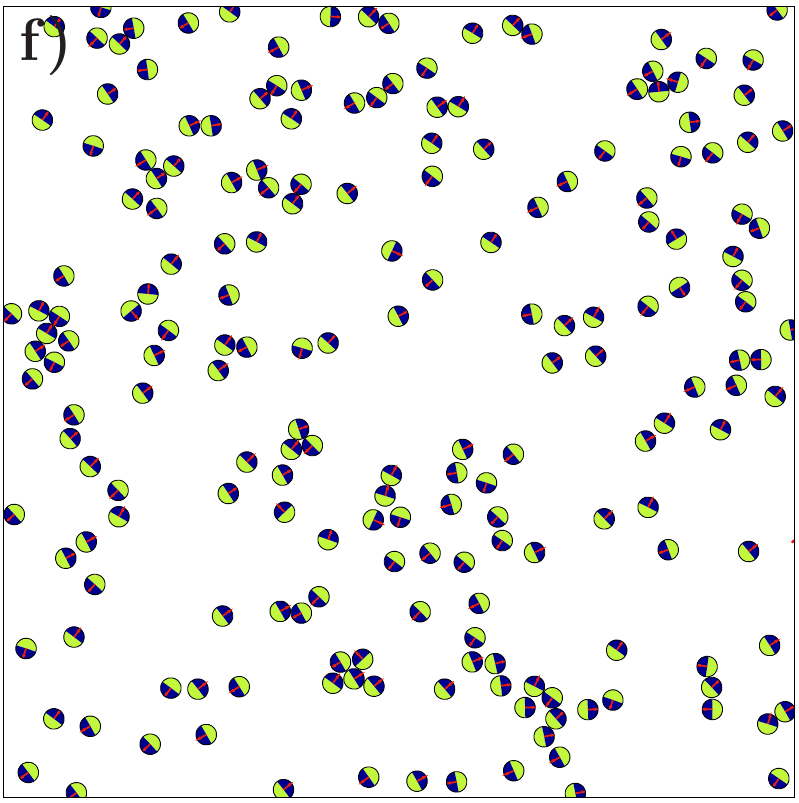}
\caption{\label{Glossary}  Snapshots of different types of self-assembly for $r_{c}=2.5 \sigma$. The attractive 
hemisphere is represented in blue, the repulsive in green and the fixed orientation vector is shown in red. (a), (b) and (c) are AP squirmer suspensions with $\xi=\{0.1,1,10\}$ respectively, while (d), (e) and (f) are WP squirmer suspensions with $\xi=\{0.1,1,10\}$ respectively.  (a) Coarsening  with $\beta=3$. (b)  
Clustering with $\beta=3$. (c) Gas system with polar order (polar gas) with $\beta=0$. (d) Chains system with $\beta=0$. (e) Trimers state with $\beta=-3$. (f) Gas case with nematic order (nematic gas) with $\beta=0$.}
\end{figure}

Fig.~\ref{Glossary} displays the different states, types of collective motion or self-assembly that WP and AP Janus  squirmers  form as  $\beta$ and $\xi$ vary. This way to classified our simulations, allow us to observe in a better way all the results.

We have identified seven different  cases: 3 gas systems, 3 clustering  systems and 1 coarsening state.
All the different cases, show different types of cluster morphologies and/or particles' alignment. The emerging structures result to be  sensitive to 
 the  patch direction (WP or AP), the hydrodynamic signature $\beta$ and the interaction range $r_{c}$ and strength $\xi$. 
The three gas systems are characterized  by the different particles' alignment:  isotropic  (with  particles randomly swimming),  polar  (with 
most of the particles swimming in the same direction) and  nematic  (with approximately half of the total amount of 
particles swimming in one direction and the other half in the opposite direction). 
The three  clustering systems can be classified as:  finite-size dynamic clusters,  chains  and  trimers. 
In the coarsening case all particles form a unique macroscopic aggregate. 

AP Janus squirmers coarsen into a single cluster 
 for $\xi \approx 0.1$, (see Fig. \ref{Glossary}-(a)) and  self-organise into  
   finite size clusters for  $\xi \approx 1$ (see Fig. \ref{Glossary}-(b)) (larger values of $\beta$ favoring 
  the formation of larger finite-size clusters, accelerating  coarsening). 
    When $\xi=10$ , 
   AP squirmers   either form a polar ordered gas  (see Fig. \ref{Glossary}-(c), for weak positive stresslets,  $0<\beta<1$) or an isotropic gas 
   (for AP pushers and pullers with $\beta > 1$).
WP Janus squirmers have a strong tendency to form 
small micelles of three or four particles  pointing at each other. For  $\xi \approx 0.1$ these micelles  coalesce 
and end up forming chains, which in most of the  cases coexist with  smaller micelles (see Fig. \ref{Glossary}-(d)). 
At $\xi=1$ the competition with the active stress ($\beta$) gives rise  to trimers and tetramers avoiding the formation of larger 
   chains of particles (see Fig. \ref{Glossary}-(e)). If activity dominates  ($\xi=10$) 
and $|\beta|<1$, even though a polar order  is not favored due to  the anisotropic interactions, 
 we  still observe a global nematic order:  a significant portion of  particles  move in one direction and the rest in 
the opposite one (see Fig. \ref{Glossary}-(f)). For other values of $\beta$, we observe an isotropic system.
{In the Supplemental Material \cite{Supp} we have included the movies of the six simulations represented in Fig. \ref{Glossary}, to better capture both the dynamical and morphological features reported in this paper.}

\subsection{Cluster size distribution}

\begin{figure}[h!]
\centering
\includegraphics[width=0.48\columnwidth]{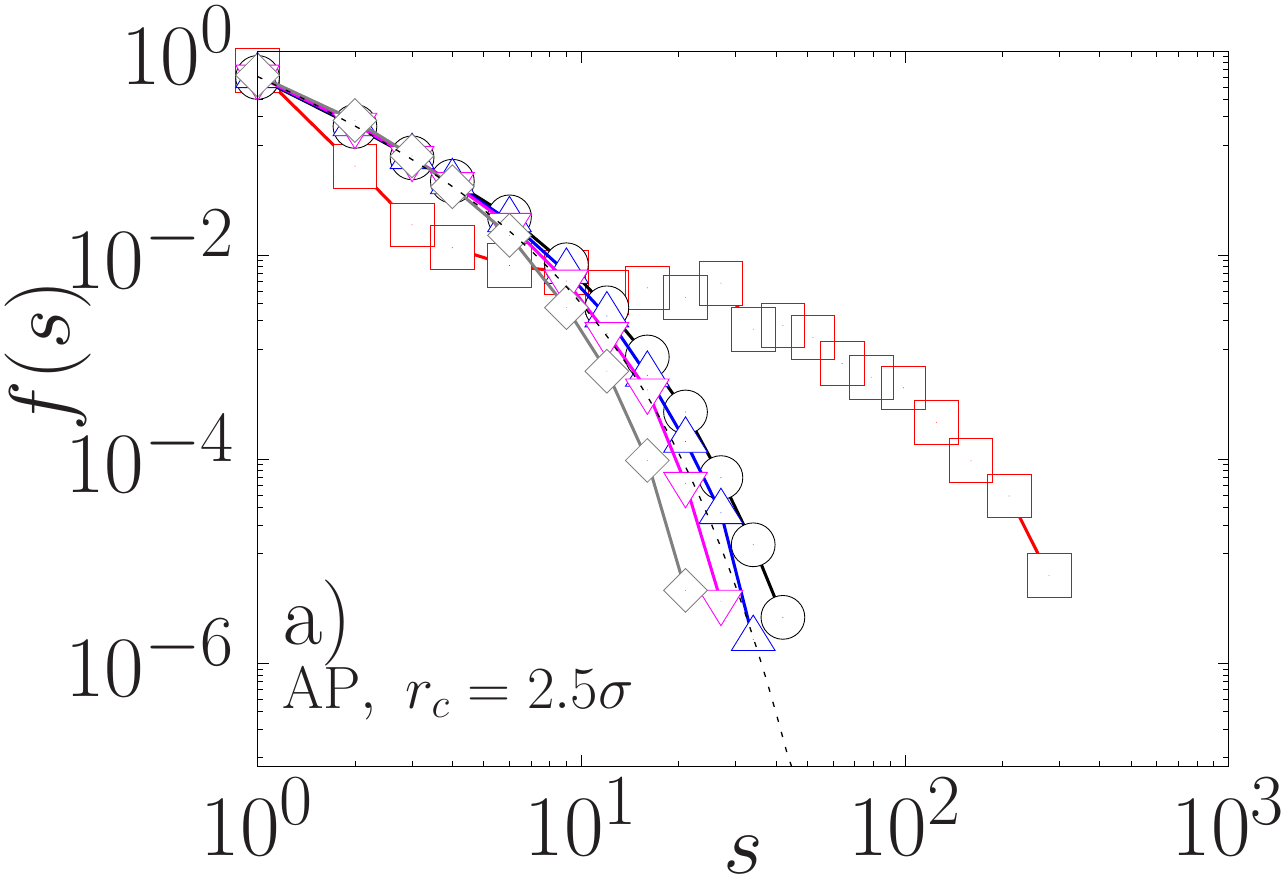}\includegraphics[width=0.48\columnwidth]{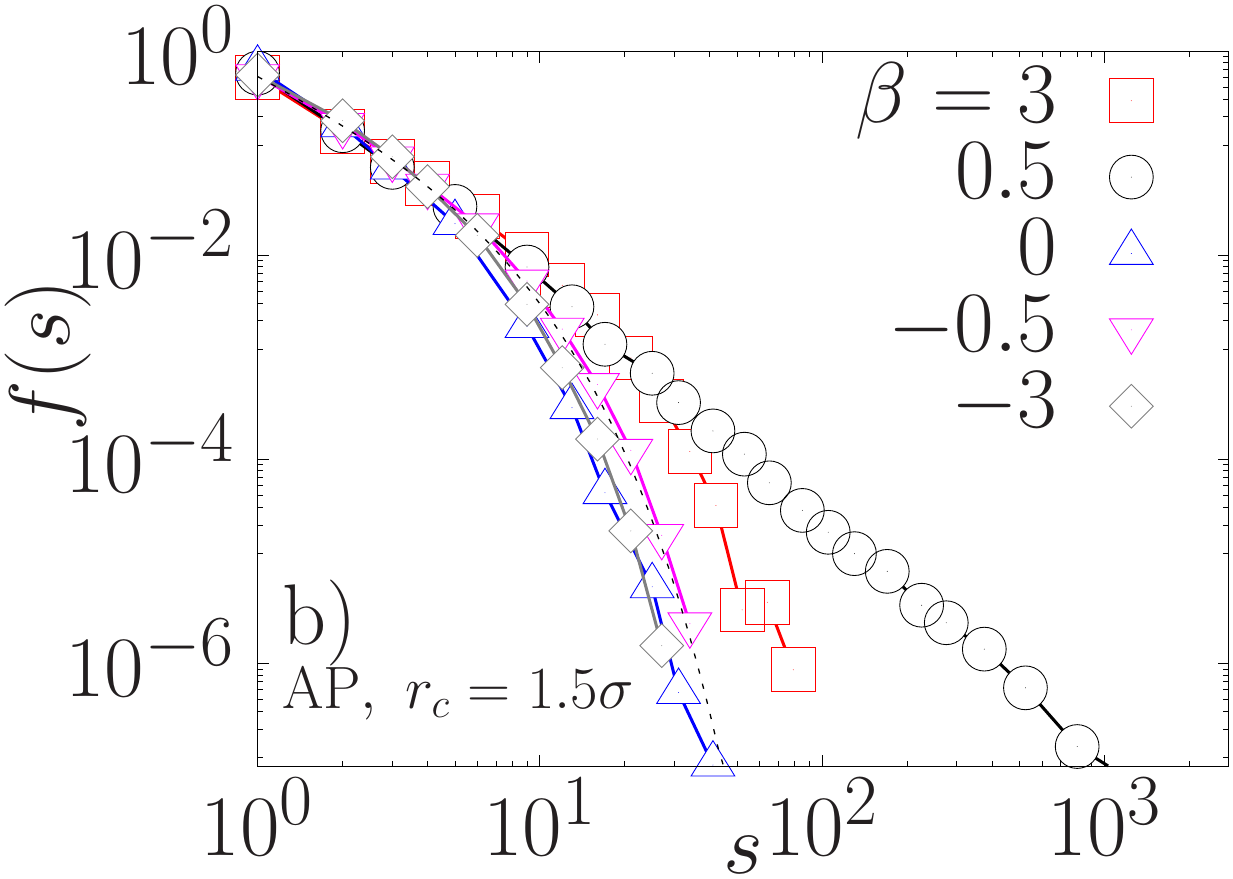}
\includegraphics[width=0.48\columnwidth]{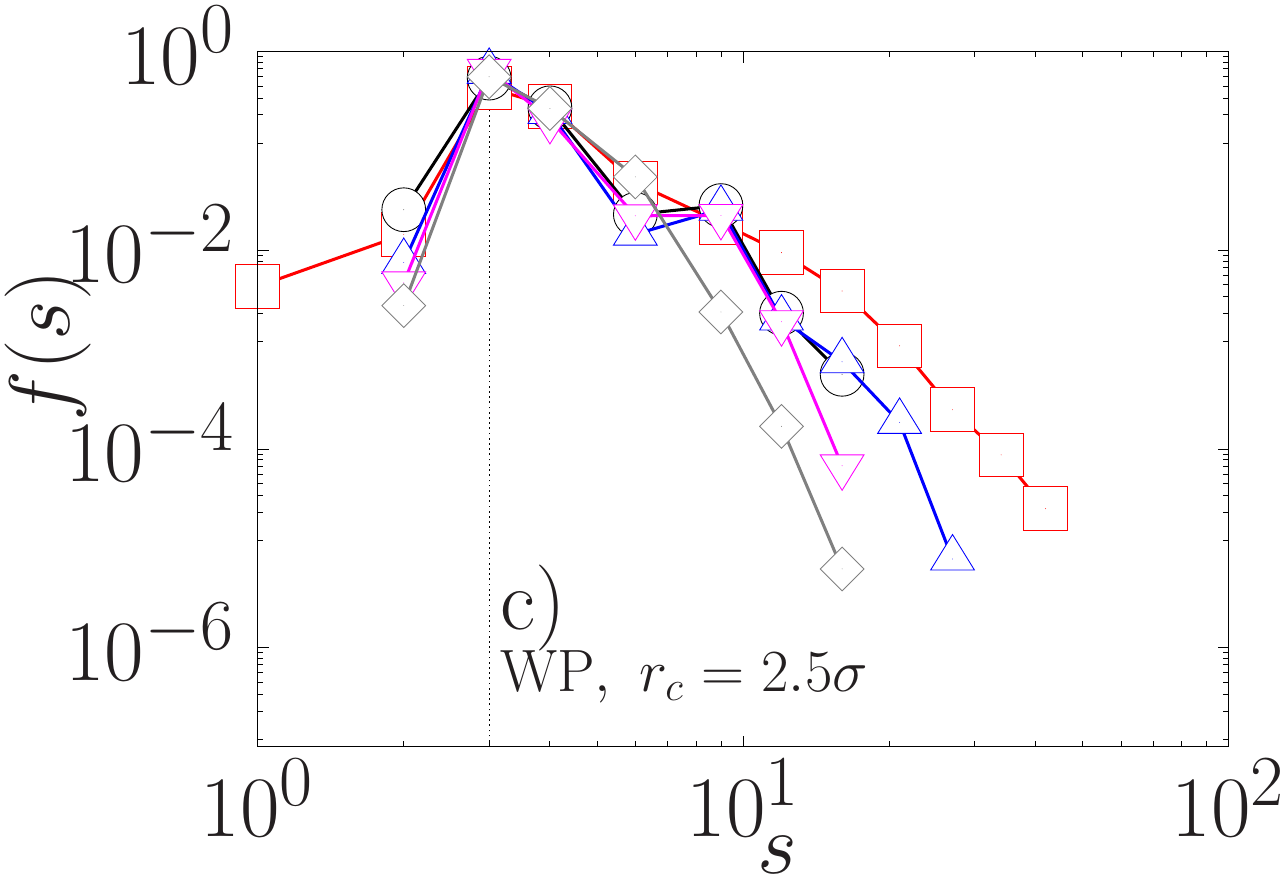}~\includegraphics[width=0.48\columnwidth]{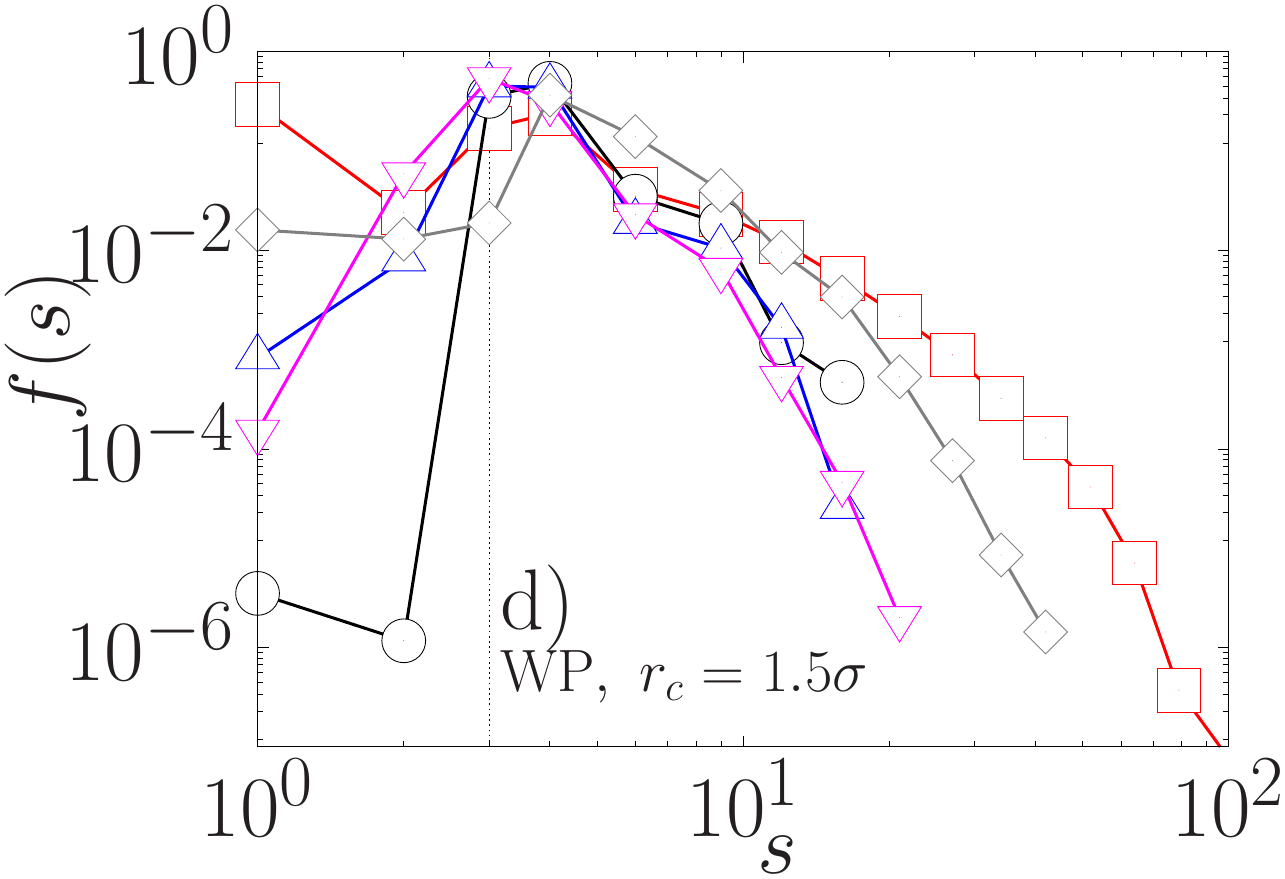}
\caption{\label{fig:csd_xi1} CSD for  $\xi=1$. AP squirmers (a) with $r_c=2.5\sigma$ and (b) $r_c=1.5\sigma$. WP squirmers (c) with $r_c=2.5\sigma$ and (d) $r_c=1.5\sigma$. 
Gray dashed line in (a) and (b) represents the analytical function for a CSD given by Eq. \ref{eq:csd_cutoff} with $B=0$, plotted as guide to the eye, with $s_0=4$, $\gamma_0=5/4$. In (c) and (d) a vertical dashed line is shown as a reference for $s=3$ (trimers). }
\end{figure}

\begin{figure}[h!]
\centering
\includegraphics[width=0.48\columnwidth]{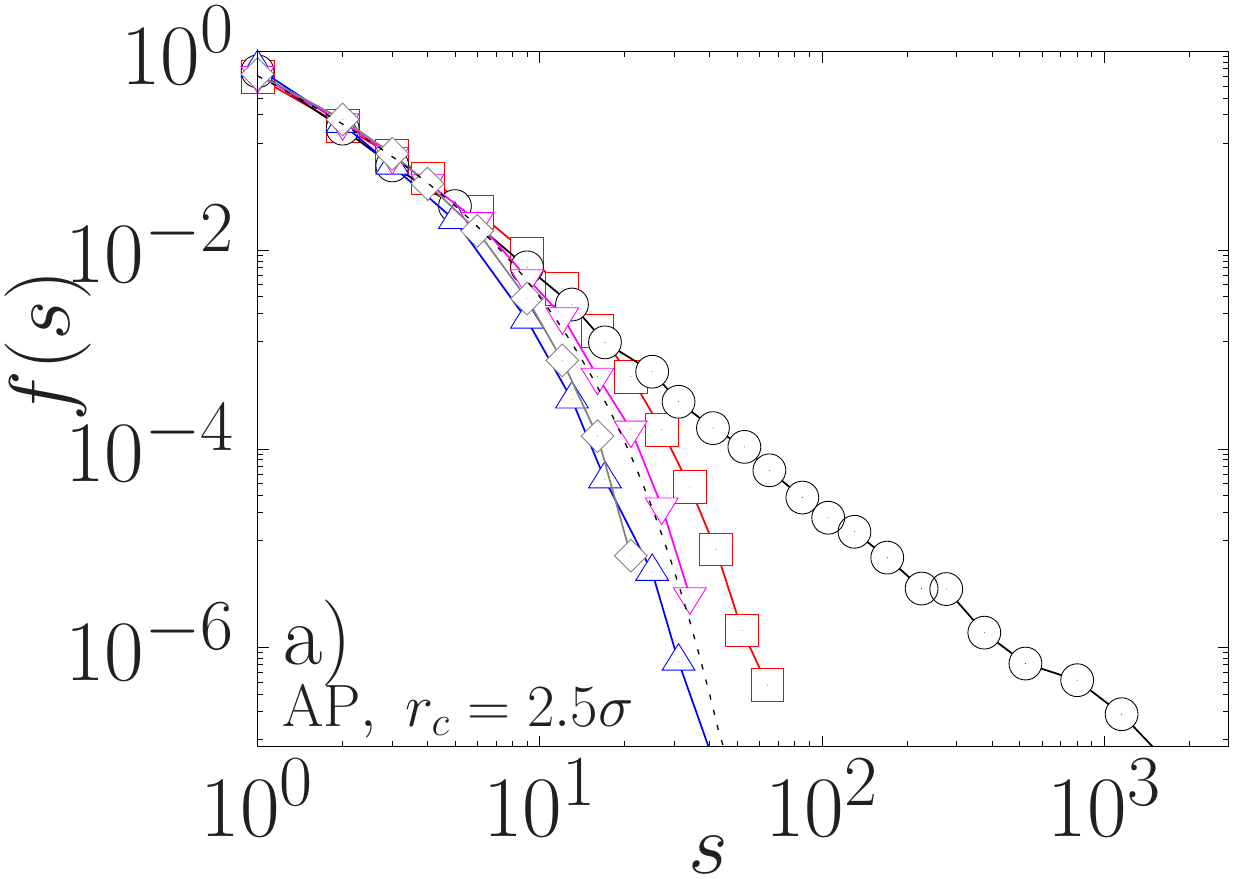}\includegraphics[width=0.48\columnwidth]{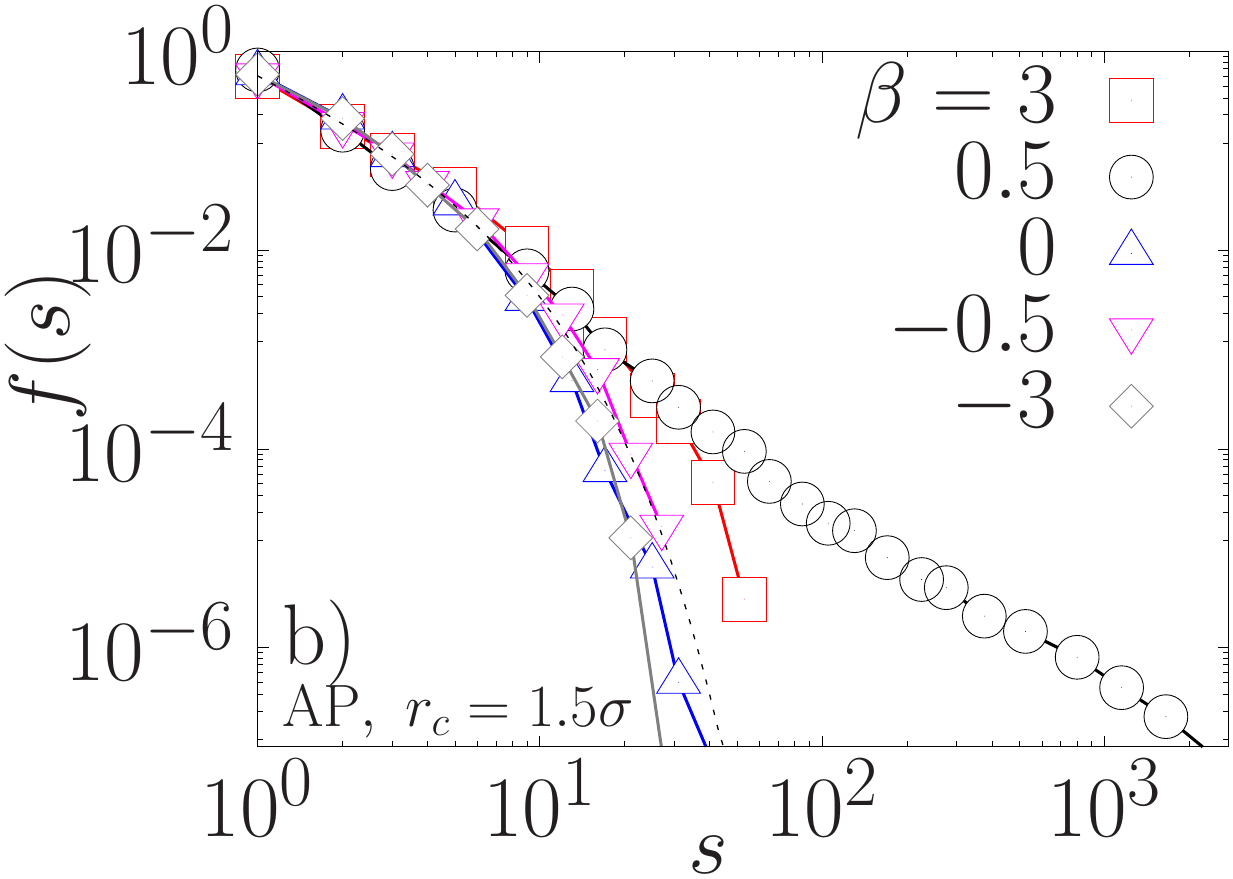}
\includegraphics[width=0.48\columnwidth]{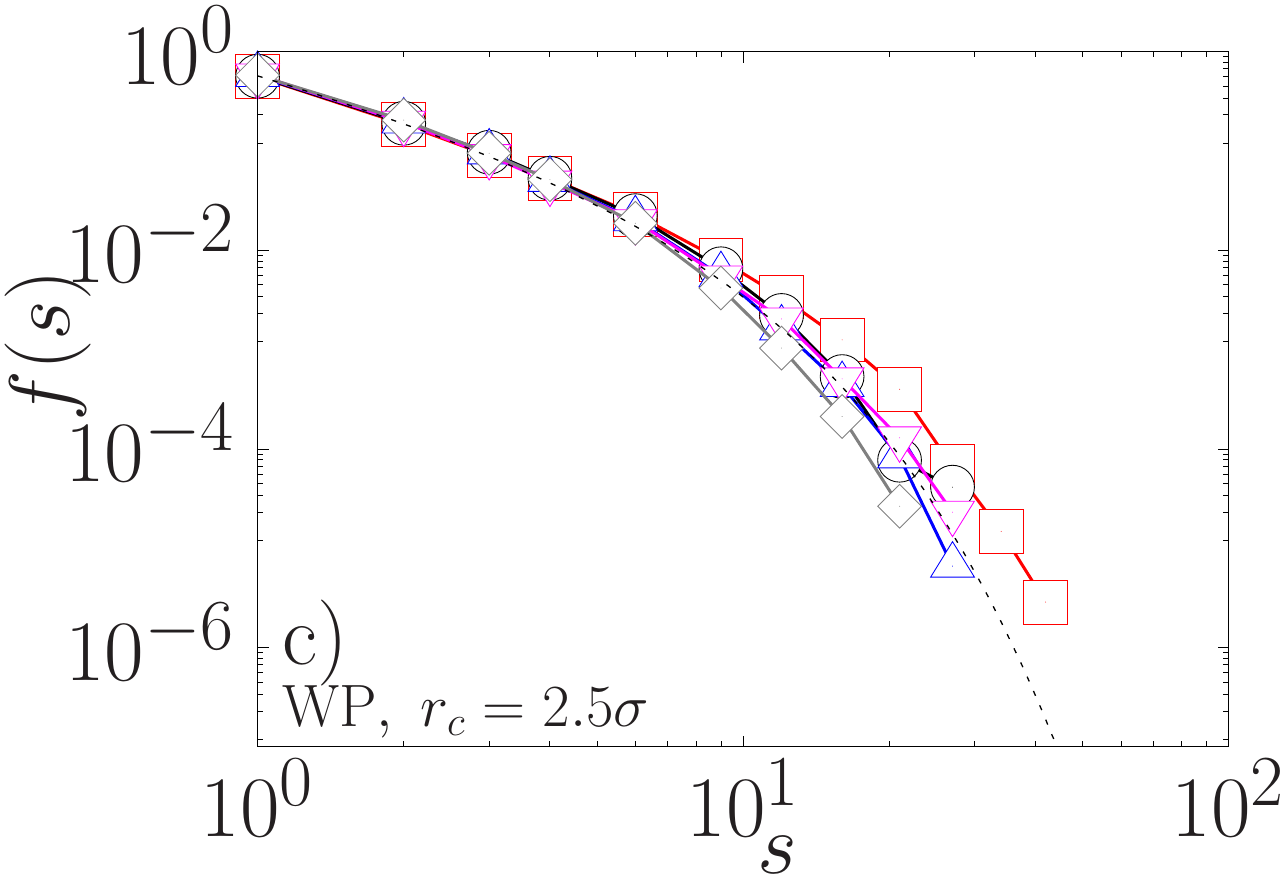}\includegraphics[width=0.48\columnwidth]{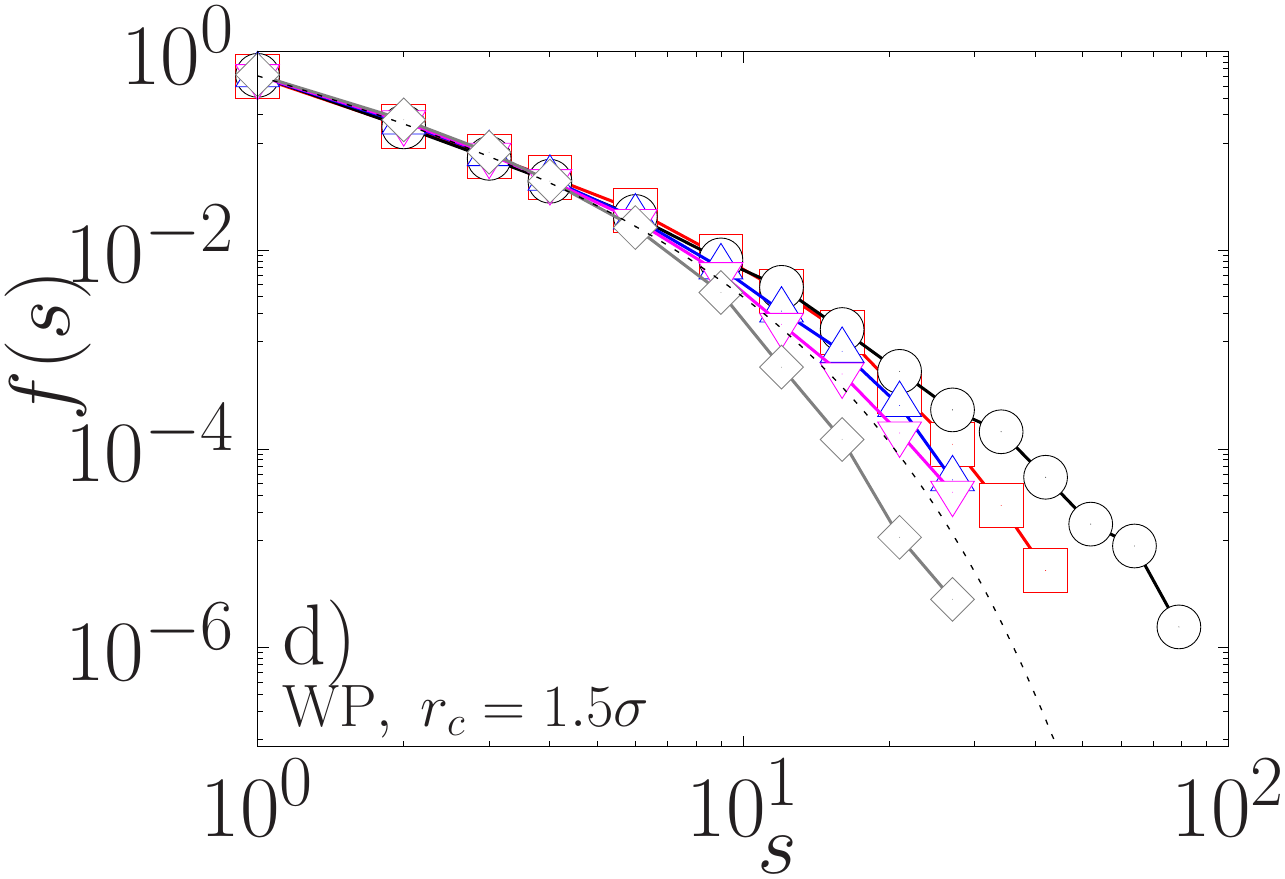}
\caption{\label{fig:csd_xi10} CSD for  $\xi=10$. AP squirmers (a) with $r_c=2.5\sigma$ and (b) $r_c=1.5\sigma$. WP squirmers (c) with $r_c=2.5\sigma$ and (d) $r_c=1.5\sigma$. 
Grey dashed line in (a) and (b) represents the analytical function for a CSD given by Eq. \ref{eq:csd_cutoff} with $B=0$, plotted as guide to the eye, with $s_0=4$, $\gamma_0=5/4$.}
\end{figure}

Additionally to the mean cluster size, we calculate the cluster size distribution (CSD) to study the statistics of the different cluster sizes found it for Janus swimmers,
  computing  the CSD for $\xi=1$ and  $\xi=10$ and 
as a function of  $\beta$,   the direction of the attractive patch and 
  the interaction  range. 
{The case when attraction dominates with respect to self-propulsion ($\xi=0.1$) for AP squirmers is analyzed in appendix \ref{Appendix:xi0_1WPrc1_5}.}
  Results for  $\xi=1$ are presented in Fig.~\ref{fig:csd_xi1} and for $\xi=10$ in Fig.~\ref{fig:csd_xi10}.  
Following previous studies \cite{SoftMatter17}, where CSD was calculated, we  use the same analytical function as a reference. 
\begin{equation}\label{eq:csd_cutoff}
\frac{f(s)}{f(1)}= A~\frac{\exp(-(s-1)/s_0)}{s^{\gamma_0}}+B~\frac{\exp(-(s-1)/z_0)}{s^{-\gamma_0}},
\end{equation}

\noindent
with $\gamma_0$, $s_0$, $z_0$ and $B$ constants such that $A=1-B$. 

Fig. \ref{fig:csd_xi1}-(a) represents  the CSDs of AP squirmers for $\xi=1$ and $r_c=2.5 \sigma$. The characteristic  distribution width 
grows with $\beta$ until $\beta < 3$. For all the explored parameters,  monomers always represent the largest contribution. 
For large enough values of $\beta$,  $\beta = 3$, a second peak emerges for cluster sizes involving a few tens of particles (around 25 for the system parameters  explored).  
Accordingly,  CSDs with $\beta < 3$ can be described by Eq. \ref{eq:csd_cutoff}, with $B=0$ 
while the CSD for $\beta=3$ requires   a non-vanishing value of $B$ in Eq. \ref{eq:csd_cutoff}  to capture the  observed maximum. 
The width of the CSDs is measured by the value of the exponential tail parameter $s_0$, thus pushers CSDs have smaller $s_0$, 
whereas the corresponding   $s_0$ for pullers is larger  until $\beta=3$. At larger values of $\beta$
 the behavior changes due to the appearance of a second peak in the CSDs.

Fig.~\ref{fig:csd_xi1}-(b) represents  the CSDs of  AP squirmers for $\xi=1$ and $r_c=1.5 \sigma$.
{CSDs of AP pushers and neutral squirmers follow the analytical behavior of a power law with an exponential tail (Eq. \ref{eq:csd_cutoff} with $B=0$: gray dashed line) with the same width. 
The squirmers with $\beta=3$ present a CSD wider than the pushers ones,
whereas weak pullers CSD follows a power law behavior instead of the power law with an exponential tail (Eq. \ref{eq:csd_cutoff}) which represents clusters of all sizes, such effect is not observed when interaction range is $2.5 \sigma$.}

Fig. \ref{fig:csd_xi1}-(c) represents the CSDs of WP squirmers for $\xi=1$ and $r_c=2.5 \sigma$.  A peak appears between 
trimers and tetramers, while the characteristic CSD width increases with the active stress $\beta$.
 Interestingly, when $\beta < 3$  monomers are not observed.  Strong WP pushers are characterized by  a suspension dominated by  trimers, while strong WP pullers form a  
 polydisperse suspension of monomers, dimers, trimers and even chains of tens of particles.

 Fig. \ref{fig:csd_xi1}-(d) represents  the CSDs of  WP squirmers for $\xi=1$ and $r_c=1.5 \sigma$, with CSDs presenting a peak between 
trimers and tetramers.  In this case, the CSD width depends on the active stress magnitude rather than in its sign: the higher the value of $|\beta|$,  the wider the distribution 
(with distributions wider than the ones  with longer  interactions range $r_c=2.5\sigma$, and with a higher number of monomers especially when $\beta=3$ ).

Therefore, CSDs present relevant differences depending on whether squirmers are AP or WP-like  and on the interaction range of the potential.
The CSDs for AP squirmers  decay monotonously as a function of cluster size, with a large tail corresponding to cases in which there is the 
formation of finite-size aggregates. On the contrary, CSDs for WP usually display a peak for trimers.
Moreover, the interaction range affects how the CSDs width depends on the hydrodynamic signature: while for large interaction range, strong AP pullers develop a wider distribution than any other $\beta$, for shorter interaction range {weak pullers, develop a power law distribution}.

In contrast, neither the range of interaction among WP squirmers nor the hydrodynamic signature affect  the shape of the CSDs,
 since CSDs with a peak on trimers is present in all cases, with the widest distribution appearing for the stronger WP or AP pullers.

 Fig. \ref{fig:csd_xi10}-(a) represents  the CSDs of AP squirmers with $\xi=10$ and $r_c=2.5 \sigma$. The  corresponding distributions are 
monotonically decreasing, particularly for $\beta=0.5$ {a power law distribution emerges while in the other cases,} 
the CSD can be approximated by Eq.~\ref{eq:csd_cutoff} with $B \neq 0$.

Fig.~\ref{fig:csd_xi10}-(b) displays AP squirmers' CSDs for $\xi=10$ and $r_c=1.5 \sigma$. 
The CSDs are essentially the same as in the previous case with longer interaction range for pushers and $\beta=0$, while for pullers the 
CSDs are lightly wider.

Fig.~\ref{fig:csd_xi10}-(c) displays  the CSDs for WP squirmers for $\xi=10$ and $r_c=2.5 \sigma$. In this case, 
hydrodynamics does not significantly affect the CSD, since  the CSDs are all monotonically decreasing.  Although $\beta=3$ is 
slightly wider  and $\beta=-3$ is slightly narrower than the others, the distributions are similar among them, with approximately the same width. 
Since  activity dominates over the interaction strength, we can infer that this independence on $\beta$  is due to two main effects: 
on the one hand, the anisotropy of the potential and on the other hand, the range of the interaction.
Since in Fig. \ref{fig:csd_xi10}-(d) for  
WP squirmers and $\xi=10$ and $r_c=1.5 \sigma$, the CSDs have indeed a different width depending 
on the hydrodynamic signature, corresponding   $\beta=0.5$ to  the widest distribution.
When activity dominates over the interaction strength, we can infer that CSDs depend mainly on hydrodynamics, but the interaction range plays an important  role too, mainly for  AP and WP  pullers.


\subsection{Radius of gyration}

The next feature we compute to study the morphology of the clusters is their radius of 
  gyration $Rg(s)$ (Eq.~\ref{eq:Rg}), as a function of  the cluster size $s$.
\begin{figure}[h!]
\centering
\includegraphics[width=0.48\columnwidth]{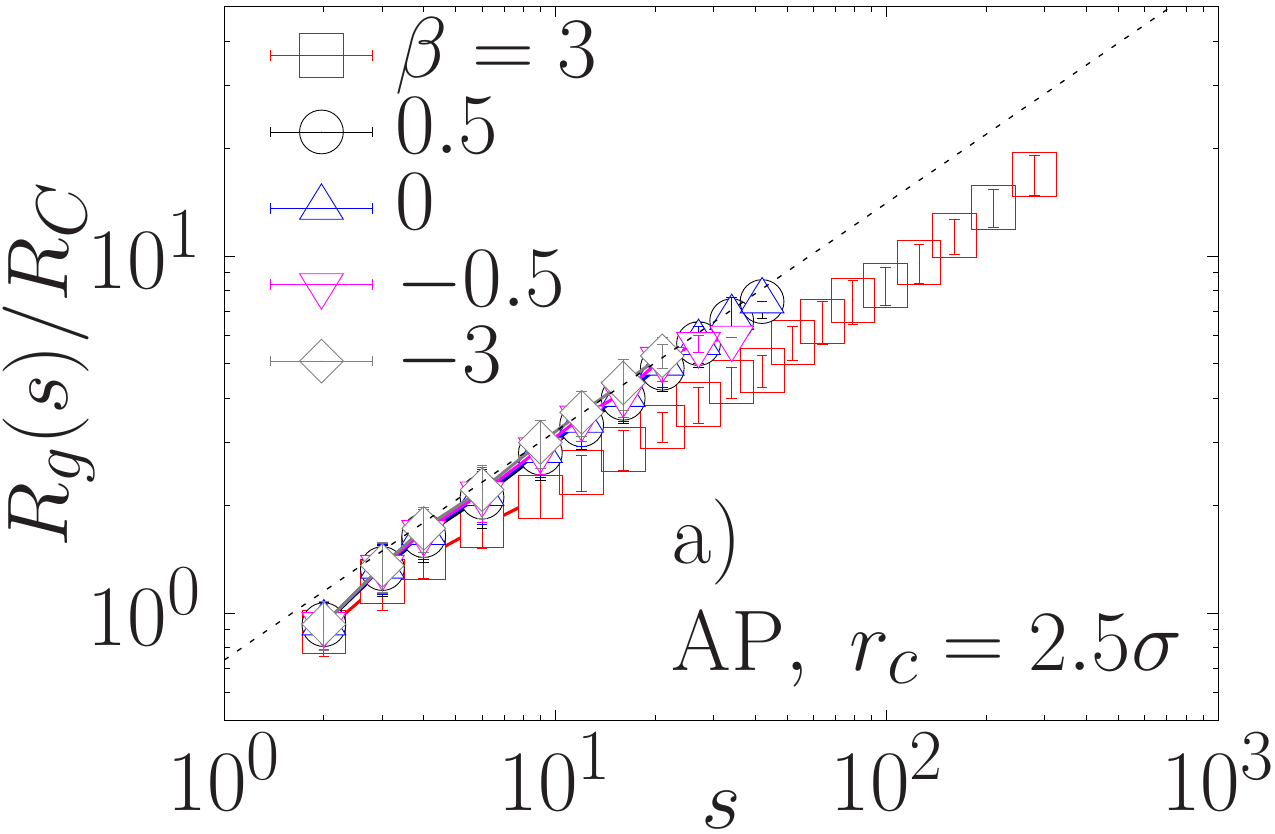}\includegraphics[width=0.48\columnwidth]{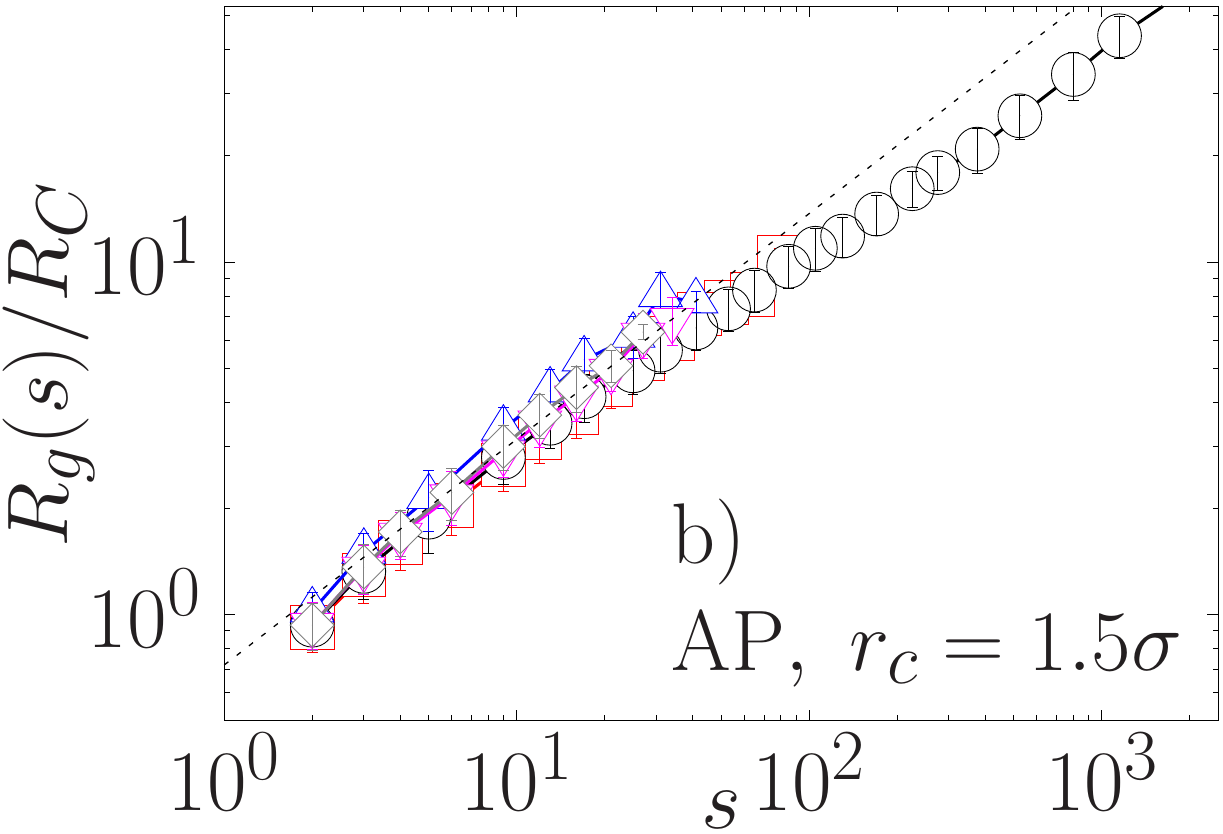}
\includegraphics[width=0.48\columnwidth]{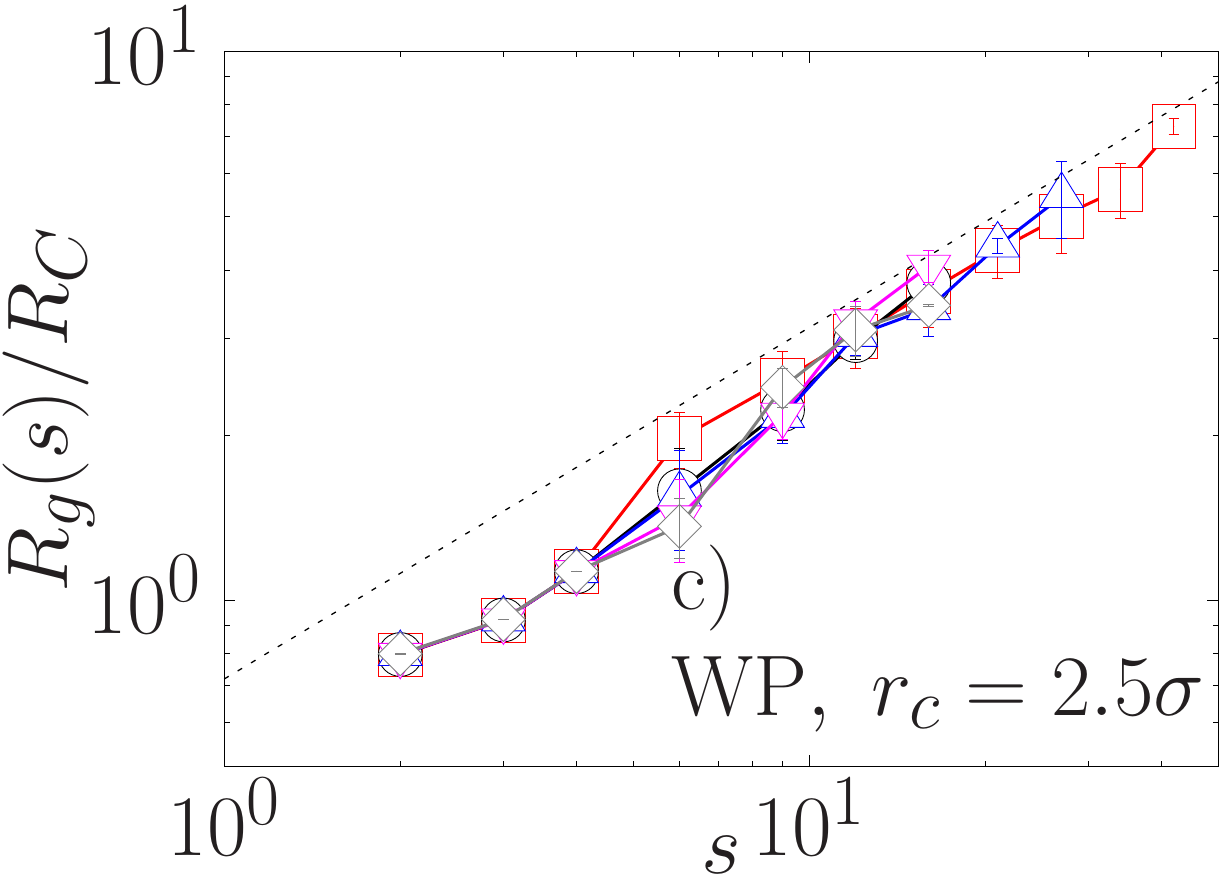}\includegraphics[width=0.48\columnwidth]{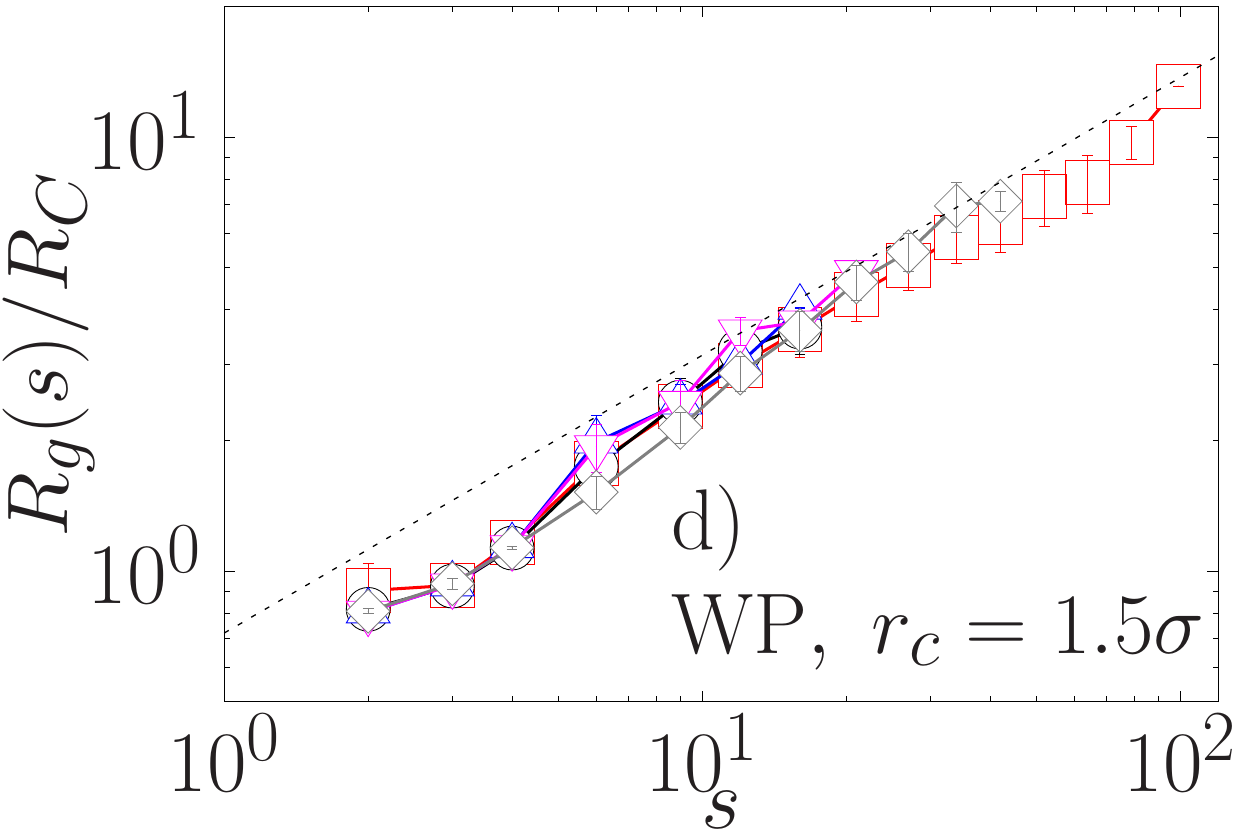}
\caption{\label{fig:Rg_xi1} Radius of gyration normalized by the particle radius as a function of the cluster size for suspensions with $\xi=1$. Top panels are AP squirmers (a) with $r_c=2.5\sigma$ and (b) $r_c=1.5\sigma$, whereas bottom panels are WP squirmers (c) with $r_c=2.5\sigma$ and (d) $r_c=1.5\sigma$. 
Grey dashed lines in all panels are the function $R_g = K s^{0.64}$.}
\end{figure}
Fig.~\ref{fig:Rg_xi1}-(a) displays  the radius of gyration $R_g(s)$ of AP squirmers  with $\xi=1$ and $r_c=2.5\sigma$. 
  We observe that $R_g\sim s^{0.64}$ for  $\beta <3$, while for $\beta=3$ (red squares in Fig. \ref{fig:Rg_xi1}-(a)),  $R_g$ starts 
following the same power law before a crossover to a lower value of   $R_g$.   
It is insightful to compare these results with  the  CSDs, since AP 
squirmers with $\beta < 3$  have monomodal CSDs, with the width of the distribution depending on the value of $\beta$. 
Nonetheless, Fig. \ref{fig:Rg_xi1}-(a) shows that clusters morphology is independent of the hydrodynamic signature when $\beta<3$, 
but once $\beta=3$ the CSD is bimodal and it turns out that the change in  $R_g$  coincides with the changes in the CSD. 
Moreover, dynamic clusters are observed when $\beta < 3$ while less dynamic ones  are formed when $\beta=3$: therefore,  attractive 
interaction is enhanced when particles are strong pullers. 

Similarly,  $R_g\sim s^{0.64}$ in Fig. \ref{fig:Rg_xi1}-(b), when the  interaction range is short  and 
for small clusters  ($s<50$); while for  larger clusters of pullers ($s>50$) the $R_g$  power law is  smaller than $0.64$. 
The slower growth of $R_g$ depends on the fact that the clusters' structures are mostly 
controlled by  hydrodynamics.
For WP squirmers, $R_g$ shows a  richer behavior: particles first aggregate  in trimers and
 then organize  in chain-like structures. As in Fig.~\ref{fig:Rg_xi1}-(c), $R_g$ presents a crossover 
to a different asymptotic regime.

Chains are observed only for WP independently of $\beta$; hence, chain
  formation is driven by the anisotropic interaction and not by hydrodynamics, and is rather insensitive to the range of the  potential, as  in Fig. \ref{fig:Rg_xi1}-(d).

\begin{figure}[h!]
\centering
\includegraphics[width=0.48\columnwidth]{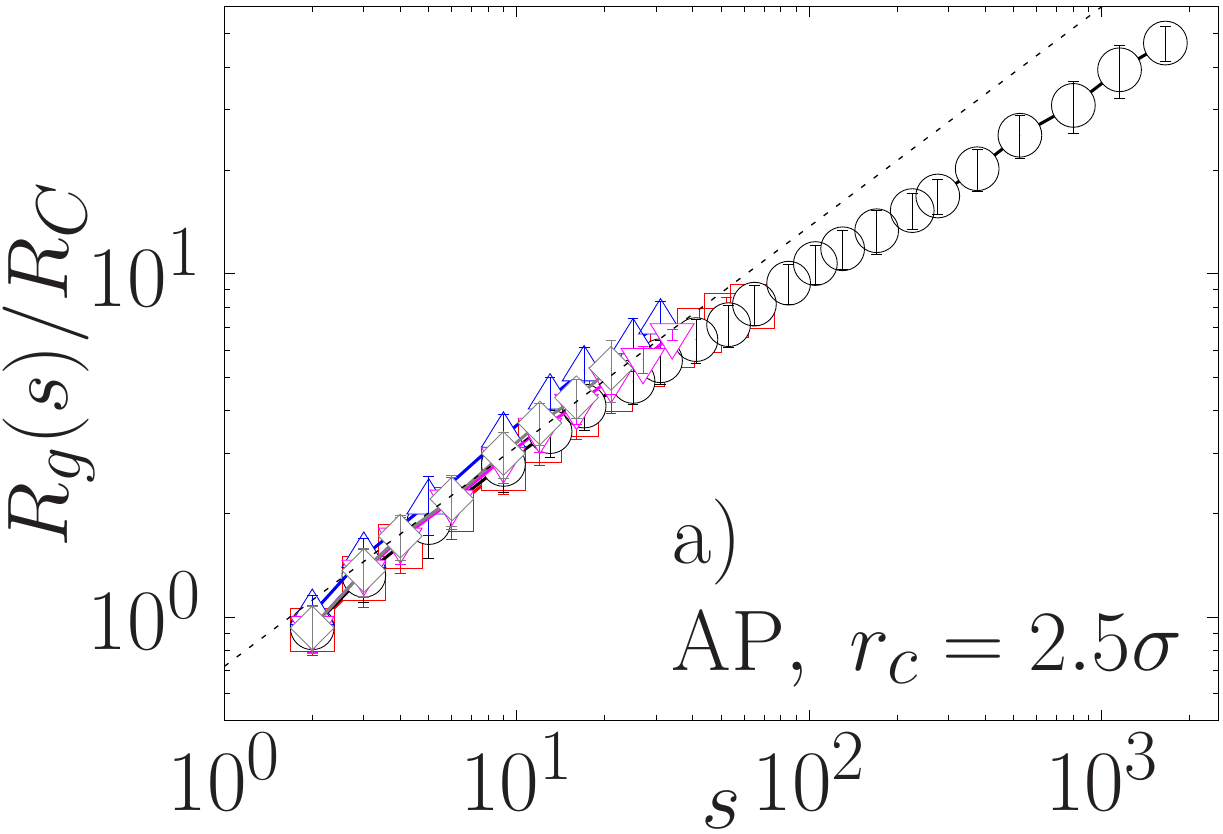}\includegraphics[width=0.48\columnwidth]{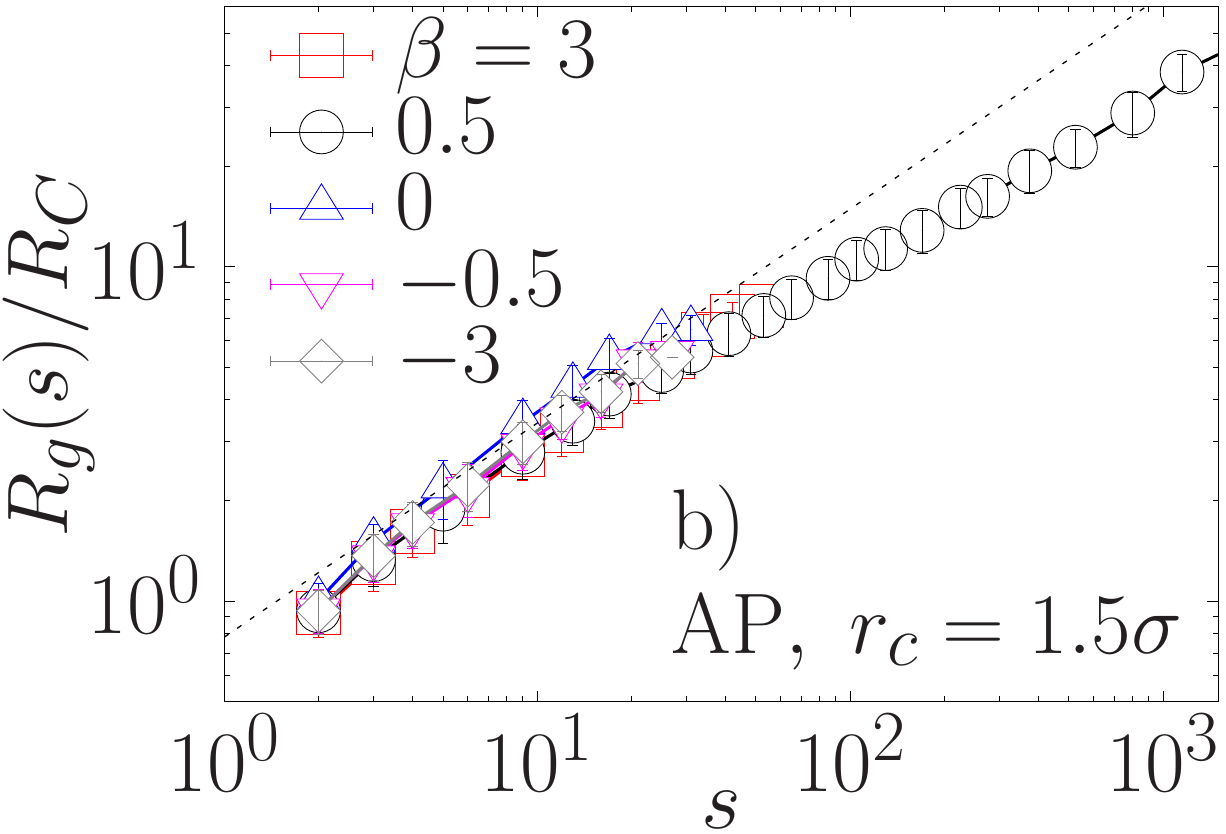}
\includegraphics[width=0.48\columnwidth]{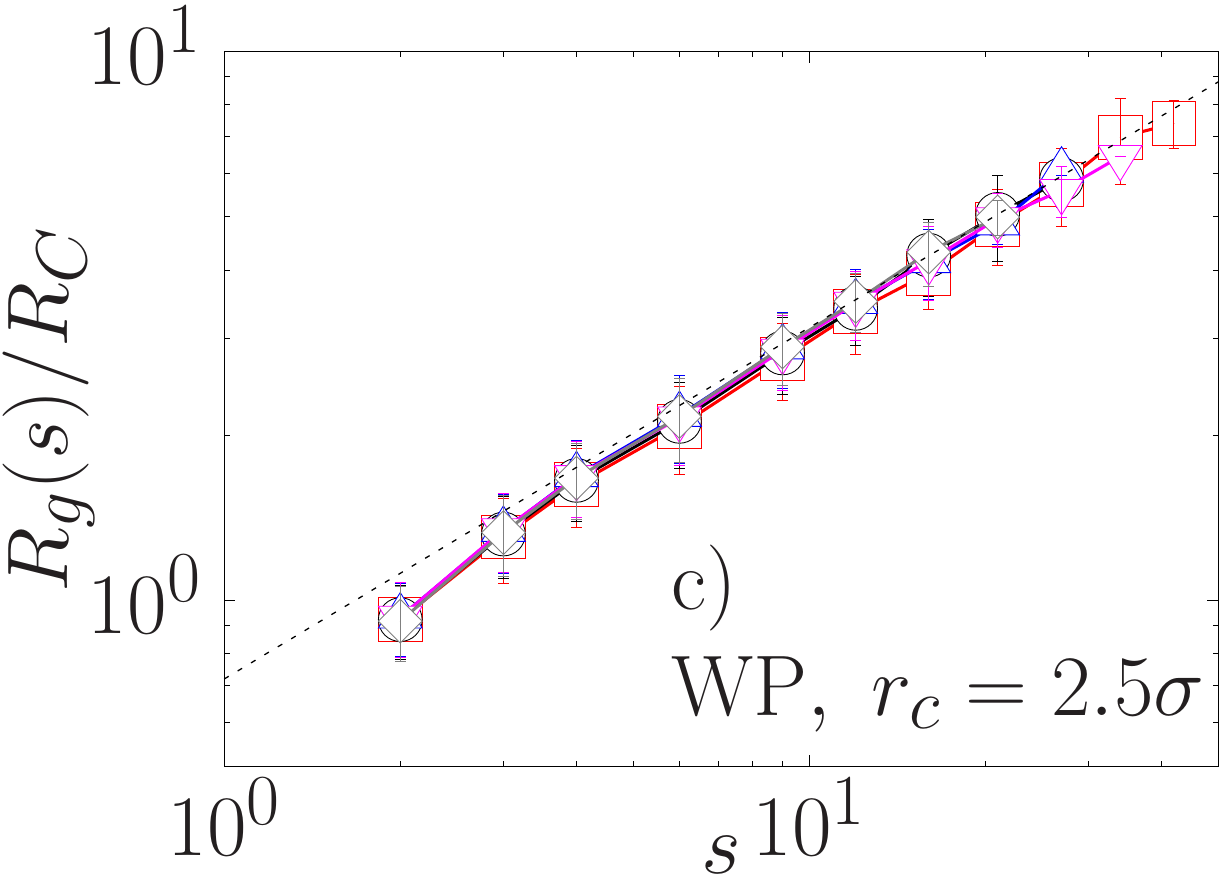}\includegraphics[width=0.48\columnwidth]{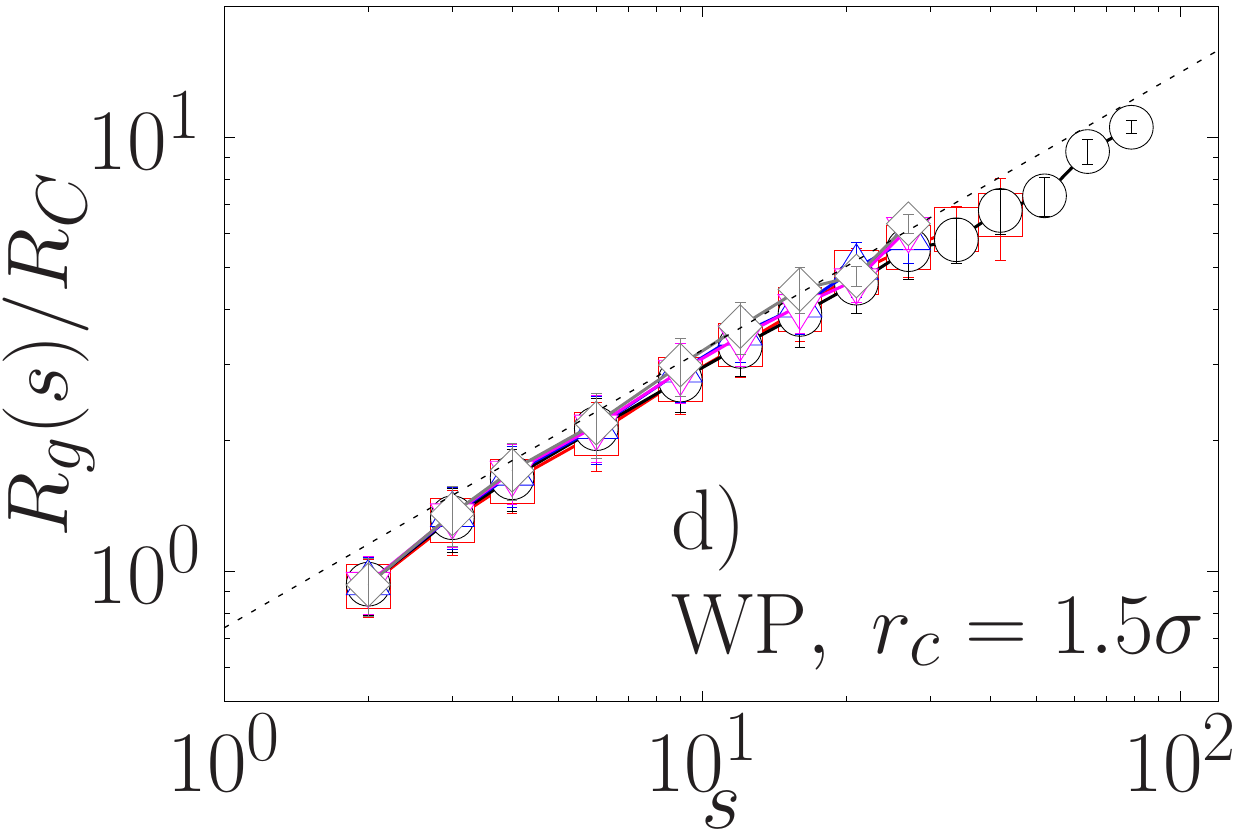}
\caption{\label{fig:Rg_xi10} Radius of gyration normalized by the particle radius as a function of the cluster size for suspensions with $\xi=10$. Top panels are AP squirmers (a) with $r_c=2.5\sigma$ and (b) $r_c=1.5\sigma$, whereas bottom panels are WP squirmers (c) with $r_c=2.5\sigma$ and (d) $r_c=1.5\sigma$. 
Grey dashed lines in all panels are the function $R_g = K s^{0.64}$ and most of the curves fit with this power law.}
\end{figure}
 
When  self-propulsion dominates over   attraction,  $R_g=ks^{0.64}$ for small clusters of AP squirmers 
 independently on $\beta$, while  for large clusters of pullers $R_g$  has a smaller exponent Fig. \ref{fig:Rg_xi10}-(a). 
The same happens when the attraction range is  $1.5 \sigma$, Fig. \ref{fig:Rg_xi10}-(b). In Fig. \ref{fig:Rg_xi10}-(c) we report the radius of gyration for the WP squirmers when $r_c=2.5 \sigma$. Given that  activity dominates over 
attraction,  particles do not assembly as chains:  $R_g$ does not present the cross-over  as  in Fig. \ref{fig:Rg_xi1}-(c), 
but rather $R_g \sim s^{0.64}$ for all $s$. Moreover,
 the  clusters' morphology is the same despite of the active stress $\beta$. However, when the interaction range is reduced to $r_c=1.5 \sigma$,  
 as in Fig. \ref{fig:Rg_xi10}-(d), clusters of pullers are more compact  when larger than 20 particles.

\subsection{Local polar and nematic order}

\begin{figure}[h!]
\centering
\includegraphics[width=0.48\columnwidth]{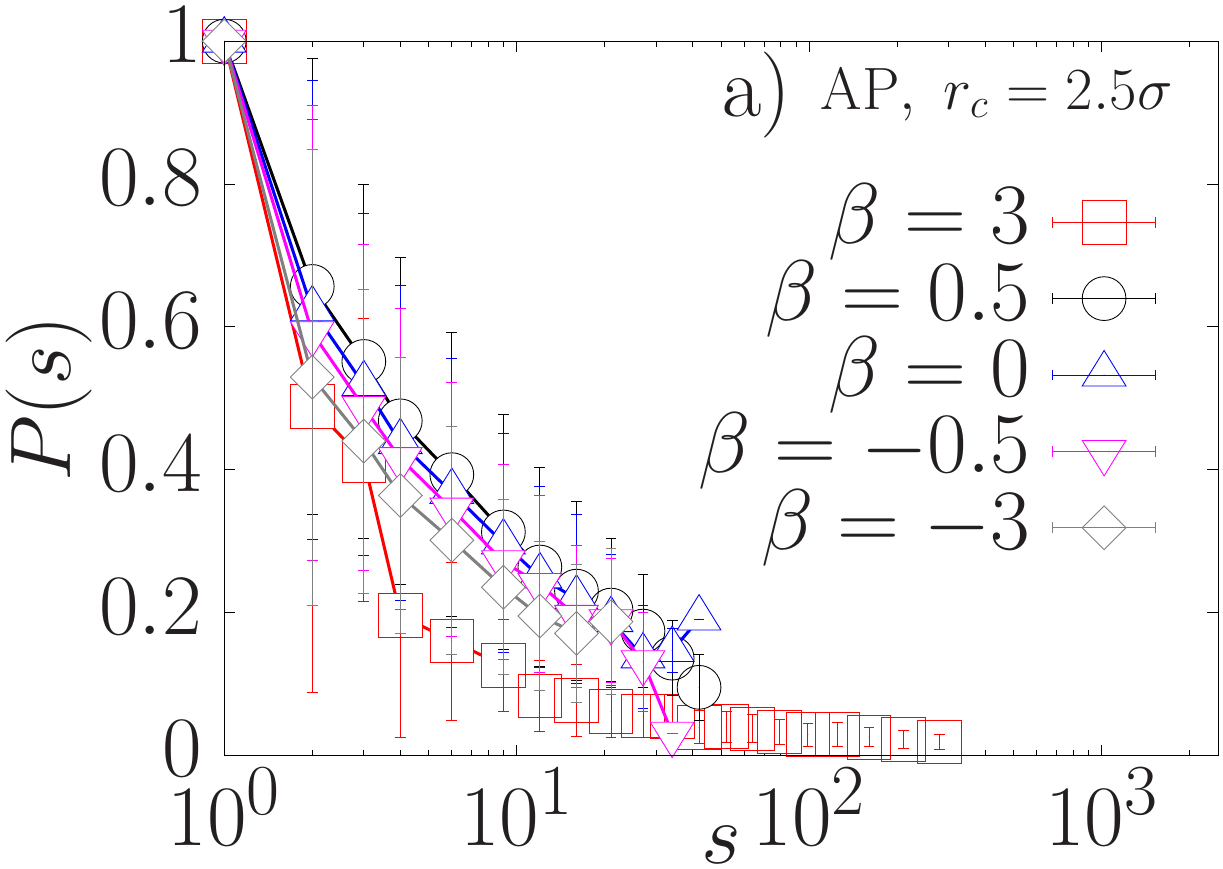}\includegraphics[width=0.48\columnwidth]{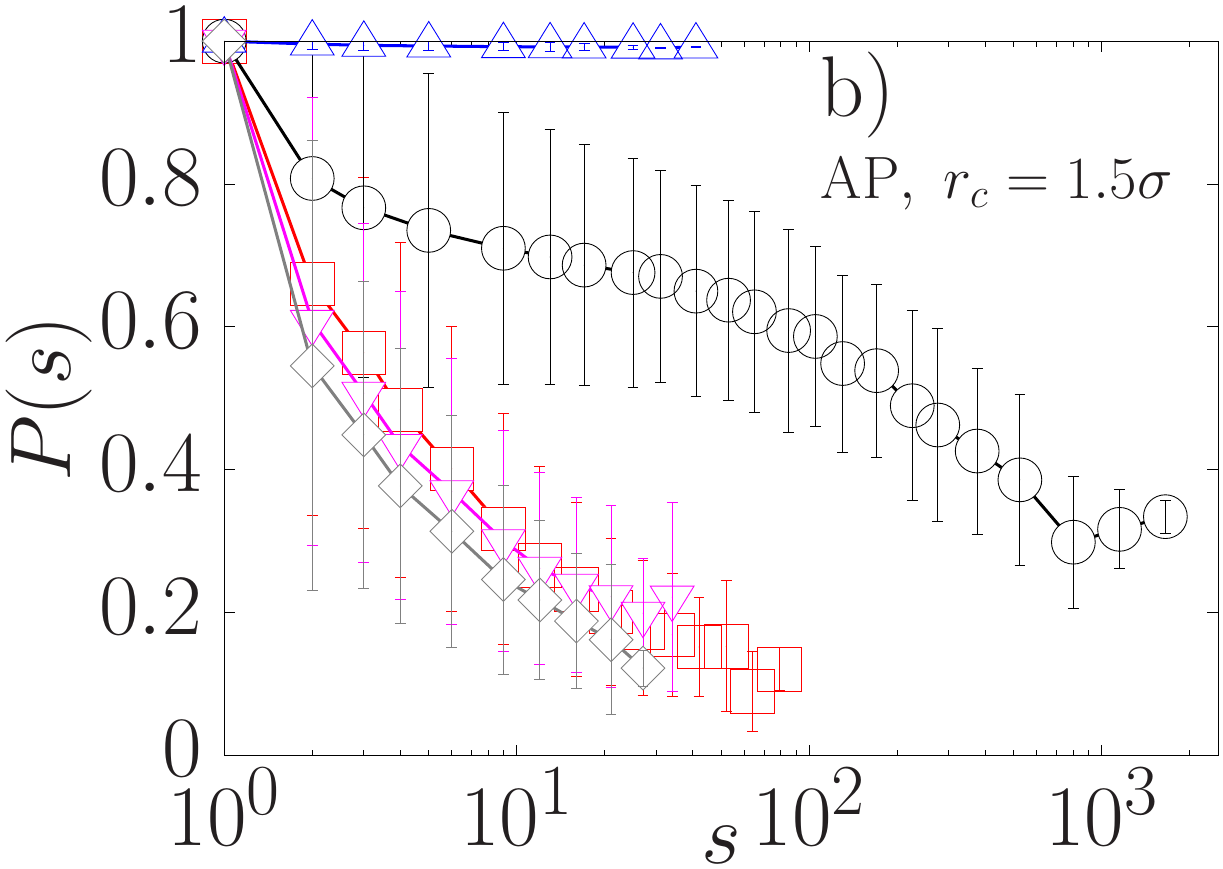}
\includegraphics[width=0.48\columnwidth]{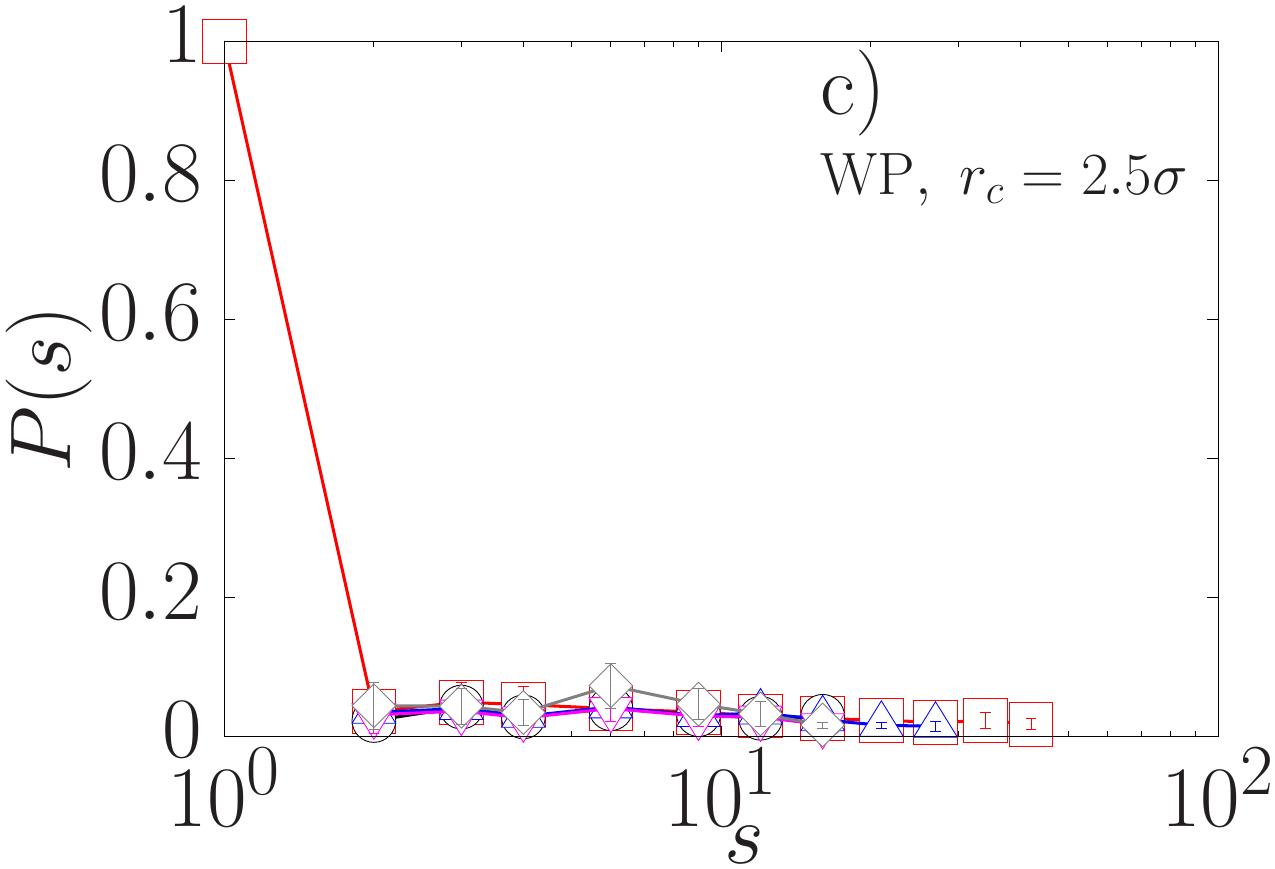}\includegraphics[width=0.48\columnwidth]{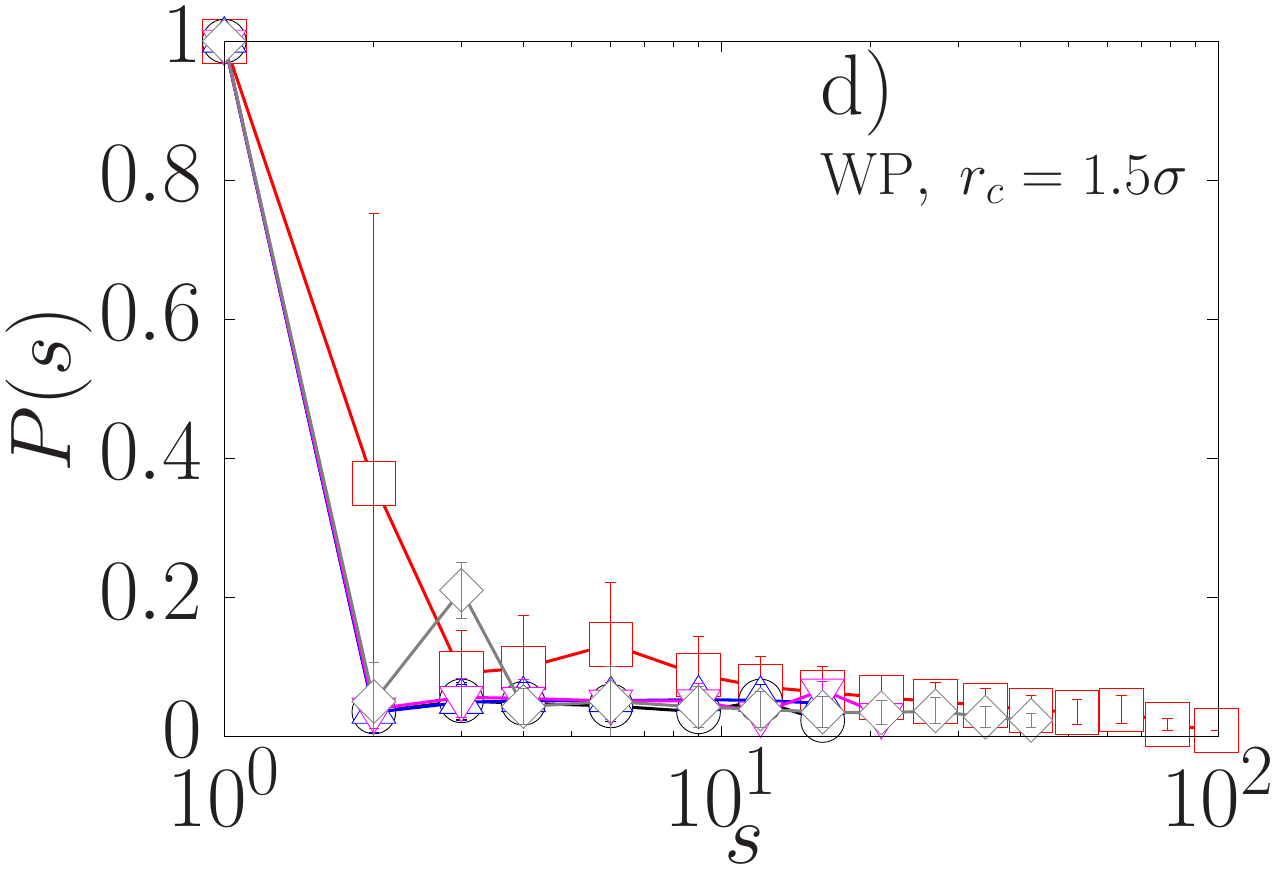}
\caption{\label{fig:Ps_xi1} Polar  order as a function of cluster size for $\xi=1$. Top panels: AP squirmers with  (a)  $r_c=2.5\sigma$ and (b) $r_c=1.5\sigma$, whereas bottom panels are WP squirmers with (c) $r_c=2.5\sigma$ and (d) $r_c=1.5\sigma$.  }
\end{figure}

We now analyze the degree of alignment as a function of the cluster size, computing in Fig. \ref{fig:Ps_xi1} 
the polar order $P(s)$ when $\xi=1$ using  Eq. \ref{eq:PolarOrderCluster}. In general, the polar order inside a cluster usually  decreases with cluster size as a power law~\cite{SoftMatter17}.

As reported in Fig. \ref{fig:Ps_xi1}-(a), increasing $|\beta|$ leads to a faster decay of the local polar order 
for AP squirmers with $\xi=1$ and  $r_c=2.5 \sigma$. 
When decreasing the interaction range,  
we observe a stronger dependence of the polar order with the  hydrodynamic signature: 
{clusters  of neutral squirmers (blue triangles in Fig.~\ref{fig:Ps_xi1}-(b)) are practically swimming in the same direction, while clusters of weak-pullers (black circles in Fig.~\ref{fig:Ps_xi1}-(b)) show different behavior depending on their size: for small and medium size (up to $100$ particles) the polar order decays with size monotonically; the polar order has a crossover for larger clusters, decreasing faster with the cluster size. In general,} weak pullers present higher degree of polar order than pushers and  
 strong pullers behave in the same way as  the corresponding pushers, 
(with  $P(s)$ decaying with the cluster-size monotonically). 

However, the results of the local polar order for WP clusters are quite different. Given that  squirmers form elongated clusters from the assembly of 
 trimers,  the local order rapidly vanishes as the cluster size grows
 (Fig. \ref{fig:Ps_xi1}-(c)). 
This tendency is less marked for short range 
  anisotropic interactions and large $\beta$  (strong pullers in Fig. \ref{fig:Ps_xi1}-(d) (red squares)).

\begin{figure}[h!]
\centering
\includegraphics[width=0.48\columnwidth]{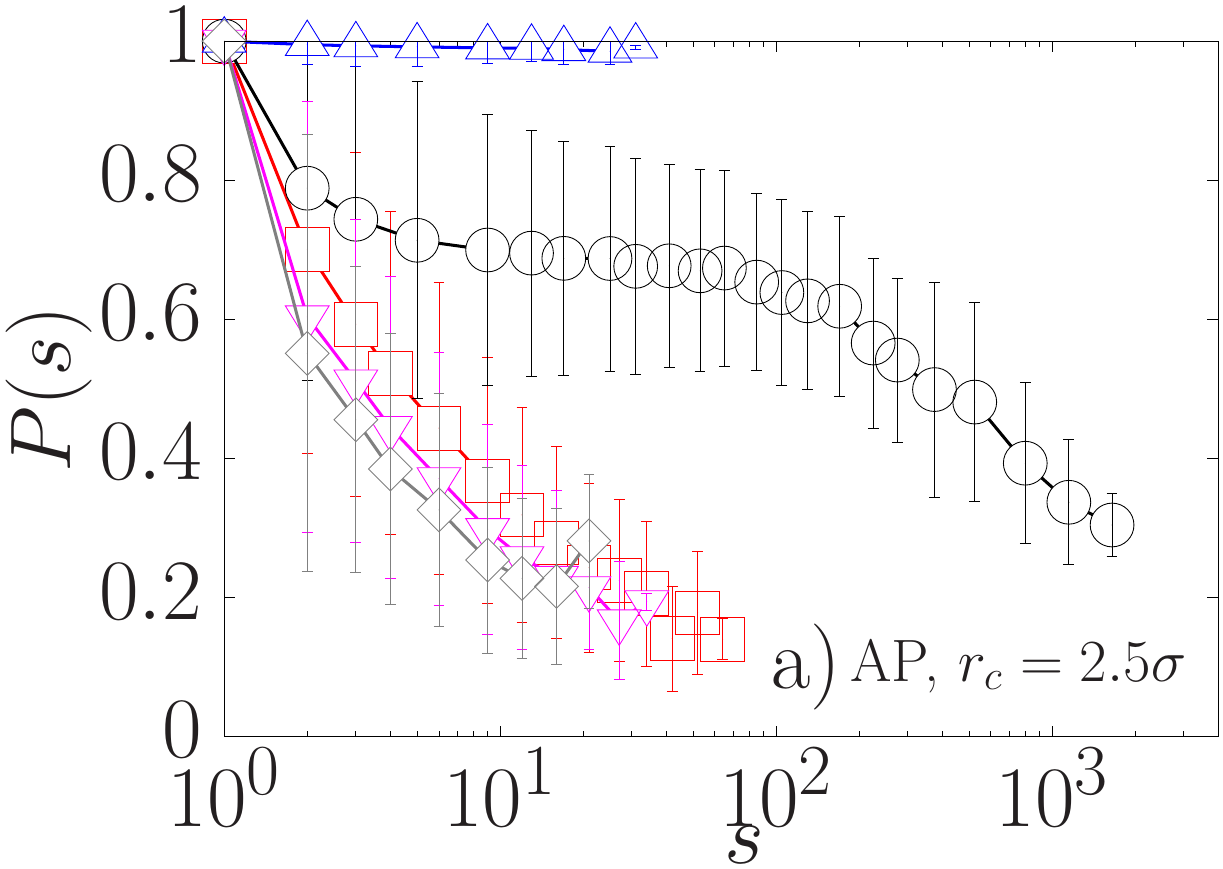}\includegraphics[width=0.48\columnwidth]{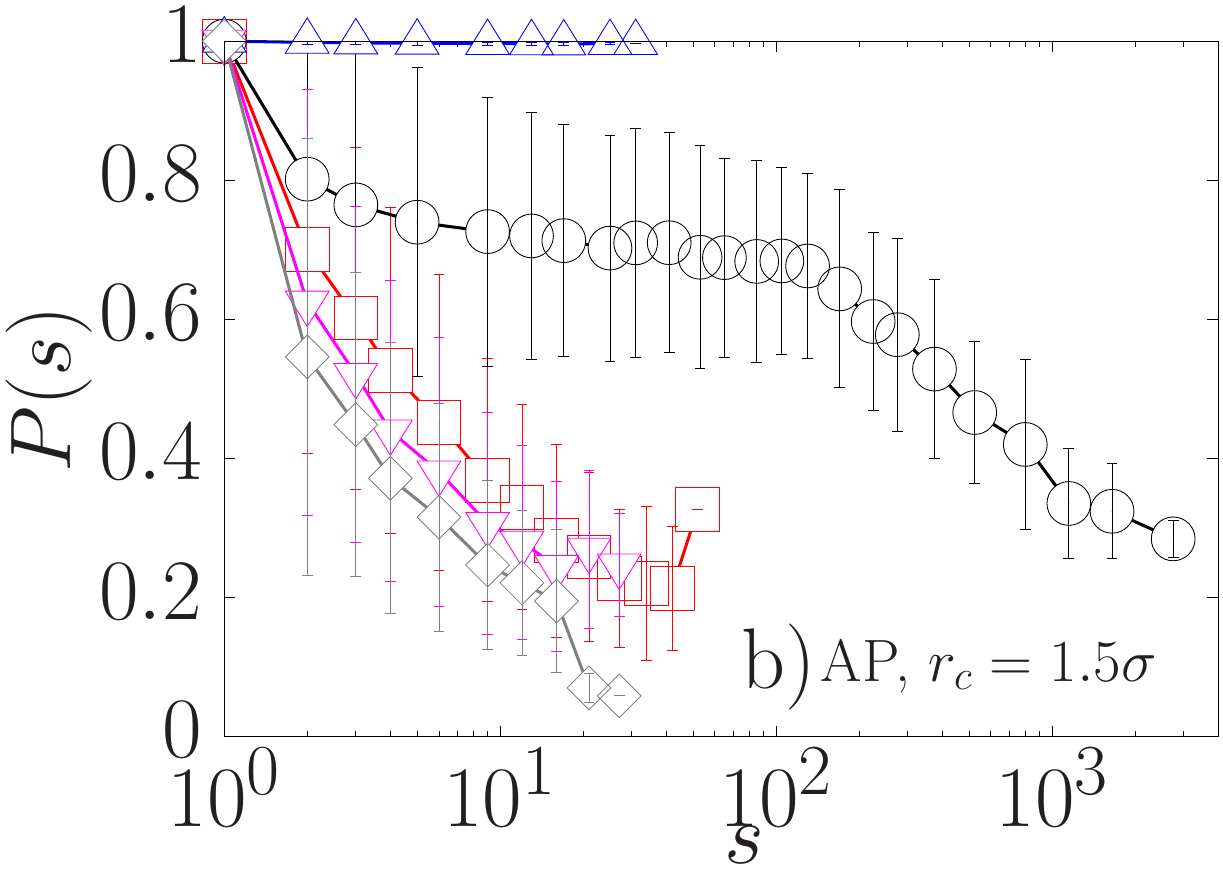}
\includegraphics[width=0.48\columnwidth]{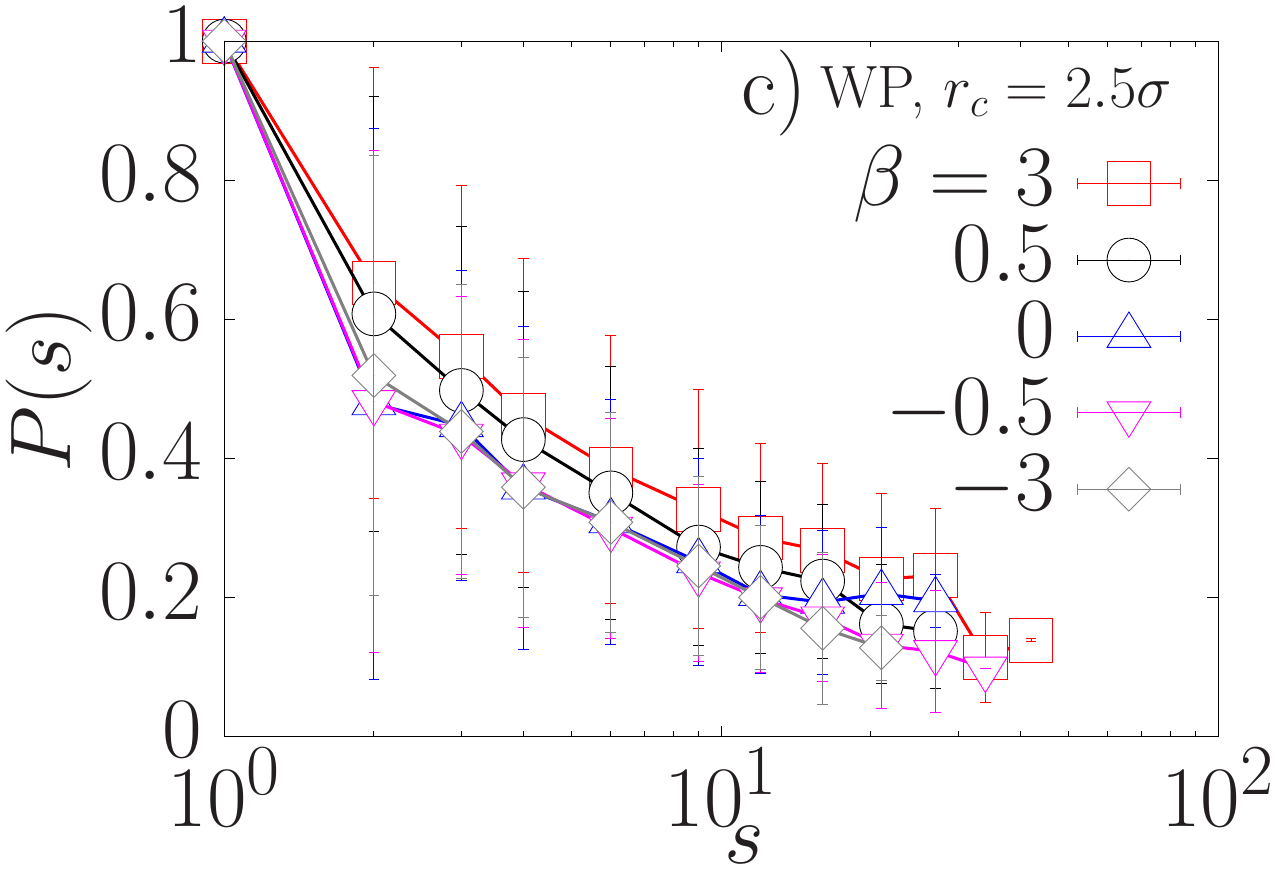}\includegraphics[width=0.48\columnwidth]{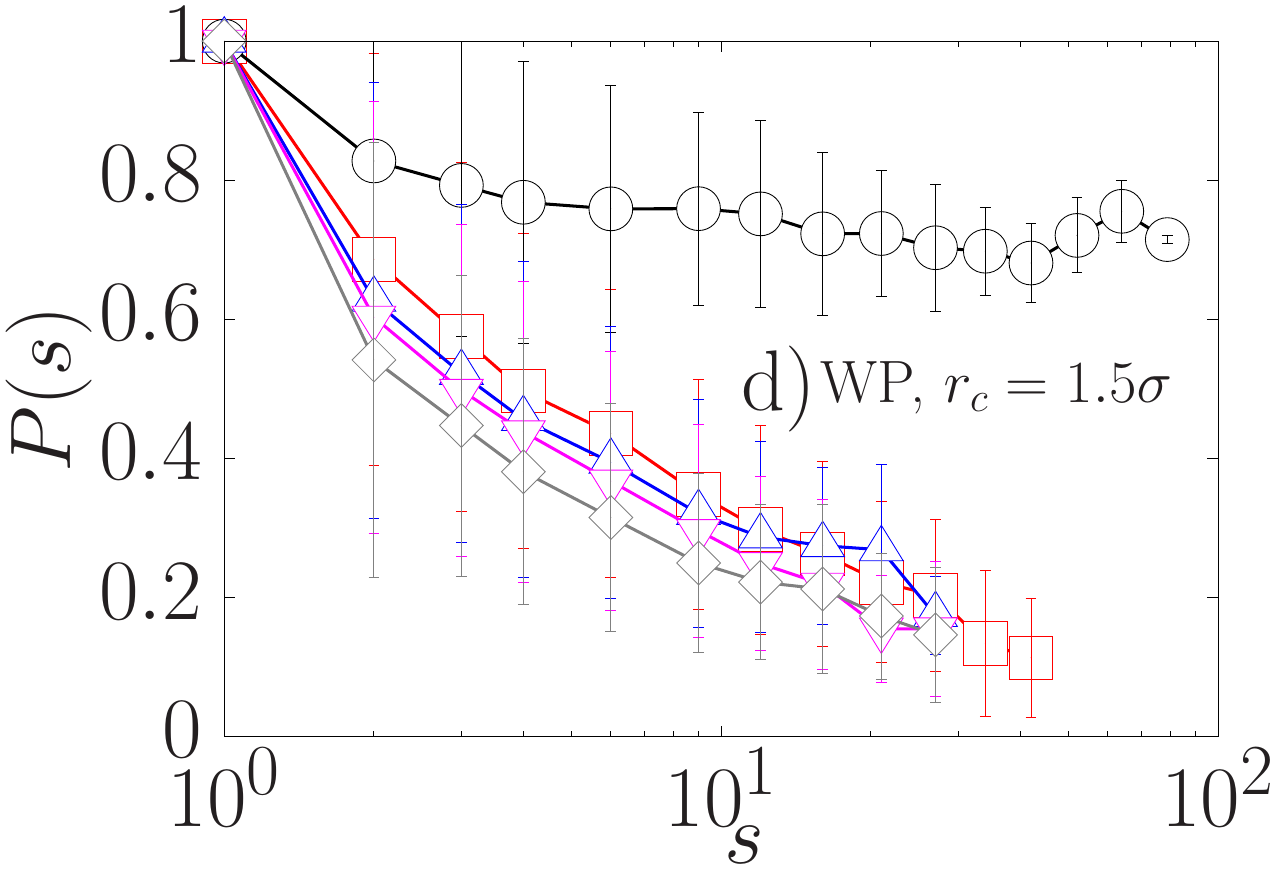}
\caption{\label{fig:Ps_xi10}Polar  order as a function of cluster size for $\xi=10$. Top panels: AP squirmers with  (a)  $r_c=2.5\sigma$ and (b) $r_c=1.5\sigma$, whereas bottom panels are WP squirmers with (c) $r_c=2.5\sigma$ and (d) $r_c=1.5\sigma$. }
\end{figure}

In Fig. \ref{fig:Ps_xi10} we report  the polar order as a function of the cluster size for both AP and WP squirmers when  
self-propulsion dominates over attraction  ($\xi=10$). 

For AP squirmers, $P(s)$ decays faster for pushers than for pullers, 
independently on the  interaction range (panels a and b of Fig. \ref{fig:Ps_xi10}). 
The degree of polar order is more 
sensitive to $\beta$ for pullers than for pushers; weak puller's clusters are more aligned than  clusters of stronger pullers, while 
neutral squirmer's clusters are the most aligned ones for all  cluster sizes. 
Panels c and d of Fig. \ref{fig:Ps_xi10} report the polar order of clusters of   WP squirmers, with a monotonous 
decay of the polar order with cluster size, weakly dependent on $\beta$ and on  the interaction 
range (except for $\beta=0.5$). 

\begin{figure}[h!]
\centering
\includegraphics[width=0.48\columnwidth]{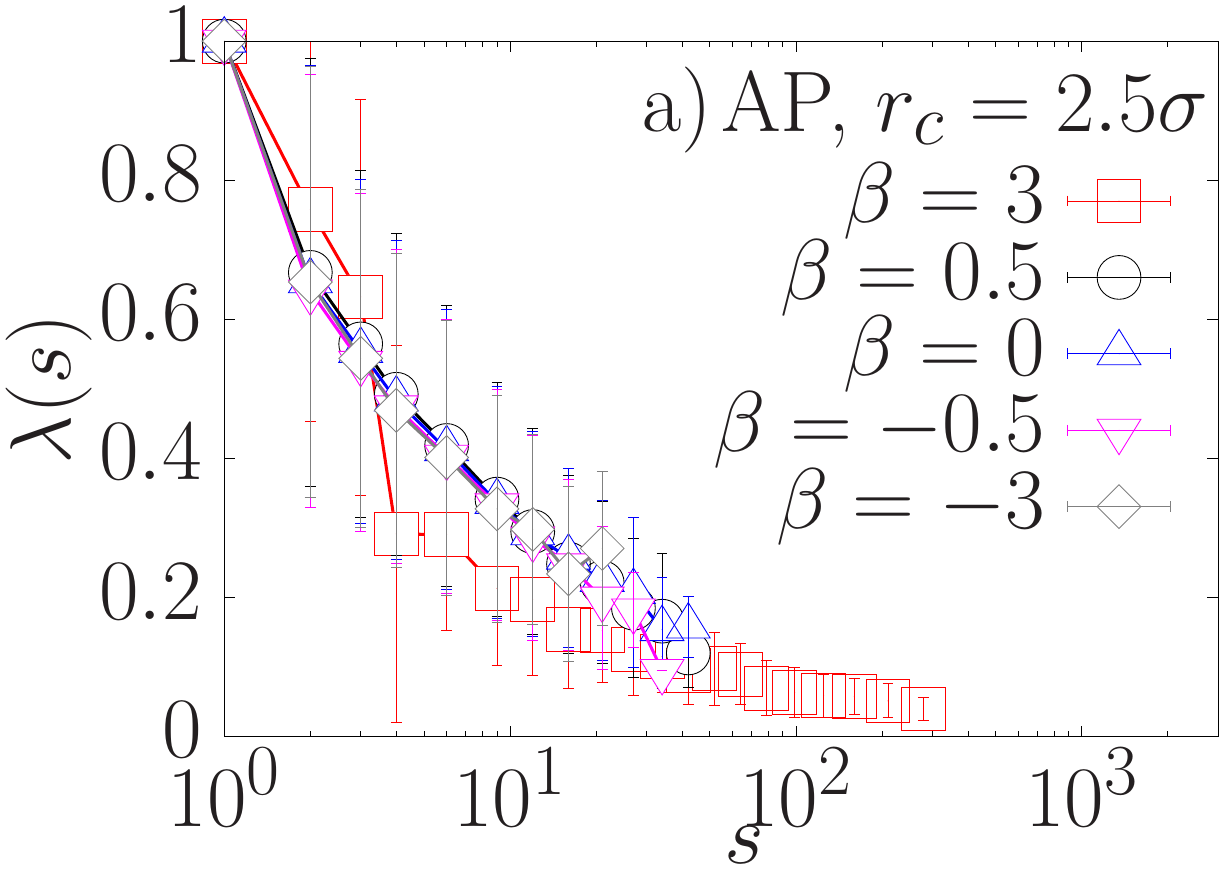}\includegraphics[width=0.48\columnwidth]{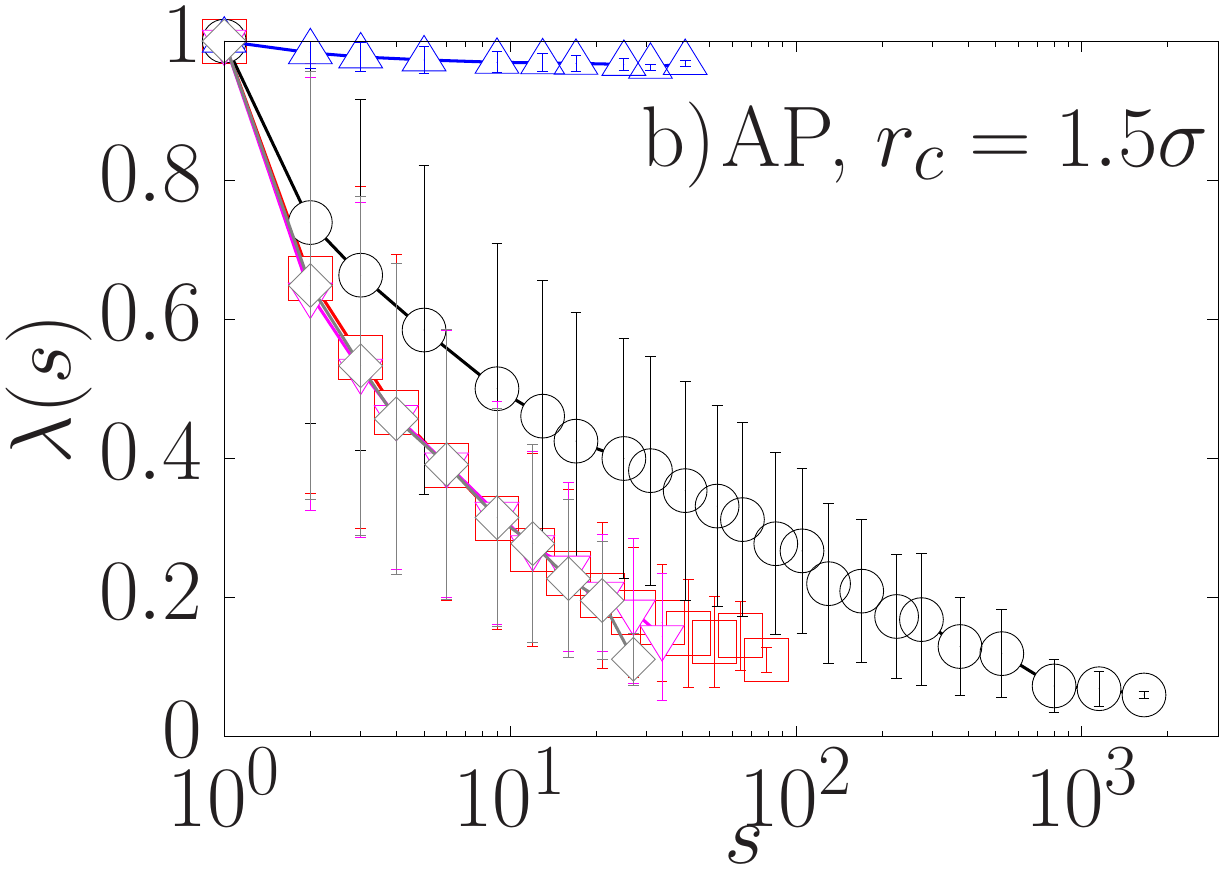}
\includegraphics[width=0.48\columnwidth]{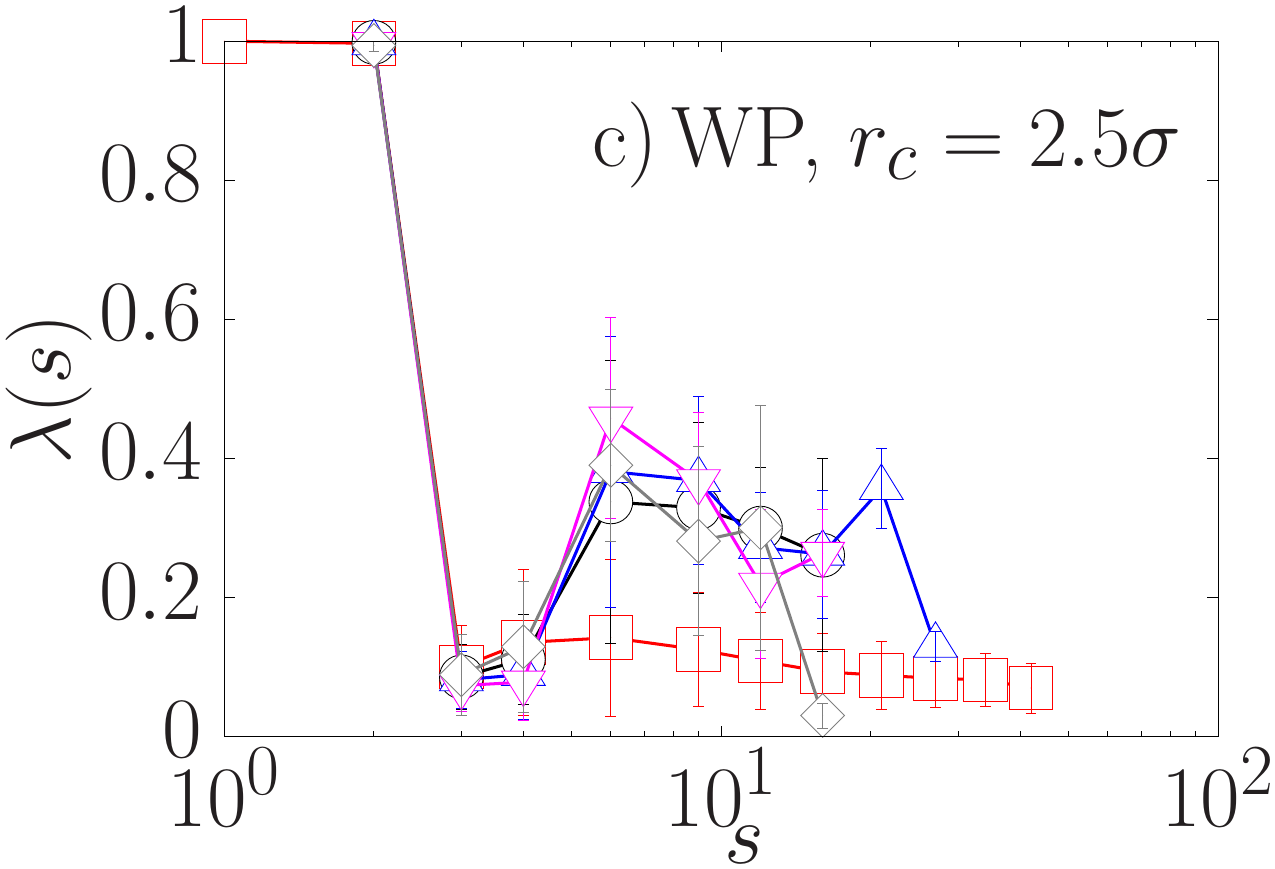}\includegraphics[width=0.48\columnwidth]{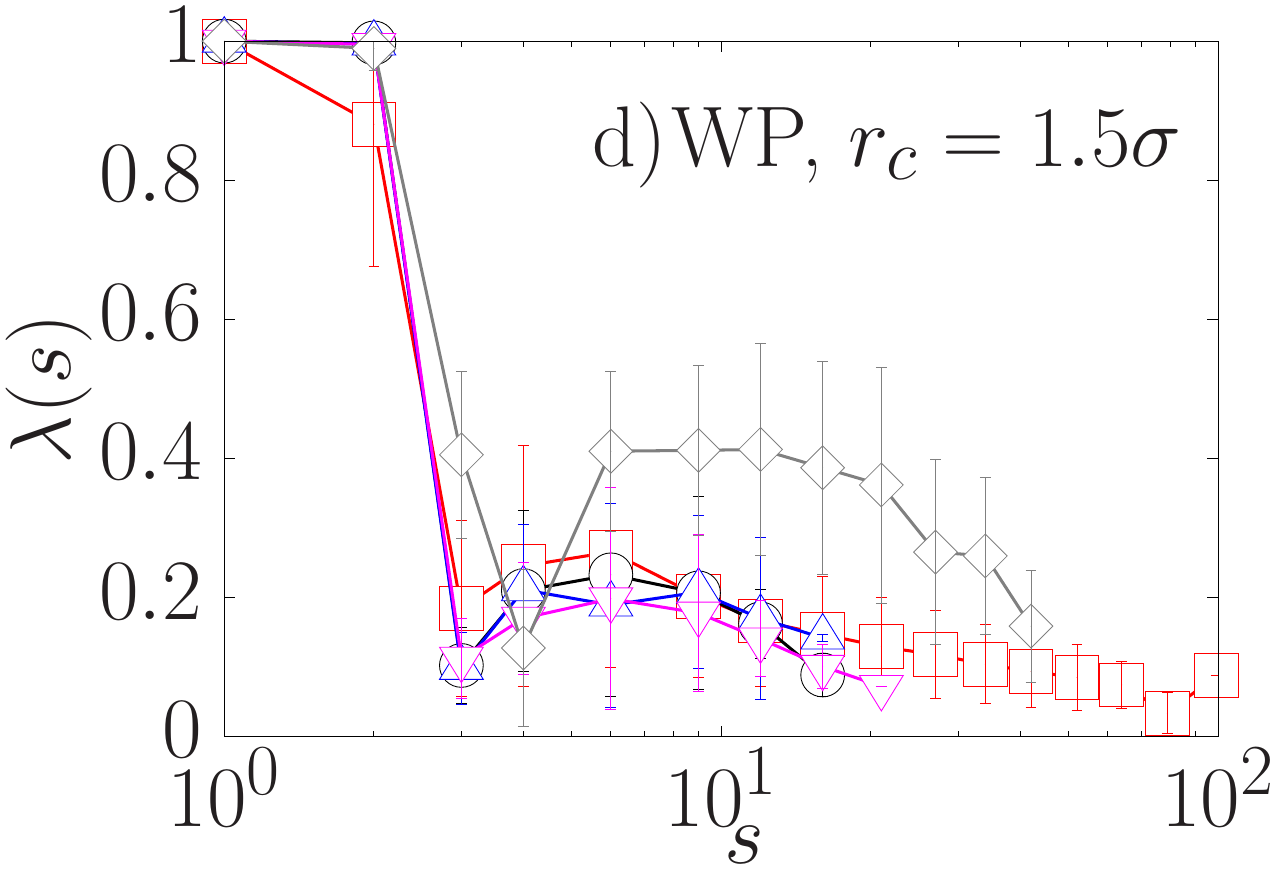}
\caption{\label{fig:Ns_xi1}  Nematic  order as a function of cluster size when $\xi=1$. Top panels represent results for AP squirmers:   (a) with $r_c=2.5\sigma$ and (b) $r_c=1.5\sigma$, whereas bottom panels are WP squirmers (c) with $r_c=2.5\sigma$ and (d) $r_c=1.5\sigma$.  }
\end{figure}

In Fig.~\ref{fig:Ns_xi1} we report results for  the nematic order $\lambda(s)$. 
 The local nematic order  for AP squirmers with $\xi=1$ decays with the cluster size $s$ 
independently on $\beta$ for $\beta < 3$ while presents  an even faster decay when $\beta=3$ (see Fig. \ref{fig:Ns_xi1}-(a)).
When  $r_c=1.5\sigma$, $\lambda(s)$ decays with $s$ in the same way as $P(s)$ does for all $\beta$.
Neutral squirmers are completely polarized (as also shown in the nematic order, blue triangles in Fig. \ref{fig:Ns_xi1}-(b)).
The high polar order observed in clusters of puller and neutral squirmers is also detected by the nematic order.

The behaviour of the local nematic order for WP clusters with  $\xi=1$ is quite striking.
The reason why $\lambda(1)=\lambda(2)=1$ is that 
when the interaction range is $2.5\sigma$, once two particles collide  they are kept together by the attractive patch. 
 Particles in  trimers and tetramers point towards 
 the center of the cluster and are kept together by the attractive patch,  reason why   they have a small nematic order (all curves
  in Fig. \ref{fig:Ns_xi1}-(c)). 
  For larger clusters, active stresses become more important: 
  hydrodynamics attract trimers so that  particles'  attractive patches   reorient in the 
  clusters forming aligned chains. This does not happen 
    for strong pullers, where   the nematic order 
  remains low and constant (red squares in Fig. \ref{fig:Ns_xi1}-(c)),  due to the fact that trimers form disordered chains 
   (made of trimers formed by  hydrodynamics). 
    When the interaction range is $1.5 \sigma$, the local nematic order behaves in the same 
  way as for $2.5 \sigma$, but in this case oriented chains are observed only for strong pushers, 
  whereas  the rest of{ squirmers form less oriented chains with a value of $\lambda(s)$ in between.}

\begin{figure}[h!]
\centering
\includegraphics[width=0.48\columnwidth]{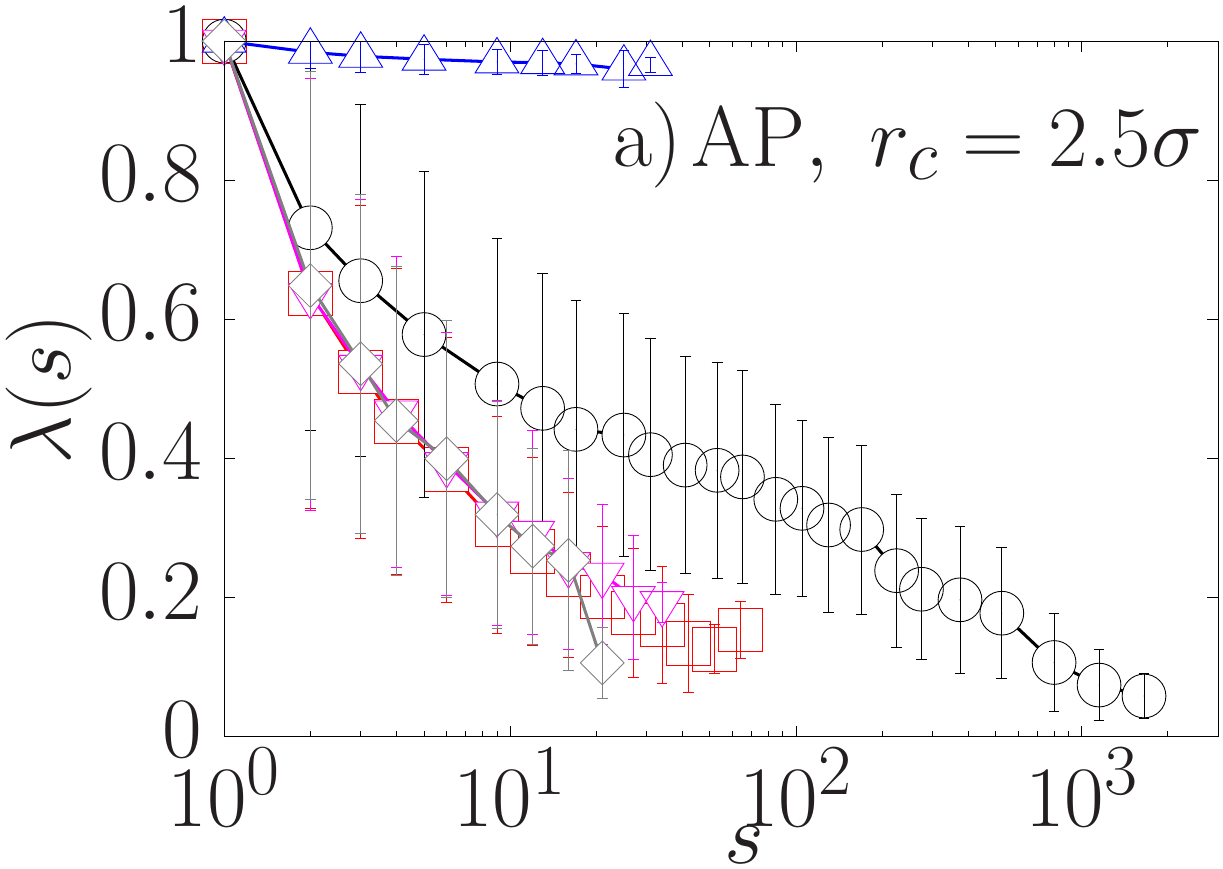}\includegraphics[width=0.48\columnwidth]{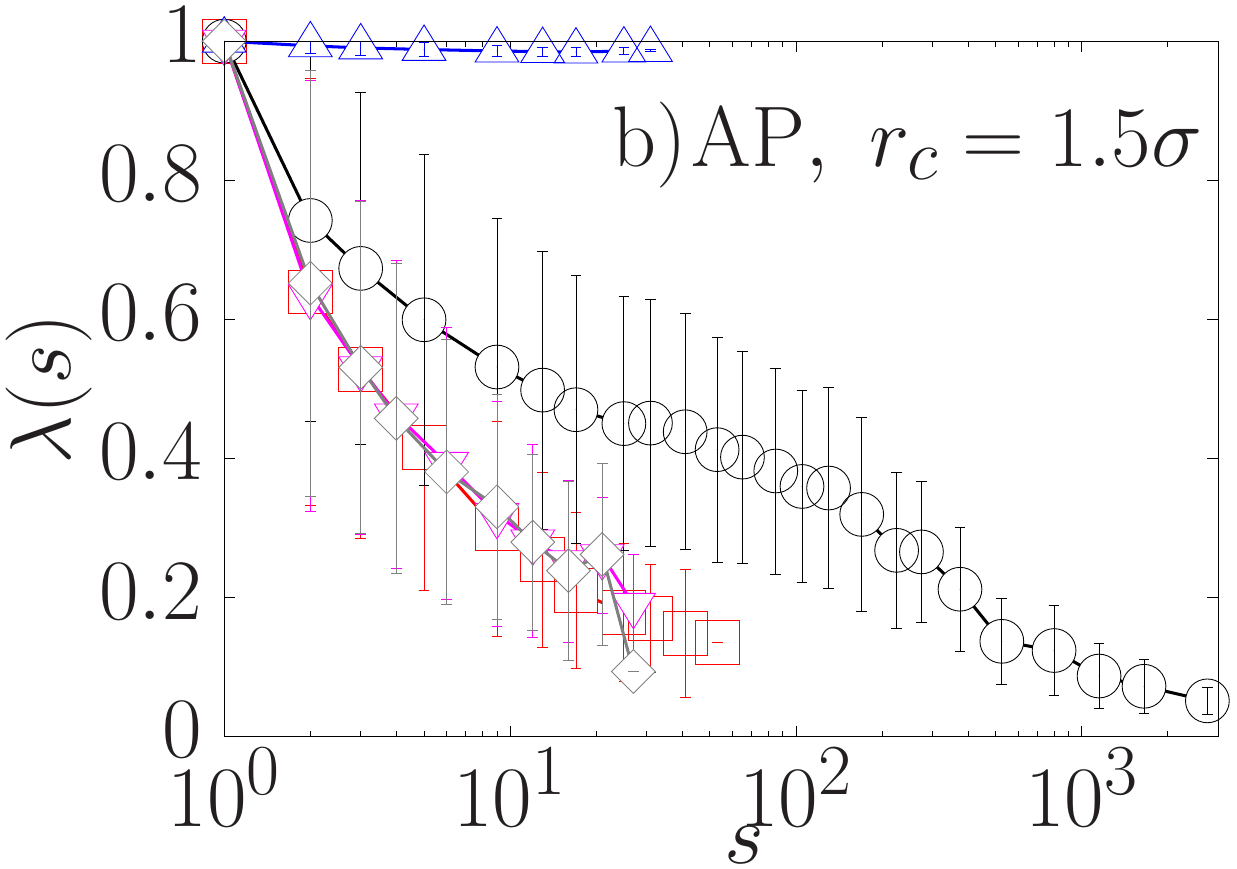}
\includegraphics[width=0.48\columnwidth]{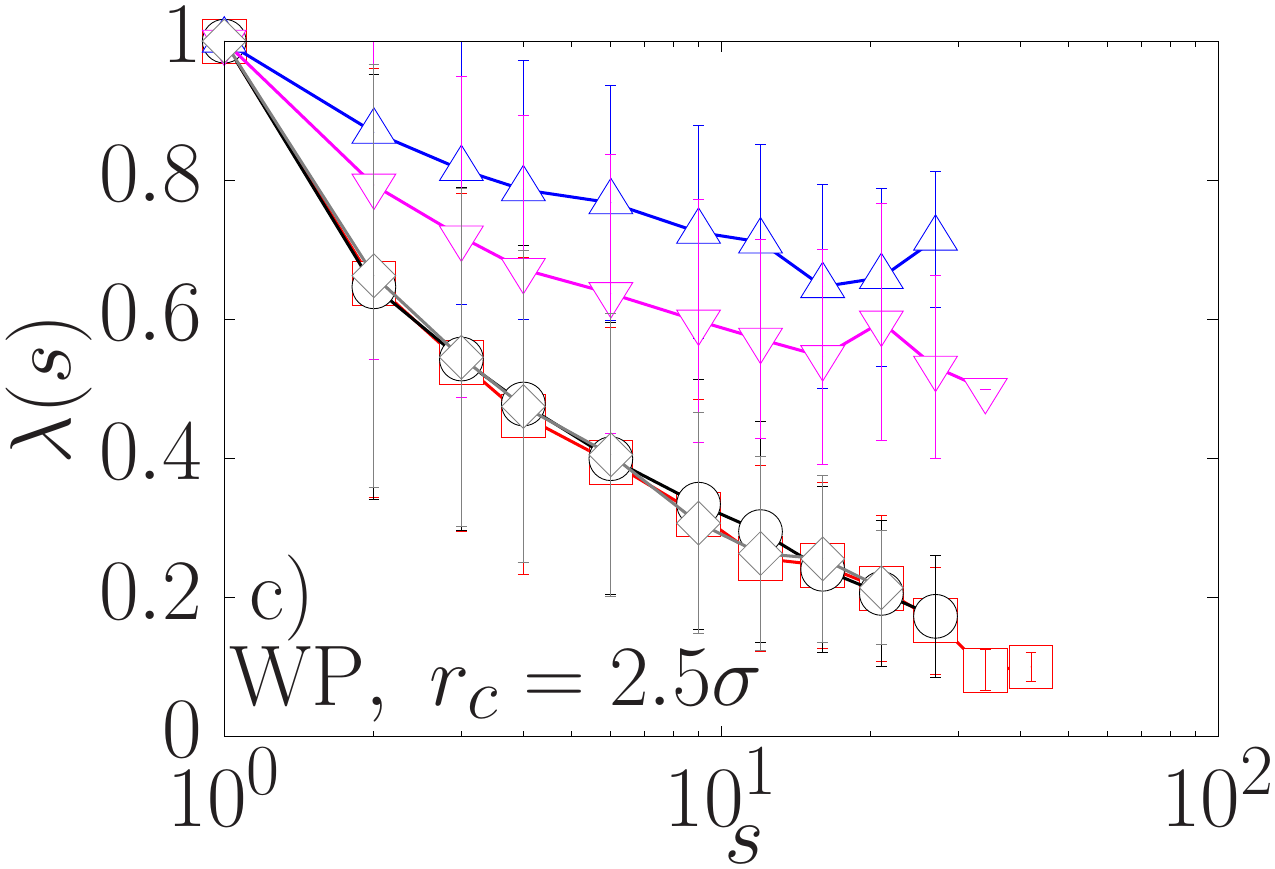}\includegraphics[width=0.48\columnwidth]{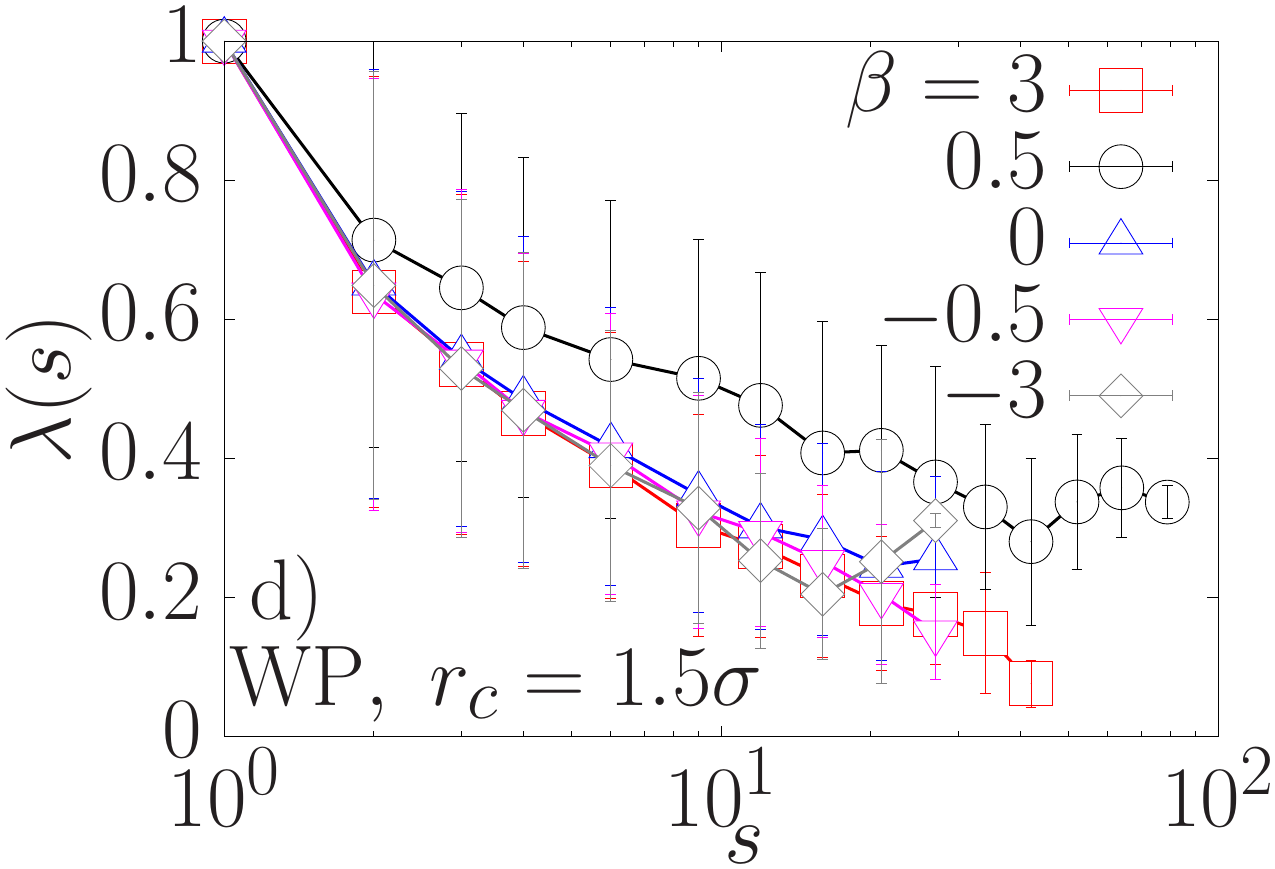}
\caption{\label{fig:Ns_xi10}Nematic order as a function of cluster size for suspensions with $\xi=10$. \textbf{(a):} AP squirmers with $r_c=2.5\sigma$. \textbf{(b):} AP squirmers with $r_c=1.5\sigma$. \textbf{(c):} WP squirmers with $r_c=2.5\sigma$. \textbf{(d):} WP squirmers $r_c=1.5\sigma$.}
\end{figure}

In Fig. \ref{fig:Ns_xi10} we report the local nematic order
 $\lambda(s)$ for  $\xi=10$ and AP squirmers  panels a and b.
 In clusters made of neutral squirmers (blue triangles), 
  particles swim in the same direction independently on the interaction range: 
  $\lambda(s) \approx 1$ for all cluster sizes. 
The nematic order for all other  squirmers 
 decays in the same way as the polar order. 
To conclude, in the case of clusters of AP squirmers,  alignment is not affected by the interaction range.

WP clusters with $\xi=10$ and long range interactions develop a high nematic order for neutral and weak pushers 
($\beta=0$ and $-0.5$) as  shown in panel c of Fig. \ref{fig:Ns_xi10}.
This nematic order is not due to a global polar order, 
(being  very low) but to the competition between the anisotropic potential and the hydrodynamic 
signature: even though activity dominates over attraction,  the  interaction range is long enough to allow for the development of  a nematic order in the system. 
Undoubtedly, the interaction range is important to develop clusters with nematic order, since $\lambda(s)$ for WP squirmers with 
short range {has a different behavior, where the nematic order only shows a non-zero value due to the polar order shown before.
}

\subsection{Global polar  and nematic order}
\begin{figure}[h!]
\centering
\includegraphics[width=0.48\columnwidth]{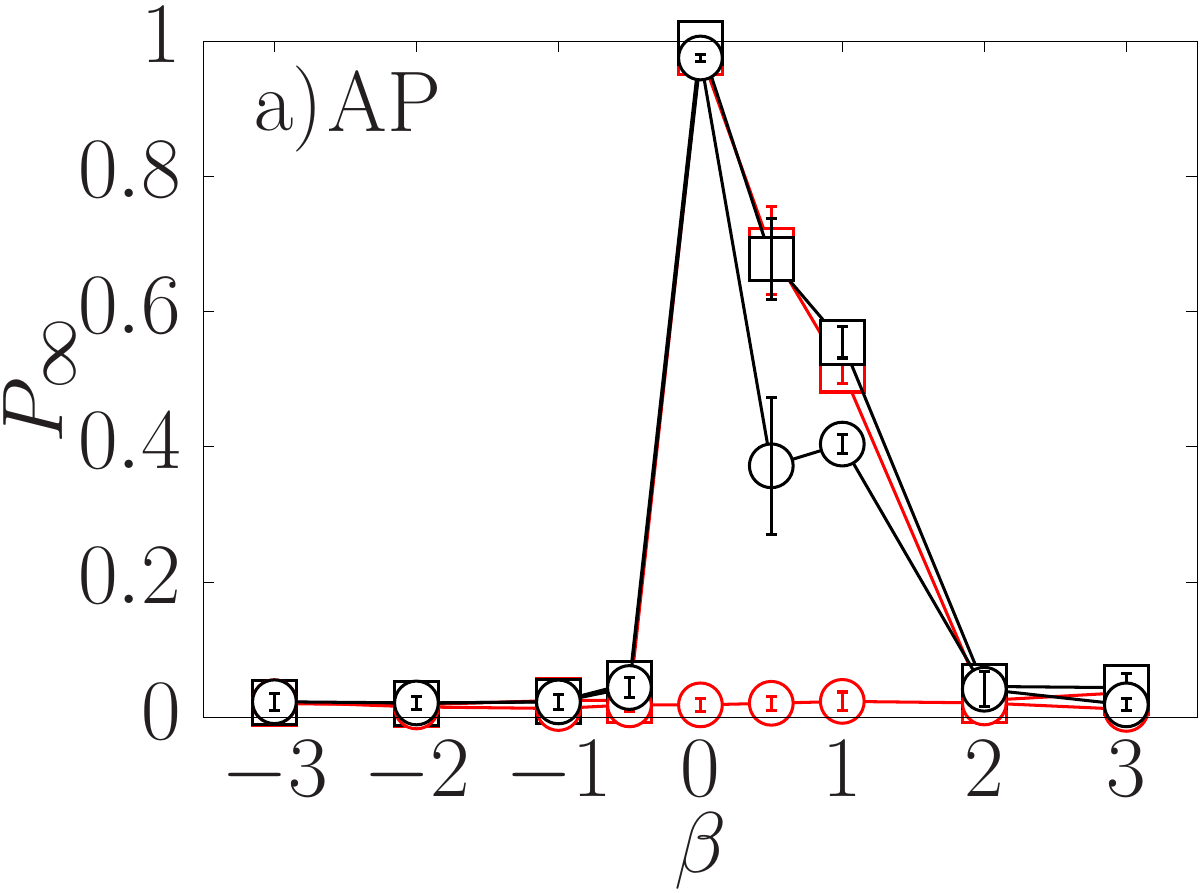}~\includegraphics[width=0.48\columnwidth]{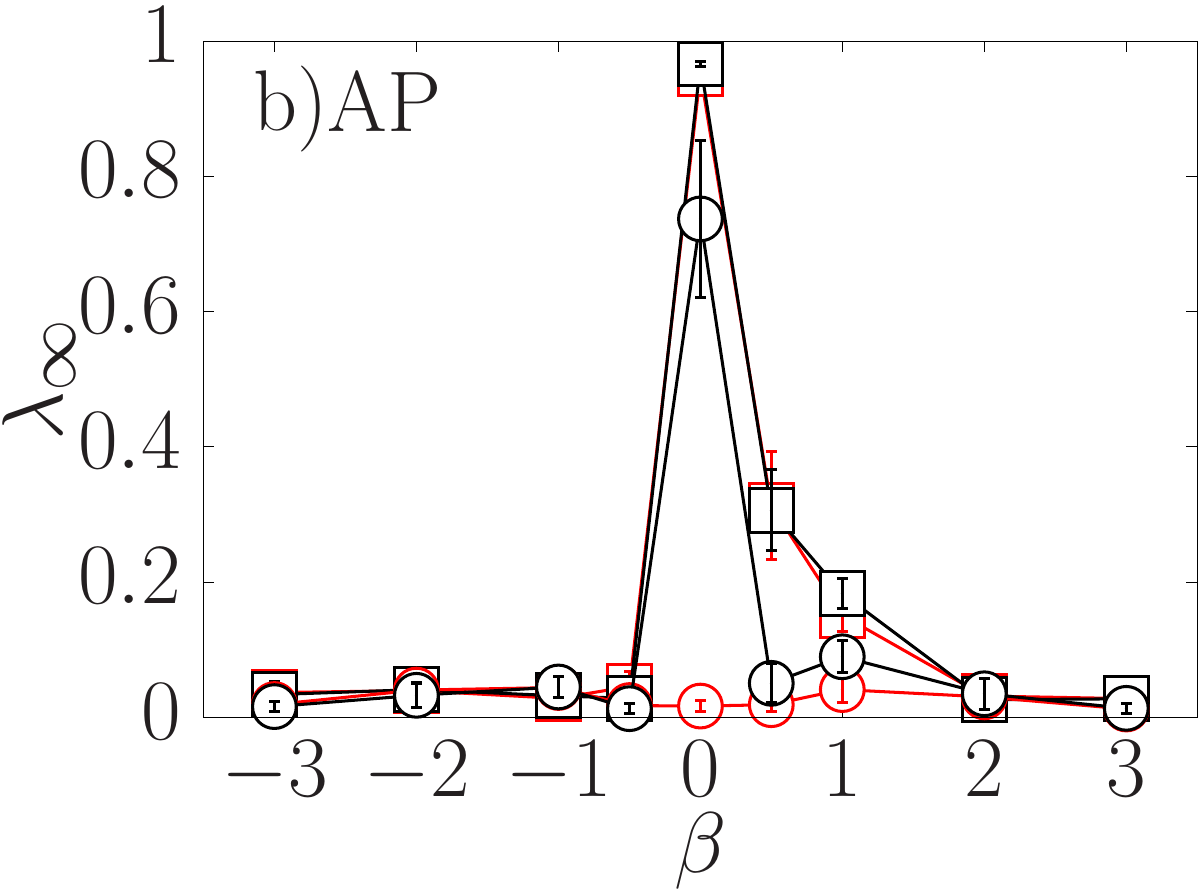}
\includegraphics[width=0.48\columnwidth]{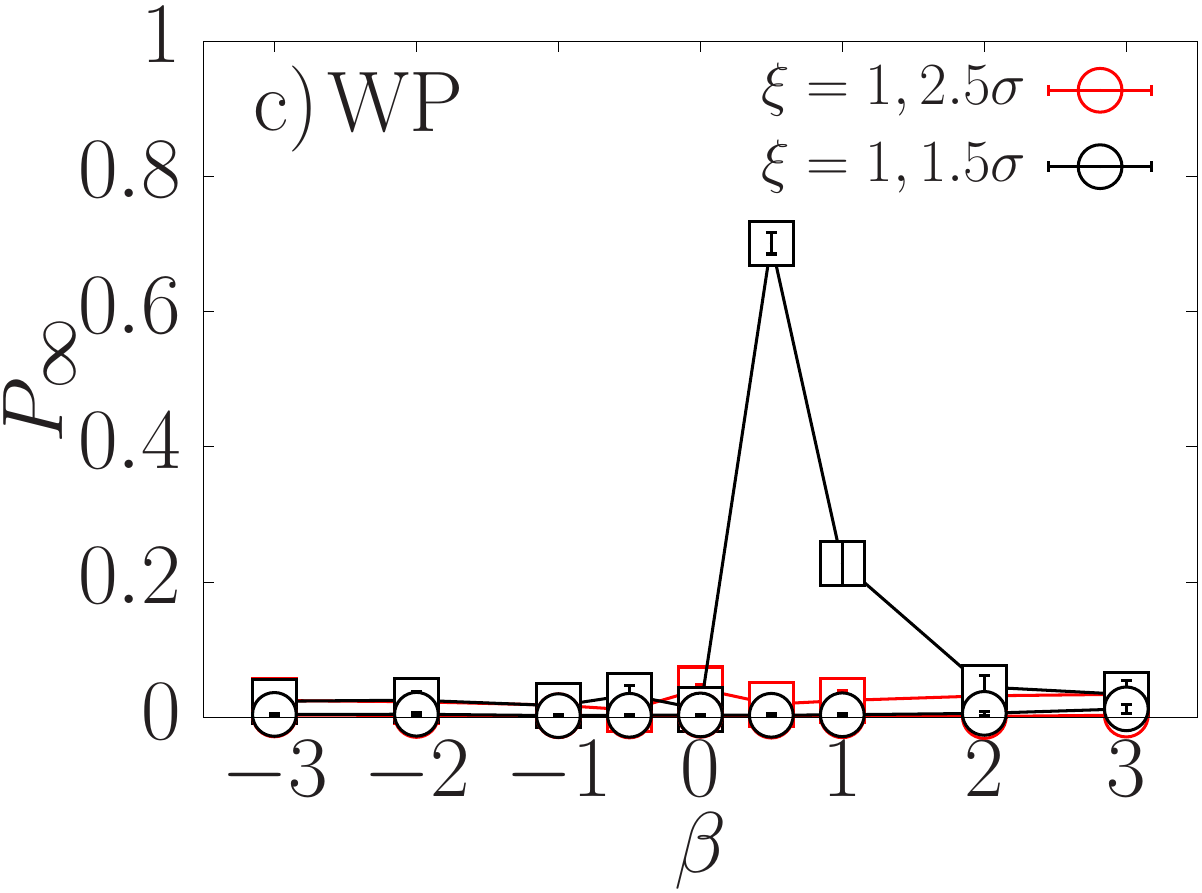}~\includegraphics[width=0.48\columnwidth]{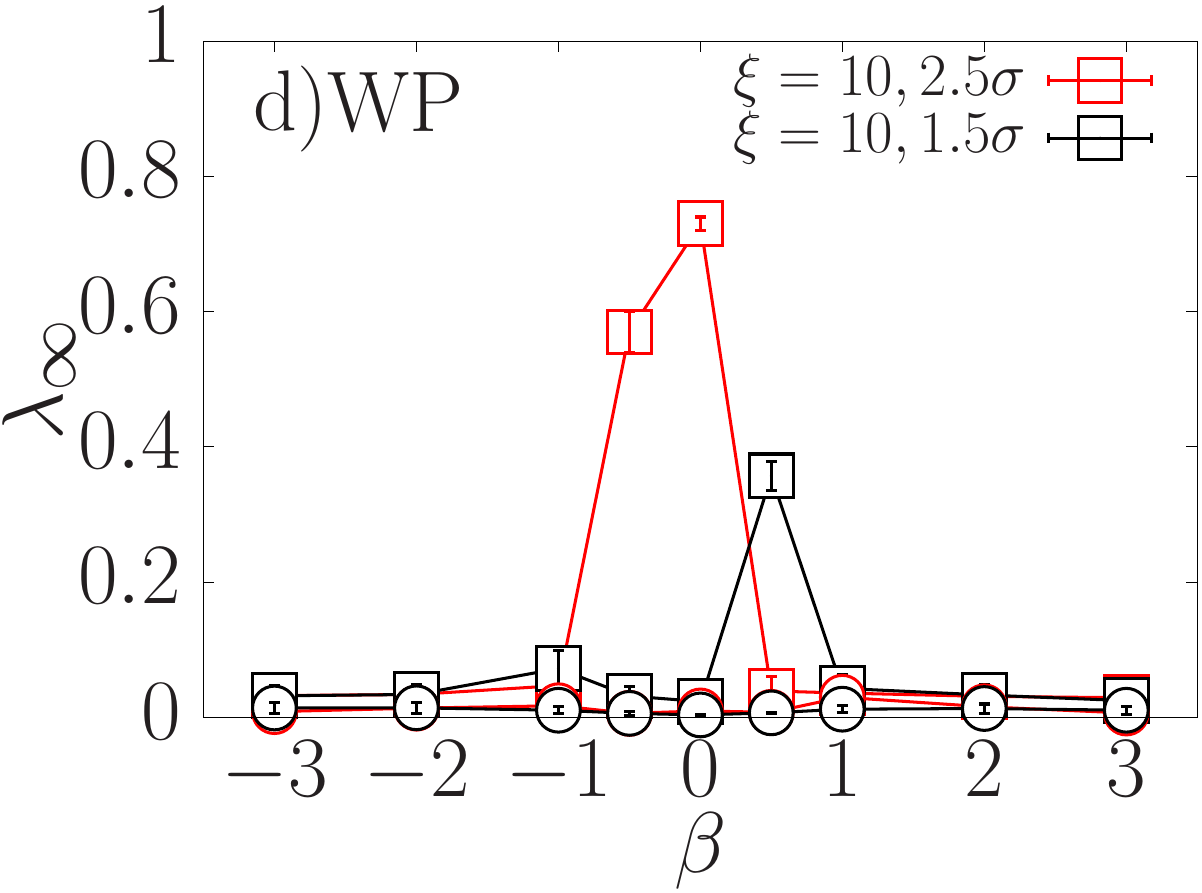}
\caption{\label{fig:PolarNematicGlobal} Global polar $P_{\infty}$ and nematic $\lambda_{\infty}$ order parameters. Top panels represent results for AP squirmers: (a) polar order parameter $P_{\infty}$, (b)  nematic order parameter $\lambda_{\infty}$. Bottom panels represent results for WP squirmers: (c) polar order parameter $P_{\infty}$, (d)  nematic order $\lambda_{\infty}$.}
\end{figure}

 To study the orientational order of the system, we compute  as in \cite{alarcon2013} the global polar order parameter
and  the nematic order tensor Eq. \ref{eq:lmbdaMtrx} \cite{Barci_2DNematic}, at    steady-state at long times.  
%
In Fig. \ref{fig:PolarNematicGlobal}-(a)  we represent the global polar order $P_{\infty}$ for AP squirmers in steady state
for  $\xi=1$ and  $\xi=10$ and both interaction ranges. 

When $\xi=10$,  activity dominates and the polar order parameter has a non-zero value for weak pullers ($0 \ge \beta \ge 1$), similarly to the 3D case 
without attractive interactions \cite{alarcon2013}. 
When $\xi=1$, the range of 
interaction affects to the formation of aligned states:  
for  $2.5 \sigma$ the system is completely isotropic despite of the value of $\beta$ (red circles in Fig. \ref{fig:PolarNematicGlobal}-(a)), whereas for $r_c=1.5 \sigma$ 
the system still has polar order for weak pullers (black circles), as   with $\xi=10$.
The Fig. \ref{fig:PolarNematicGlobal}-(b), shows that the nematic order $\lambda_{\infty}$  follows  the polar order: for AP squirmers there is not an additional orientation 
besides the polar order. 

In Fig. \ref{fig:PolarNematicGlobal}-(c)
we represent the global polar order $P_{\infty}$ for WP squirmers in steady state
for  $\xi=1$ and  $\xi=10$ and both interaction ranges. 
For strongly anisotropic interactions ($\xi=1$) the 
suspension does not develop any significant  degree of polar ordering. For weakly interacting WP squirmers ($\xi=10$) and $r_c=1.5 \sigma$, 
the polar order emerges in the same range of $\beta$'s (black squares) as the AP squirmers.
In Fig. \ref{fig:PolarNematicGlobal}-(d) we report the nematic order parameter $\lambda_{\infty}$.  
When $\xi=1$ WP squirmers do not develop any nematic 
order (red and black circles): either polar or nematic order are absent for WP squirmers with strongly anisotropic interactions. 
A non-zero nematic order is observed for $\xi=10$, when $r_c=2.5 \sigma$ and $r_c = 1.5 \sigma$: in the latter,  the nematic order 
 is due to the polar order (black squares), while in the former mid-range interactions create a new orientational order (red squares). 
Interestingly, the nematic order parameter for WP squirmers with weak interactions shows a non-zero nematic order  when the polar order parameter is zero.

\section{\label{sec:Conclusions} Discussion and Conclusions}
In this paper, we  report a numerical study of the self-assembly  of Janus squirmers 
 in a quasi two dimensional  semi-dilute  suspension.  
 
{Even though we are aware of the fact that experimental catalytic active colloids might present other types of interactions, like phoretic ones, the nature of such interactions is going to depend on the particular system under study \cite{JCP_Benno2019}. As in the case of active dipolar Janus particles \cite{Dipolar_Granick} where electrophoresis is effectivelly modelled by the interaction between two imbalanced charges on each particle, in our work we suggest that amphiphilic active Janus colloids \cite{Gao_OTSJanusExp} can be modelled by means of an effective anisotropic interaction. For this 2D system, we study the effect of the hydrodynamic interactions in the self-assembly of such particles, hopefully shedding some light on experiments with active Janus colloids where the fluid flow field can be measured \cite{Campbell_FlowFieldJanus} or in experiments where the self-propulsion direction can be tuned \cite{Brown_Poon_SaltJanus}.}
 
In a previous work, we have studied aggregation in  a semi-dilute quasi two dimensional 
suspension of isotropically  attractive squirmers \cite{SoftMatter17}, demonstrating that 
both clusters' morphology and  alignment within clusters  depended on  particles' 
 interaction strength  and  hydrodynamic signature.  
 In the current work,  by changing the symmetry of the inter-particle interaction 
 (from isotropic to anisotropic) 
  we have observed that the interaction anisotropy, {modulated by its range and strength}, compete with the active stress {to re-orient the particles and therefore,} giving rise to a  rich aggregation phenomenology. 
  
  {The study of two different interaction ranges was originated from previous results \cite{SoftMatter17, SoftMatter_43, SoftMatter_26, SoftMatter_27,Fillion_AABP_3D} where $r_c=2.5 \sigma$ has been used for isotropic potentials, while in \cite{Stewart_17, pre09, Pu_ActiveJanus2017} an interaction range of $r_c=1.5 \sigma$ is used for anisotropic potentials. Both cutoff values are typical in the context of attractive Brownian colloids either passive or active.}

 As shown in all studied cases, 
 the amount of activity versus attraction has a strong effect on the aggregation dynamics, affecting not only the 
  characteristic cluster size $\langle  s \rangle$, but also 
  the time scale at which the steady state (if any)
is reached. 

{When the attractive interaction dominates over propulsion, e.g. at $\xi=0.1$, AP squirmers coarsen and the velocity of coarsening will depend on the hydrodynamic stresses for both interaction ranges, similarly to the isotropic case with lower interaction strength ($\xi=0.7$)~\cite{SoftMatter17}, where
the system coarsens for pullers and weak pullers and form clusters for strong pushers. As opposed to AP, WP squirmers develop more strinking effects, for long range interactions squirmers coarsen very slowly as far as our calculation time is concerned.
 While for short range, WP squirmers have reached a clustering steady state, where clusters are self-assembly from small chains or trimers. WP cases at $\xi=0.1$ are explained deeply in appendix \ref{Appendix:xi0_1WPrc1_5}.
}


{
AP squirmers display a behavior closer to that of  squirmers interacting with an isotropic potential when interaction competes with activity and when activity dominates over attraction \cite{SoftMatter17}.
{
The competition between interaction and activity always in terms of $\xi$.
}
 For example, the fluctuation of the mean cluster size for weak pullers observed in panel b of Fig. \ref{fig:nct_xi1_0} and panels a and b of Fig. \ref{fig:nct_xi10} or the power law behavior of the CSD for weak pullers in panel b of Fig. \ref{fig:csd_xi1} and panels a and b of Fig. \ref{fig:csd_xi10}.
In contrast, WP squirmers have a stronger tendency to  develop novel collective behavior,   from polydisperse suspension of monomers, dimers, trimers and even chains of tens of particles.
}
{Our interpretation here is that, since two AP squirmers interact  when their attractive patches are within their range, this effect is reduced due to their propulsion in opposite directions. 
Thus the contribution to the torque is mostly given by hydrodynamic stresses rather than by the anisotropic interactions, as in the isotropic case.}

In order to perform a  statistical study of the clusters, we  focus on  $\xi = 1$ and $\xi=10$ when the system is in  steady state. 
We first compute the cluster size distribution (CSD). 
 On the one side,   CSD for AP squirmers can be described by a power law with an exponential tail.
  When $\xi=1$, CSDs are characterized by a  cutoff-algebraic shape except for strong pullers, which develop a bimodal distribution
 not present for short interaction range ($1.5 \sigma$).  In this case, the CSD's width follows a  behaviour reminding that of  isotropic 
 squirmers  with $\xi>1$, i.e. wider CSDs for weak pullers.
  When $\xi=10$,   the effect of the interaction range on the CSDs is 
 not observed  for any value of  $\beta$. 
 On the other side, the CSDs for WP squirmers has a peak for trimers  when $\xi=1$. 
  The shape of the CSDs is approximately the same independently on the attraction range. 
When $\xi=10$, 
  even though the CSD is described by a  power law with an exponential tail, when $r_c=2.5\sigma$ the larger the value of $\beta$ 
  the wider the distribution; whereas  when $r_c=1.5\sigma$ the shape of the CSD is mainly dictated by 
    hydrodynamics, with weak pullers showing a wider distribution. 

 


    

Next, we compute the radius of gyration as a function of the cluster size. 
When  $\xi=1$ and $r_c=2.5\sigma$, the morphology of AP squirmers follows the same behaviour as the one  detected in  the isotropic case \cite{SoftMatter17},  
with more compact clusters for strong pullers; differently from the case when 
 $r_c=1.5\sigma$, where clusters  are more compact  for weak pullers. 
When $\xi=10$ the clusters' $R_g$ does not change, independently on 
 the interaction range or the hydrodynamic signature. 
When $\xi=1$, the $R_g$ for WP squirmers varies  with the clusters sizes since, for both interaction 
ranges,  trimers and tetramers gives rise to the formation of chains, which lead to different exponents. 
 In contrast, when $\xi=10$ and $r_c=2.5 \sigma$, the $R_g$ for WP squirmers  is the same despite $\beta$; however,   
when $r_c=1.5 \sigma$ pullers form more compact clusters.

{It is worth mentioning that these chains resemble the chains observed with amphiphilic active brownian particles \cite{Stewart_17}, but clearly their nature is different since WP squirmers form chains due to their stresslet, while the Brownian case depends on the activity and the size of the hydrophobic patch (see appendix \ref{Appendix:JanusOrient}). Moreover, it seems that hydrodynamic interactions cancel the formation of chains for AP squirmers, since we have observed chains just for WP squirmers.} 

To establish a potential alignment within clusters, we compute both  polar and nematic order.   
When $\xi=1$, the polar order of the clusters for AP squirmers strongly depends  on $r_c$:  when $r_c=2.5\sigma$, $P(s)$  quickly decays 
 with the cluster size (quicker for pushers  than for pullers);  whereas when $r_c=1.5 \sigma$, 
  $P(s)$  corresponds to high polar order for large clusters with $\beta=0$ and weak pullers. 
 The same features can be observed when  $\xi=10$, independently on $r_c$.  
$P(s)$ for WP squirmers vanishes for all cluster sizes, independently 
of $\beta$ and $r_c$ when $\xi=1$, since the formation of trimers and chains assemble the particles head to head, thus it avoids 
the polar order. However, when $\xi=10$, the polar order decays with $s$: for $r_c=1.5\sigma$ weak pullers form clusters with high polar order, 
while   for $r_c=2.5\sigma$ $P(s)$ decays faster for pushers than for 
pullers.  In the latter case, 
 the WP squirmers  show another type of alignment, 
 with particles with a high nematic order in the range of $\beta=0$ and weak pushers.

 During aggregation, we have identified 
 seven different cases depending on whether  the  attractive patch is oriented against the propulsion direction (AP squirmers) 
 or directed towards it  (WP squirmers): 
 3 gas states, 3 clustering cases and 1 coarsening.
 On the one side,  AP squirmers
  coarsen isotropically if the interaction is strong enough  $\xi \approx 0.1$. 
    When the interaction strength competes with   self-propulsion ($\xi =1$), dynamic 
  clusters emerge whose  mean size  depend on  hydrodynamic stresses.  When activity dominates ($\xi =10$), 
  particles form a gas and depending on the value of $\beta$ this gas can be isotropically oriented or polarly oriented. 
  On the other side,  
  WP squirmers  form chains of particles if  the interaction is high enough  ($\xi \approx 0.1$), or a suspension of trimers and tetramers if the
   interaction competes with the active propulsion ($\xi =1$). Whenever activity dominates  ($\xi =10$),
   particles form a gas and depending on the value of $\beta$ this gas can be isotropic, polar or even nematically oriented.

Therefore, on one hand  anisotropy drives the formation of structures of trimers and tetramers for 
 WP squirmers; on the other hand anisotropy  modulates the sensitivity of the hydrodynamic signature: 
 while WP squirmers are more sensitive to $r_c$ when $\xi=10$, 
AP squirmers are more sensitive to the interaction 
range ($r_c$) when $\xi=1$.
To conclude, the rich morphology of the detected squirmers' aggregates is the result of 
the anisotropic interactions, characterized by the angular attractive potential and its interaction range, 
in  competition with the active stress, that can be pointing towards or against the attractive patch.  

\section*{Acknowledgments}
We thank to S.A. Mallory and A. Cacciuto for helpful discussions. This work was possible
thanks to the access to MareNostrum Supercomputer at Barcelona Supercomputing Center (BSC). This article is based upon work from COST Action MP1305, supported by COST (European
Cooperation in Science and Technology). This work was supported by 
FIS2016-78847-P of the MINECO and the UCM/ Santander PR26/16-10B-2. IP acknowledges MINECO and DURSI for financial
support under projects FIS2015-67837- P and 2017 SGR-884, respectively. FA acknowledges
funding from Juan de la Cierva-formaci\'on program.

%
\clearpage
\appendix

\section{\label{Appendix:attra}Attractive part of the potential}

\begin{figure}[!ht] 
\scalebox{0.75}{\includegraphics{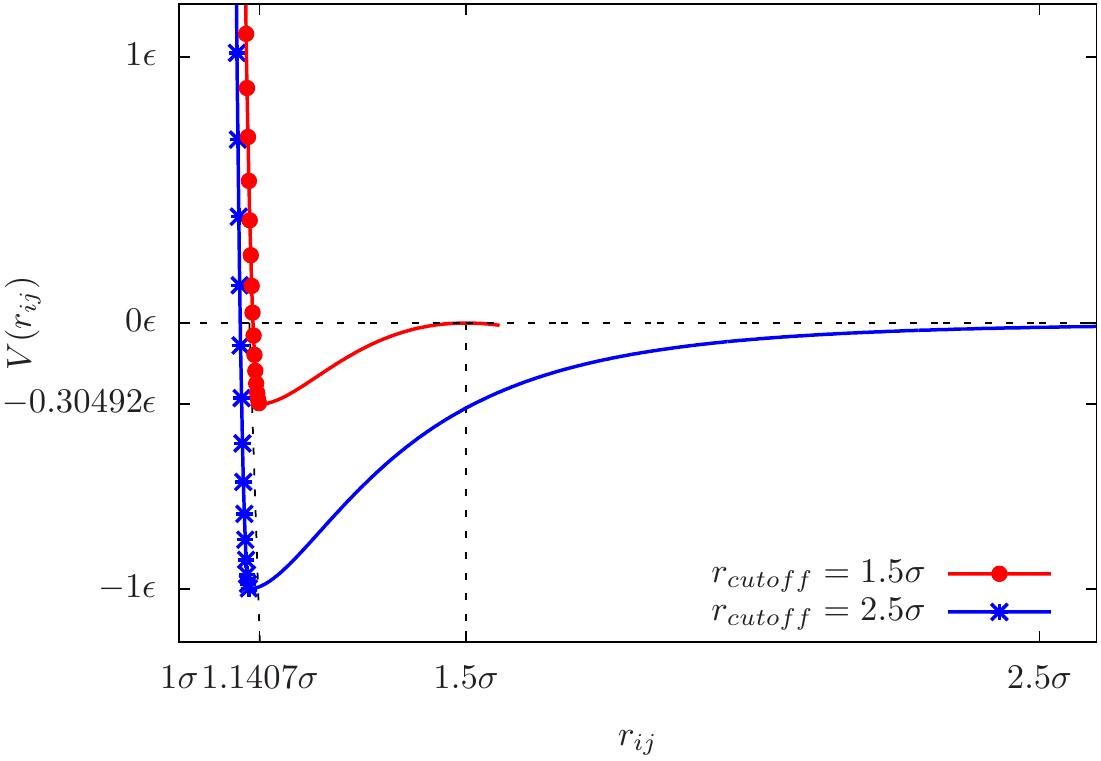}} 
 \caption{Graphic representation of the total potential $V(r_{ij})$ with $\phi=1$, using the Lennard-Jones potential with a smooth transition function 
 given in Eq. \ref{eq:VLJshortrange} as the attractive part of $V(r_{ij})$ (red curve), the repulsive soft sphere potential at short separation distance 
 with $h_0=(2.2^{1/6}-1)\sigma$ (red points) is using with this cut-off potential. The blue curve is using Eq. \ref{eq:VLJmidrange}, with the 
 repulsive soft sphere potential at short separation distance with $h_0=(2^{1/6}-1)\sigma$ (blue asterisks).}\label{Fig:Vsmooth}
\end{figure}

 Fig.~\ref{Fig:Vsmooth} compares the radial part of the total short range potential given by Eq. \ref{eq:VLJshortrange} 
(red curves) and the classical LJ potential Eq. \ref{eq:VLJmidrange} and cut-off distance of $2.5 \sigma$ (blue curves), $\sigma$ is the particles' diameter. Both curves in Fig~\ref{Fig:Vsmooth} are the sum of the repulsive soft-sphere potential (Eq. \ref{eq:SSPotential}) and the respective attractive potential.

\section{\label{Appendix:JanusOrient}Different types of Janus interactions}
There are different models to simulate the amphiphilic behavior of Janus colloids, in this appendix we have compared two different models for the angular potential $\phi$ in equation \ref{eq:JanusGralPotential}. The angular potential used in this manuscript (equation \ref{eq:phi}) has been used to simulate spherical particles with one hydrophobic hemisphere and charged on the other \cite{langmuir08} where both hemispheres have the same size, thus lets call it symmetric potential $\phi_s(\theta_i,\theta_j)$.

\begin{figure}[t]
\scalebox{0.70}{\includegraphics{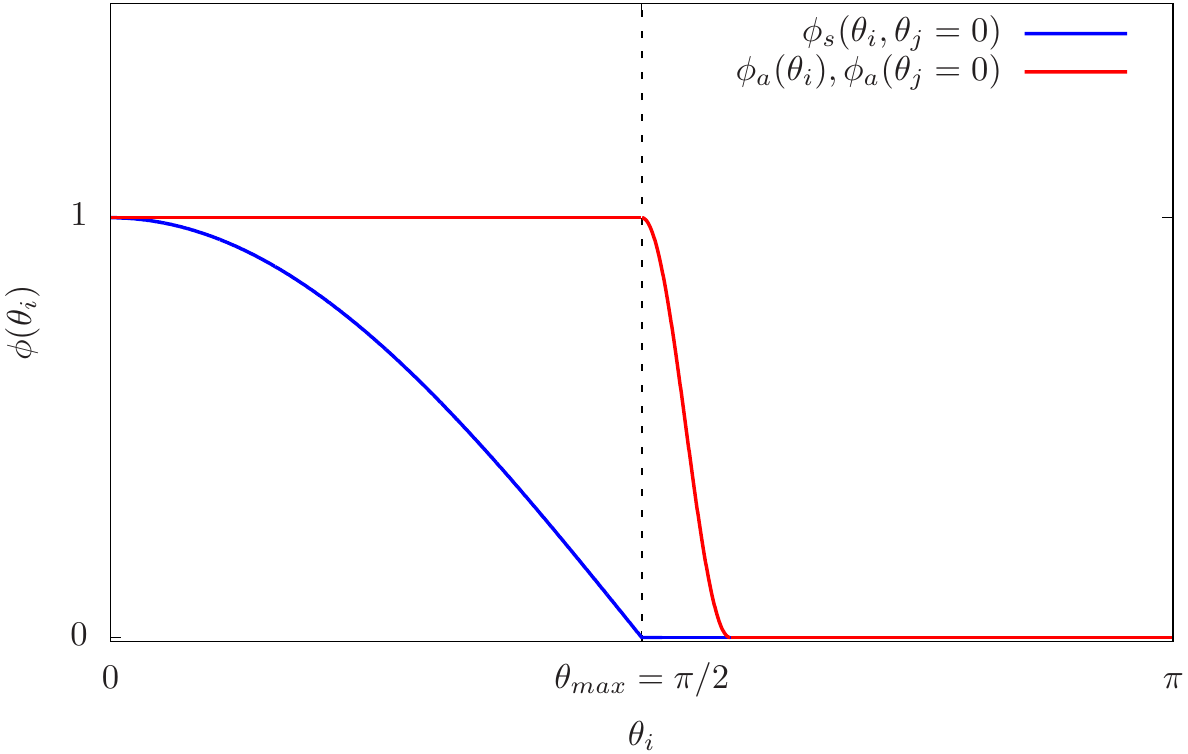}}
\caption{\label{fig:TwoJanusOrientations} Graphical representation of the angular dependence of two different Janus potentials. In blue the symmetric potential used in our simulations given by equation \ref{eq:phi}, while
in red the asymetric potential used in \cite{pre09, Stewart_17}  and depicted in equation \ref{eq:asymetric_orient}. In order to compare both angular functions, we set $\theta_j=0$ in both cases and $\theta_{max}=\pi/2$ which corresponds to the symmetric case.}
\end{figure}

On the other hand, in reference \cite{pre09}, they developed an asymmetric potential, where the size of the hydrophobic region is tuned, in order to study 3D Janus colloid suspensions in equilibrium and more recently, some of us have used this potential to study the self-assembly of Active amphiphilic Janus particles with attractive patches of different sizes \cite{Stewart_17}. Analogously to the symmetric potential above, the 
the orientation depending term is given by
\begin{eqnarray}\label{eq:asymetric_orient}
\begin{split}
&\phi_a \left(\theta_i \right) = \\
&\left\{ \begin{array}{lc}
1  & \mbox{$\theta_i \leq \theta_{max}$}\\
\cos ^2 \left( \frac{\pi \left(\theta_i - \theta_{max} \right)}{2 \theta_{tail}} \right)  & \mbox{$\theta_{max} \leq \theta_i \leq \theta_{max} + \theta_{tail} $}\\
0 & \mbox{otherwise.} \end{array} \right.
\end{split}
\end{eqnarray}

The orientational part of the potential $\phi_a$, is a a smooth step function that modulates the angular dependence of the potential, the smoothness is modulated by the parameter $\theta_{tail}$, which it is set to $\theta_{tail}=0.2618$ in \cite{pre09} for 3D Janus colloids while in \cite{Stewart_17} was tuned in $\theta_{tail}=0.436$  to generate a sufficiently smooth potential at the Janus interface and $\theta_{max}$ tunes the size of the hydrophobic region, for a patch coverage of 50$\%$ then $\theta_{max} = \pi/2$.

By the analysis of the curves shown in Fig. \ref{fig:TwoJanusOrientations}, it is clear that interactions among the Janus particles will depend on the angular function we choose, despite of the fact that $\theta_{max}=\pi/2$ in the asymmetric potential.

 In terms of the interaction strength, it is clear that $\phi_s$ will give us a lower attraction than $\phi_a$ for any direction of $\theta_i$. An important remark is that $\phi_s=0$ when two particles are parallel while $\phi_a \neq 0$ in the same configuration, therefore $\phi_a$ enhance the emergence of chain-like aggregates \cite{Stewart_17}. 
 
 To understand the relevance of the orientational part in the potential and the hydrodynamic interactions, we are working in a model of amphiphilic squirmers using $\phi_a$ in order to compare with the 'dry' case of Ref. \cite{Stewart_17}, this work is in progress.

\section{\label{Appendix:Clusters} Identifying clusters in the suspension}
In order to determine the criterion to characterize whether a particle belongs or not to a cluster, we computed first the radial distribution function of our swimmer suspensions.

\begin{figure}[t]
\includegraphics[width=0.48\columnwidth]{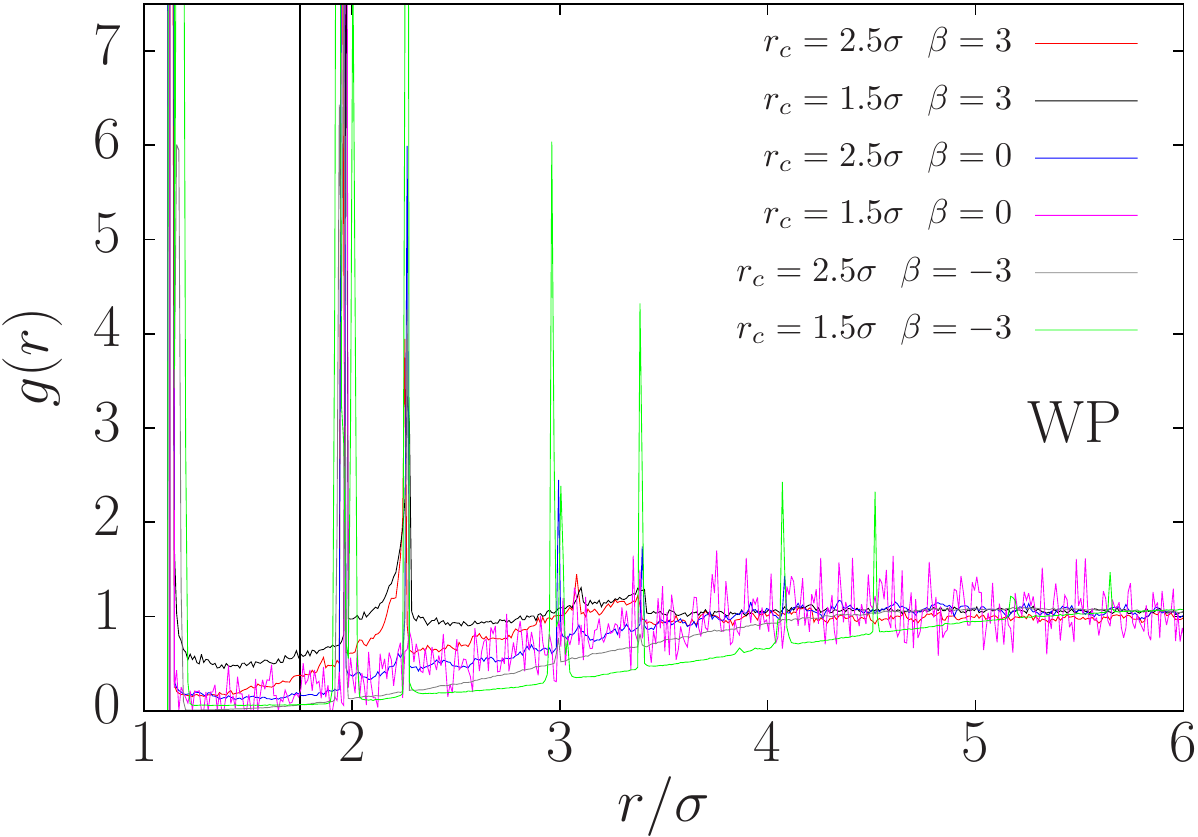}
\includegraphics[width=0.48\columnwidth]{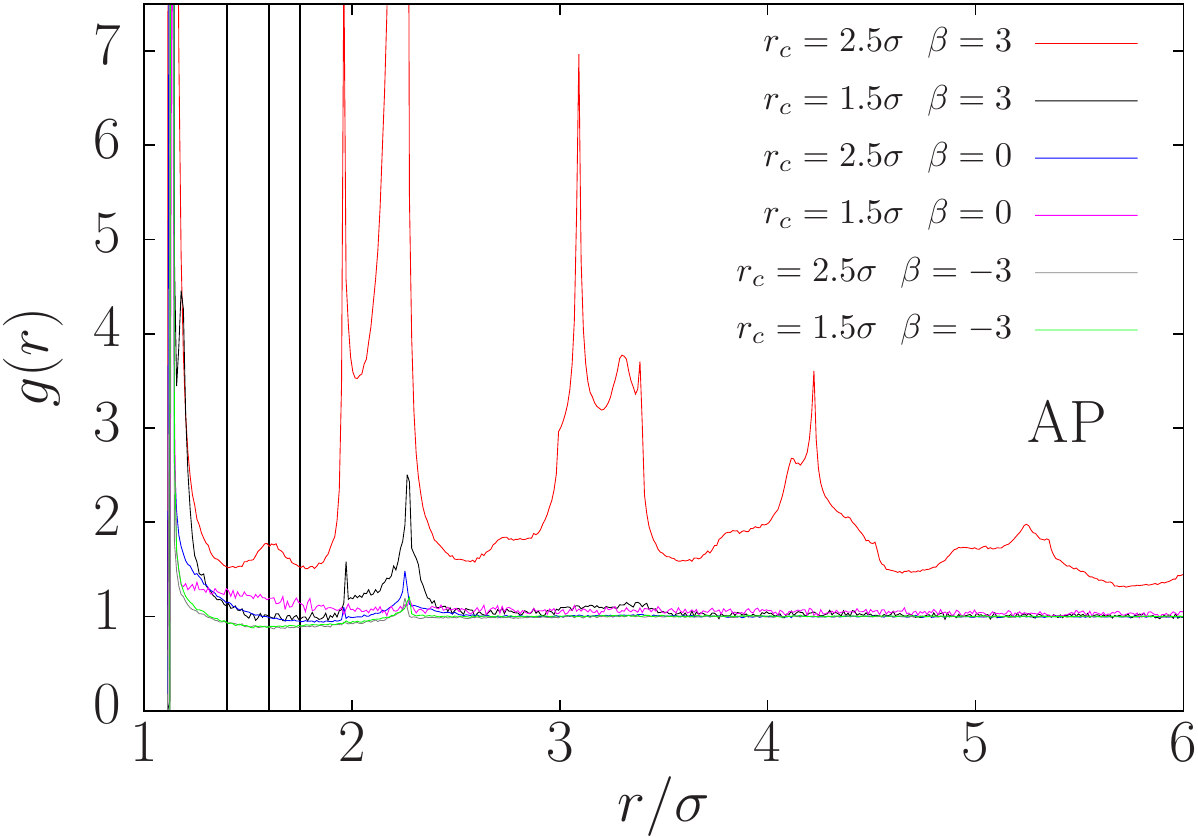}
\caption{Radial distribution functions $g(r)$ with $\xi=1$ for aligned with the patch interaction(left) and aligned against the patch interaction (right).}
\end{figure}

We observe a local peak at $r=1.6\sigma$ for the $g(r)$ corresponding to $\beta=3$ with $r_c=2.5\sigma$ with AP interaction. We took $r_{cl}=1.75\sigma$ for all cases, although for this case it is evident the first minimum cannot be located there. We analyze how the mean cluster size and the cluster size distribution vary for this troublesome case with the choice of $r_{cl}$, comparing $r_{cl}=1.75\sigma$ to a cutoff in the peak $r_{cl}=1.6\sigma$ and a cutoff lying in the first valley $r_{cl}=1.4\sigma$.

\begin{figure}[t]
\includegraphics[width=0.48\columnwidth]{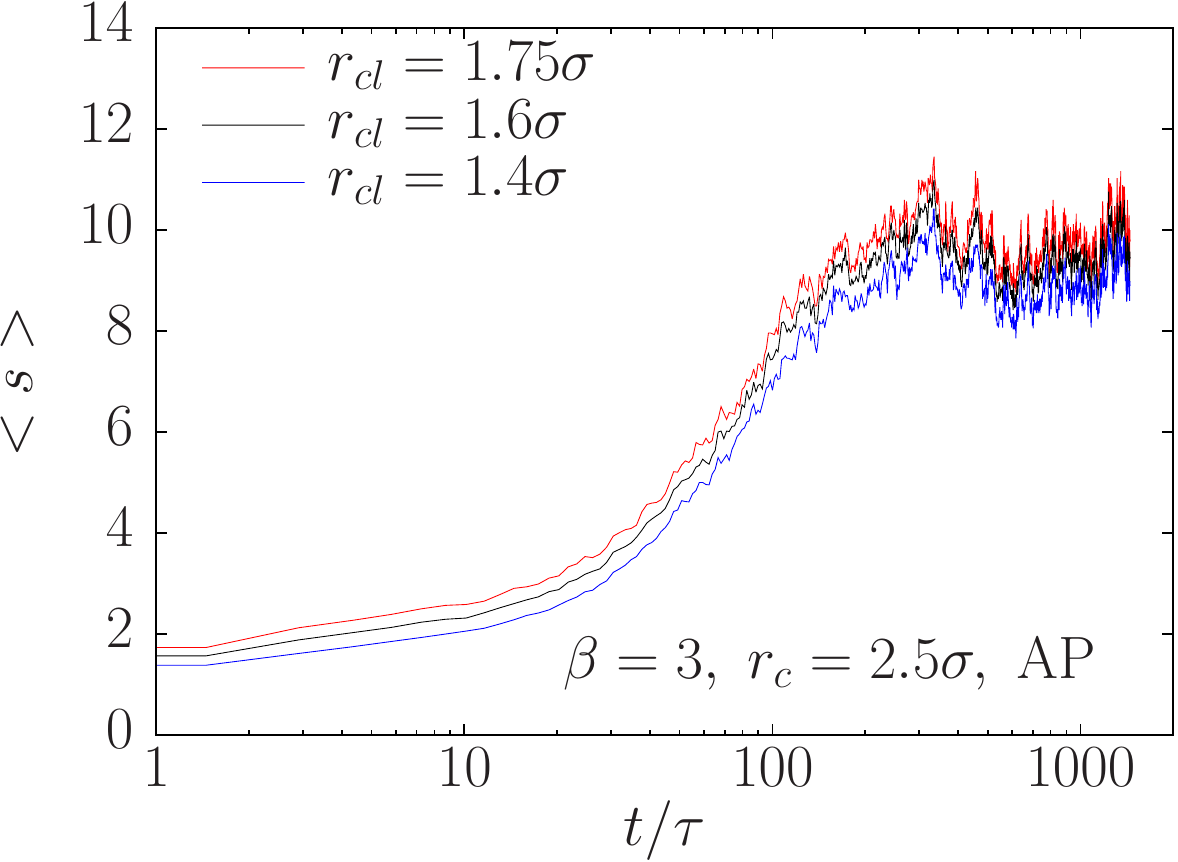}
\includegraphics[width=0.48\columnwidth]{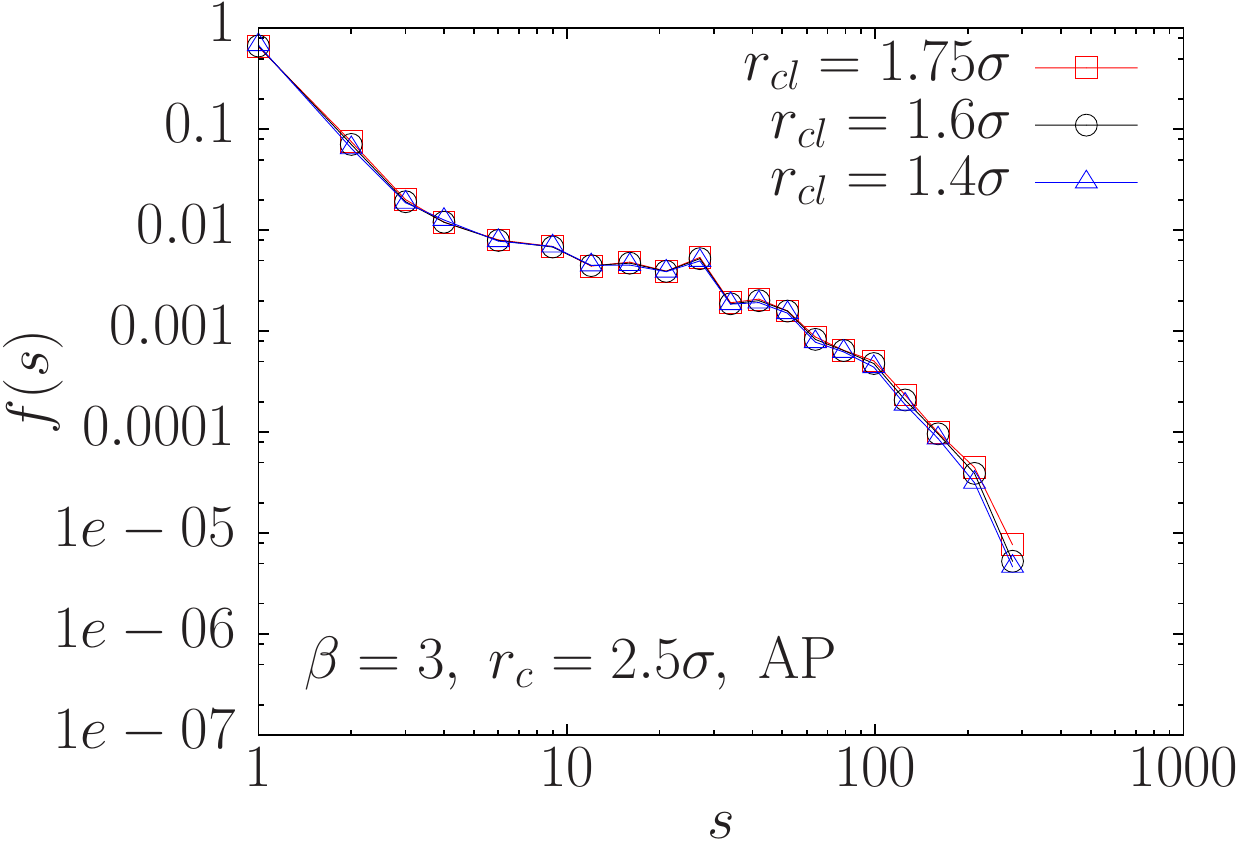}
\caption{Left: time evolution of the mean cluster size. Right: cluster size distribution.}
\end{figure}

We observe that as we reduce $r_{cl}$, the mean cluster size is reduced, being the difference between $r_{cl}=1.75\sigma$ and $r_{cl}=1.4\sigma$ of 0.9 units ($\approx 10\%$). As for the cluster size distribution, we find that all three of them are very similar. Even if the frequencies for the largest observed clusters are reduced with $r_{cl}$, all distributions are cut at the same bin (clusters sized between 240 and 318 particles), with a frequency $7.7 \cdot 10^{-6}$ for $s=279$ and $r_{cl}=1.75\sigma$, and $4.6 \cdot 10^{-6}$ for the same size with $r_{cl}=1.4\sigma$. The monomer frequency goes from 0.680 for $r_{cl}=1.4\sigma$ to 0.656 for $r_{cl}=1.75\sigma$, and all three curves have a very similar shape, sharing even local  maxima and minima along the whole profile.

\section{\label{Appendix:xi0_1WPrc1_5} Clustering analysis for WP Janus with $\xi=0.1$.}
\begin{figure}[h!]
\centering
\includegraphics[width=0.45\columnwidth]{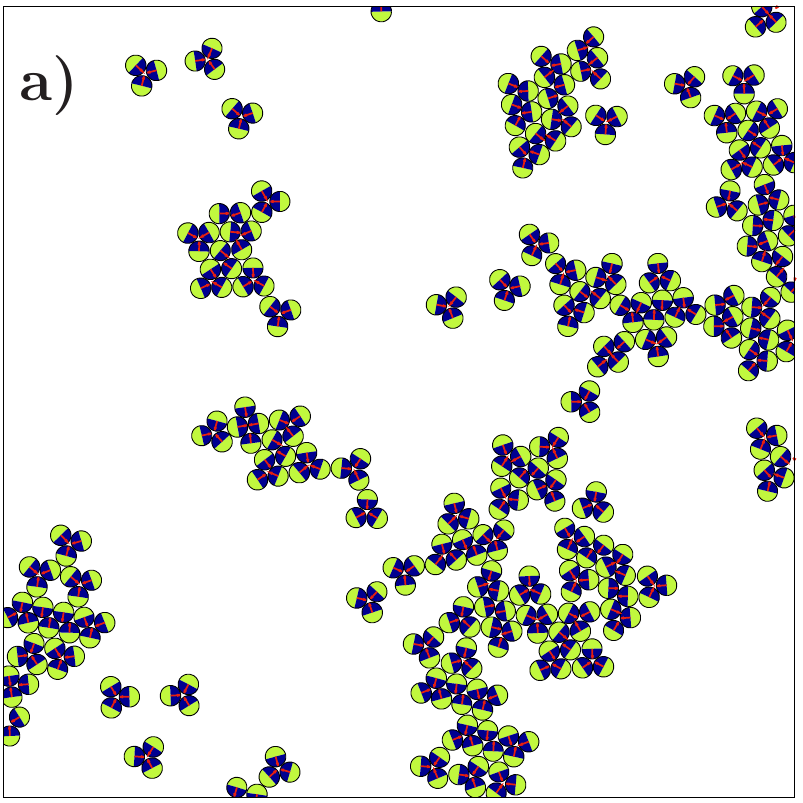} \includegraphics[width=0.45\columnwidth]{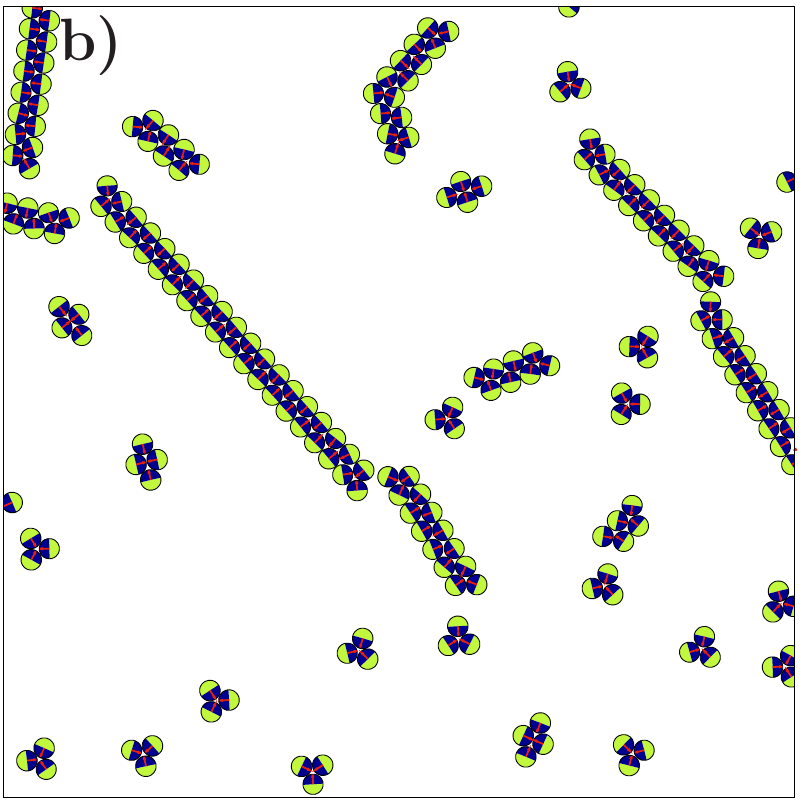}
\caption{\label{Samples_rc2_5xi0_1} Simulation zoomed snapshots of WP squirmers with $\xi=0.1$ and $r_c=2.5 \sigma$ snapshots are taken at the last timestep of the simulations. The attractive hemisphere is represented in blue, while pure repulsive in green and the fixed orientation vector is shown in red. (a) $\beta=3$ and (b) $\beta=-3$.}
\end{figure}

Janus squirmers with interaction strength $\xi=0.1$ develop very interesting dynamics, on one hand, for strong pullers with $\beta=3$, particles form mainly trimers and tetramers that eventually aggregate due to hydrodynamic interactions (panel a of Fig.~\ref{Samples_rc2_5xi0_1} and Fig.~\ref{Samples_xi0_1}). The other case is the chains system, where particles self-assembly in chains of different sizes, such chains are form by two lines of squirmers with the attractive patch pointing towards the other line of squirmers that form the chain (panel b of Fig.~\ref{Samples_rc2_5xi0_1} and Fig.~\ref{Samples_xi0_1}). The chain-length depends on the value of $\beta$. This chains system has been observed for pushers and weak pullers as well.

\begin{figure}[h!]
\centering
\includegraphics[width=0.48\columnwidth]{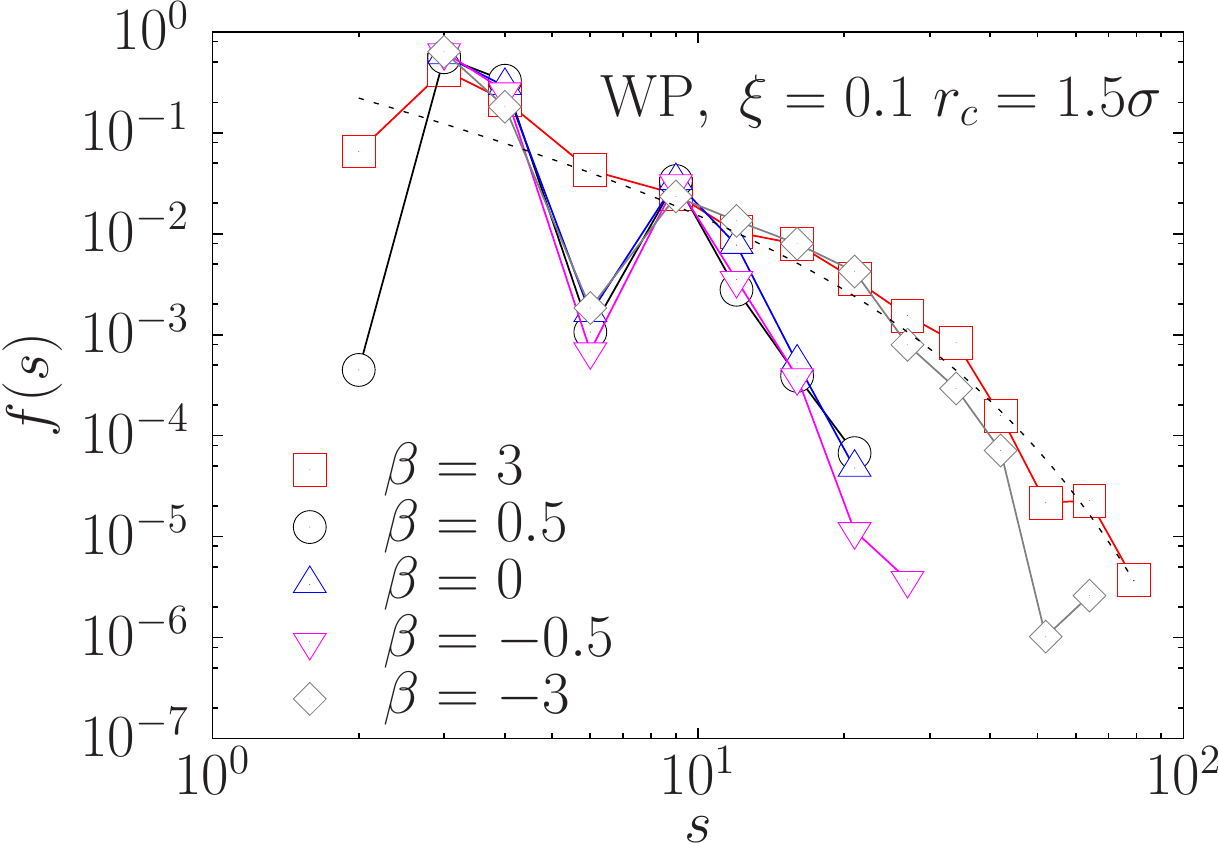} \includegraphics[width=0.48\columnwidth]{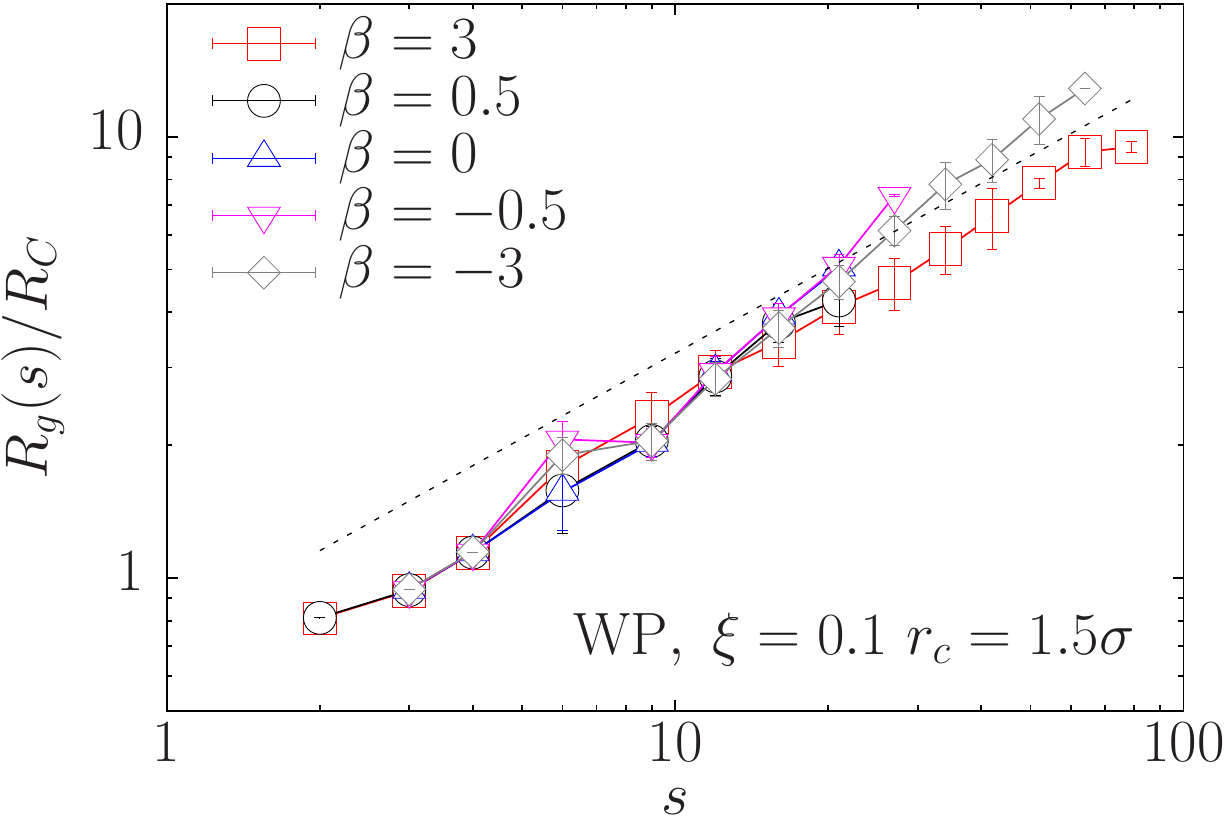}
\includegraphics[width=0.48\columnwidth]{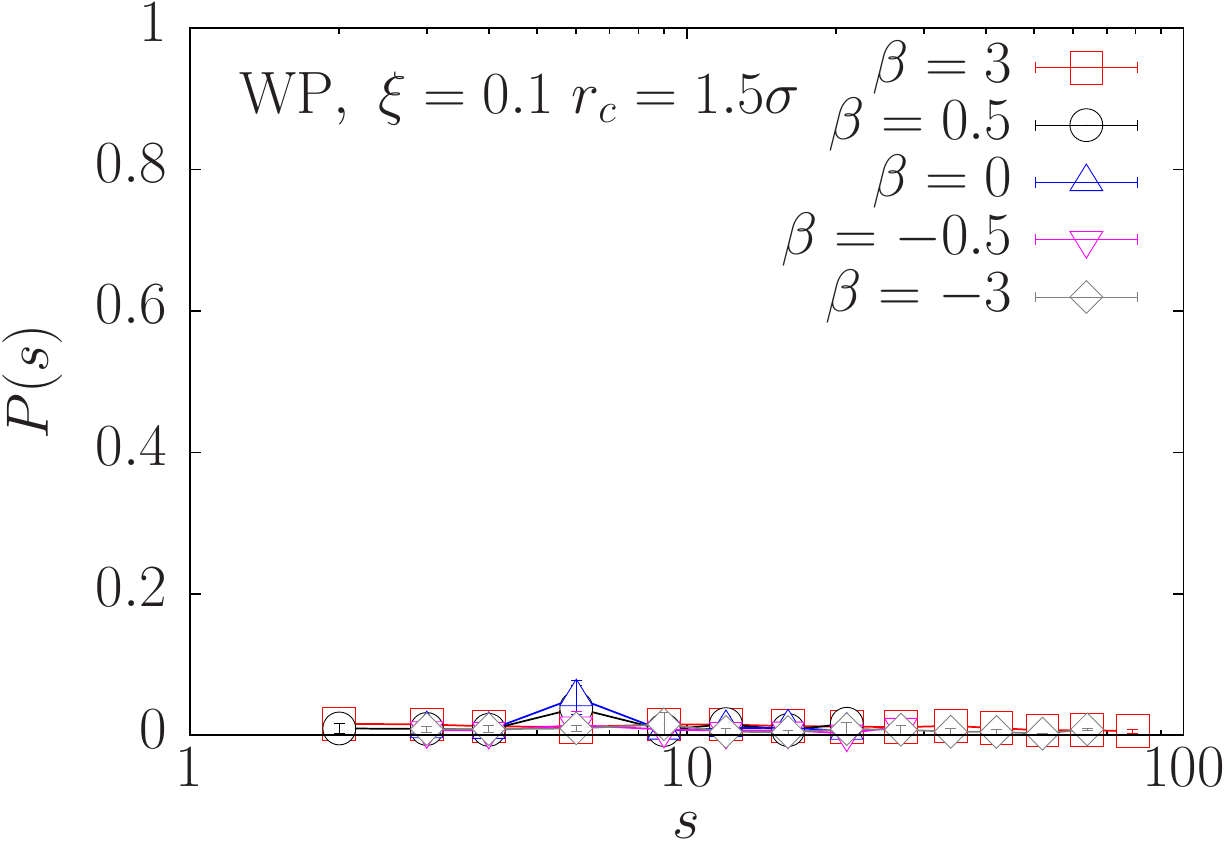} ~ \includegraphics[width=0.48\columnwidth]{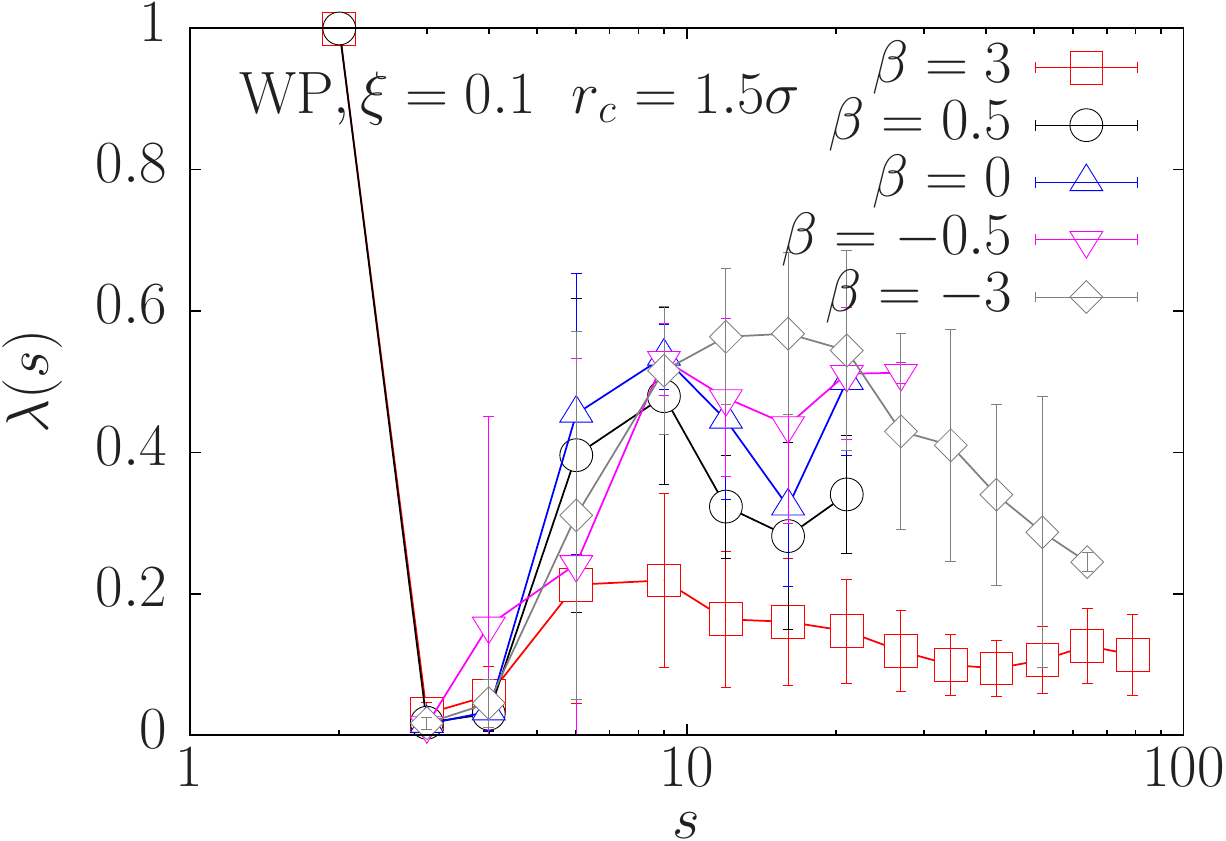}
\caption{\label{fig:clusters_xi0_1WP} Cluster analysis of WP squirmers with $\xi=0.1$ and $r_c=1.5\sigma$. (a) Cluster size distribution. (b) Radius of gyration as a function of cluster size. (c) Local polar order and (d) local nematic order parameters. Pullers curves are red and black, while pushers are pink and gray and $\beta=0$ in blue. Note the different scale on the y-axis.}
\end{figure}

When interaction range is $r_c=2.5\sigma$, the clusters either trimers or chains, keep growing slowly, as far as our calculation capacity could reach, we have observed coarsening for all $\beta$. In contrast with WP squirmers with interaction range of $r_c=1.5\sigma$. In this case WP squirmer suspensions reach a clustering state for all hydrodynamic signature $\beta$. We have found two general types of clustering: 

\begin{figure}[h!]
\centering
\includegraphics[width=0.45\columnwidth]{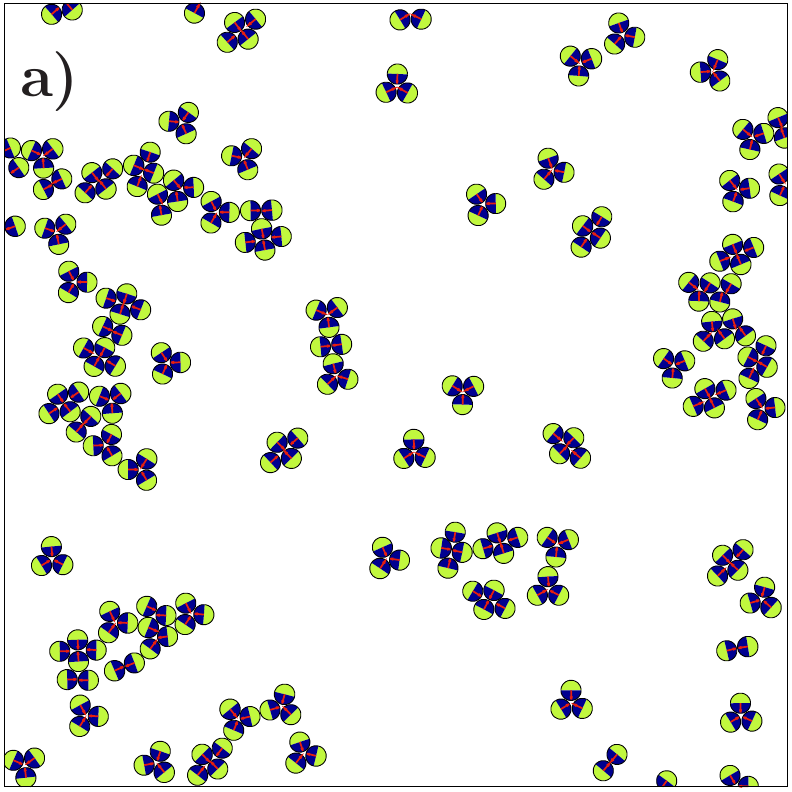} \includegraphics[width=0.45\columnwidth]{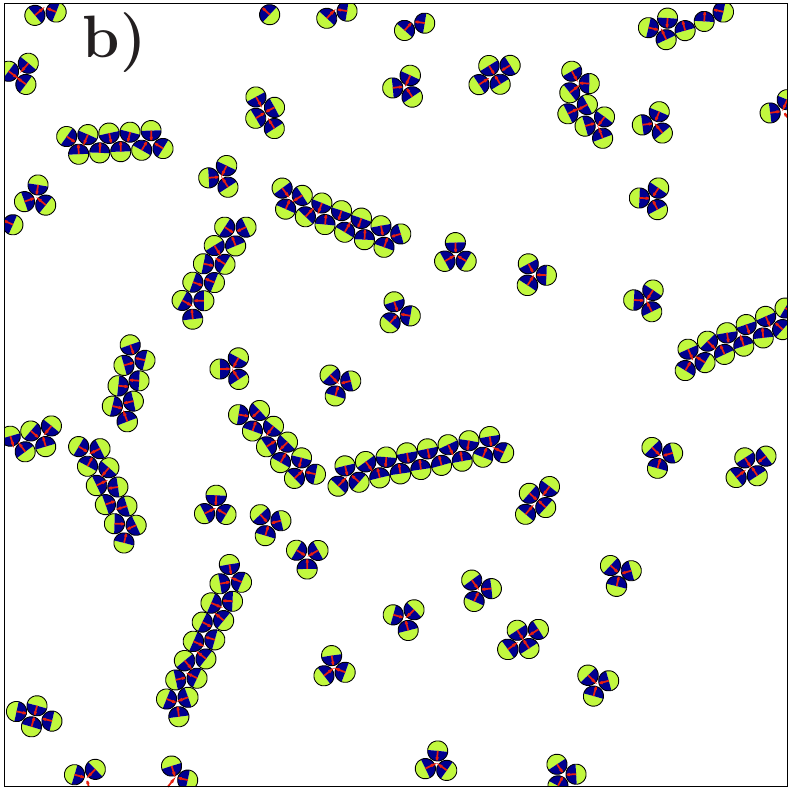}
\caption{\label{Samples_xi0_1} Simulation zoomed snapshots of WP squirmers with $\xi=0.1$ and $r_c=1.5 \sigma$ snapshots are taken once the simulations reach a steady state given by the mean cluster size. The attractive hemisphere is represented in blue, while pure repulsive in green and the fixed orientation vector is shown in red. (a) $\beta=3$ and (b) $\beta=-3$.}
\end{figure}
We have characterized the clusters observed for these WP squirmers. In (panel a in Fig. \ref{fig:clusters_xi0_1WP}) we show the cluster size distribution $f(s)$. In general, $f(s)$ has two main peaks, one for trimers and another one in the bin of cluster-size around 10 particles for pushers and weak pullers, while strong pullers (red squares) have just one peak for trimers. Moreover, none of these distributions show monomers and just pullers show dimers in their distributions. Another interesting feature of the cluster size distributions are the fact that both strong pushers and pullers have bigger clusters than the rest of squirmers (gray diamonds and red squares respectively) and both distributions follow the analytical shape of equation (\ref{eq:csd_cutoff}) (dotted line) for large clusters.

Even though that both cluster size distributions are similar for strong pushers and pullers, the morphology of such clusters are not the same. As we can observe in Fig. \ref{fig:clusters_xi0_1WP}-panel b, where despite that the radius of gyration for small clusters ($s \leq 10$) follows the same behavior for all $\beta$, the large clusters for strong pullers (red squares) are more compact (small slope) than clusters of the rest of squirmers, in fact, clusters of pushers and weak pullers are chains, thus we can observe a greater slope for large clusters (gray diamonds and pink triangles).

We have also calculated both polar and nematic order parameters at each cluster, we have observed that clusters have no polar order at any value of $\beta$ (see panel c of Fig. \ref{fig:clusters_xi0_1WP}), but an alignment, given by a non-zero nematic order is observed (see panel d of Fig. \ref{fig:clusters_xi0_1WP}). For pullers, all dimers are aligned by a head-to-head configuration and then the nematic order is 1, particles in trimers and tetramers in all cases are pointing out to the center of the clusters therefore, the nematic order is zero. But, in the case of pushers to weak pullers all clusters have a large non-zero nematic order, since particles are forming chains of two lines of particles aligned to the center of the chain giving rise to a nematic order. In contrast, for strong pullers (red squares) large clusters are formed by trimers gather together by hydrodynamic interactions, thus particles are pointing in different directions and nematic order parameter is reduced. In fact, the magnitude of nematic order parameter for large clusters is given by the $\beta$ value. The lower the $\beta$ the greater the nematic order.

\end{document}